\documentclass[12pt,twoside,letterpaper]{article}
\usepackage{epsfig,dcolumn}

\usepackage{color}
\usepackage{eso-pic}

\newcommand{\met}{E\!\!\!/_T}
\newcommand{\ptpartjet} {p_T^{particle}}
\newcommand{\ptcaljet} {p_T^{jet}}
\newcommand{\ptpartjeti} {p_{T,i}^{particle}}

\newcommand{\ptparton} {p_{T}^{parton}}

\newcommand{\gevcc}[1]{\ensuremath{#1~\mathrm{GeV}/c^{2}}}

\newcommand{\nt}{0-tag}
\newcommand{\st}{1-tag}
\newcommand{\stl}{1-tag(L)}
\newcommand{\stt}{1-tag(T)}
\newcommand{\dt}{2-tag}

\newcommand{\pb}{pb$^{-1}$}

\newcommand{\ttb}{t\bar{t}}

\newcommand{\mtop}{M_{top}}

\newcommand{\wjj}{W\to jj}

\newcommand{\mjj}{m_{jj}}

\newcommand{\mtopdef}{\ensuremath{\mtop} = \gevcc{178}}

\def\D0{D\0}

\newcommand{\rd} {{\rm d}}

\begin{document}
\lefthyphenmin=2
\righthyphenmin=3


\vspace{0.5in}

\begin{center}
  \begin{Large}
  {\bf Determination of the Jet Energy Scale at the Collider Detector at Fermilab}   
  \end{Large}
\end{center}

\font\eightit=cmti8
\def\r#1{\ignorespaces $^{#1}$}
\hfilneg
\begin{sloppypar}
\begin{center}
A. Bhatti\r {19}, F. Canelli\r {2}, B. Heinemann\r {10}\\
\vspace{0.4cm}
\end{center}
\end{sloppypar}
J. Adelman\r {4}, D. Ambrose\r {16},  J.-F. Arguin\r {13}, A. Barbaro-Galtieri\r {9}, H. Budd\r {18}, Y.S. Chung\r {18}, K. Chung\r {3}, B. Cooper\r {11}, C. Currat\r {9}, M. D'Onofrio\r {8}, T. Dorigo\r {14}, R. Erbacher\r{5},  R. Field\r {6}, G. Flanagan\r {12}, A. Gibson\r {9}, K. Hatakeyama\r {19}, F. Happacher\r {7}, D. Hoffman\r{4}, G. Introzzi\r{15}, S. Kuhlmann\r {1}, S. Kwang\r {4}, S. Jun\r {3}, G. Latino\r {17},  A. Malkus\r{4}, M. Mattson\r {20}, A. Mehta\r {10}, P.A. Movilla-Fernandez\r {9}, L. Nodulman\r {1}, M. Paulini\r {3}, J. Proudfoot\r {1}, F. Ptohos\r {7}, S. Sabik\r {13}, W. Sakumoto\r {18}, P. Savard\r {13}, M. Shochet\r {4}, P. Sinervo\r{13}, V. Tiwari\r {3}, A. Wicklund\r {1}, G. Yun \r {18}

\vskip .026in

\begin{center}
\r {1}  {\eightit Argonne National Laboratory, Argonne, Illinois 60439} \\
\r {2}  {\eightit University of California at Los Angeles, Los Angeles, California  90024} \\
\r {3} {\eightit Carnegie Mellon University, Pittsburgh, Pennsylvania  15213} \\
\r {4} {\eightit Enrico Fermi Institute, University of Chicago, Chicago, Illinois 60637} \\
\r {5} {\eightit Fermi National Accelerator Laboratory, Batavia, Illinois 60510} \\
\r {6} {\eightit University of Florida, Gainesville, Florida  32611} \\
\r {7} {\eightit Laboratori Nazionali di Frascati, Istituto Nazionale di Fisica
               Nucleare, I-00044 Frascati, Italy} \\
\r {8} {\eightit University of Geneva, CH-1211 Geneva 4, Switzerland} \\
\r {9} {\eightit Ernest Orlando Lawrence Berkeley National Laboratory, Berkeley, California 94720} \\
\r {10} {\eightit University of Liverpool, Liverpool L69 7ZE, United Kingdom} \\
\r {11} {\eightit University College London, London WC1E 6BT, United Kingdom} \\
\r {12} {\eightit Michigan State University, East Lansing, Michigan  48824} \\
\r {13} {\eightit University of Toronto, Toronto, Canada M5S~1A7} \\
\r {14} {\eightit University of Padova, Istituto Nazionale di Fisica 
          Nucleare, Sezione di Padova-Trento, I-35131 Padova, Italy} \\
\r {15}{\eightit University of Pavia, Istituto Nazionale di Fisica Nucleare, Sezione di Pavia, I-27100 Pavia, Italy} \\
\r {16} {\eightit University of Pennsylvania, Philadelphia, Pennsylvania 19104} \\   
\r {17} {\eightit Istituto Nazionale di Fisica Nucleare Pisa, University of Pisa, Siena and Scuola
               Normale Superiore of Pisa, I-56127 Pisa, Italy} \\
\r {18} {\eightit University of Rochester, Rochester, New York 14627} \\
\r {19} {\eightit The Rockefeller University, New York, New York 10021} \\
\r {20} {\eightit Wayne State University, Detroit, Michigan  48201} \\
\end{center}

\newpage

\begin{abstract}

A precise determination of the energy scale of jets at the Collider
Detector at Fermilab at the Tevatron $p\bar{p}$ collider is
described. Jets are used in many analyses to estimate the energies of
partons resulting from the underlying physics process. Several
correction factors are developed to estimate the original parton
energy from the observed jet energy in the calorimeter. The jet energy
response is compared between data and Monte Carlo simulation for
various physics processes, and systematic uncertainties on the jet
energy scale are determined. For jets with transverse momenta above
50~GeV the jet energy scale is determined with a $3\%$ systematic
uncertainty.

\end{abstract}

\thispagestyle{empty}

\clearpage
\thispagestyle{empty}
\tableofcontents

\thispagestyle{empty}

\clearpage


\section{\label{Intro}Introduction}
\pagenumbering{arabic}

Measurements of hard scattering processes in $p\bar{p}$ collisions
often depend on the determination of the four-momenta of quarks and
gluons produced in the hard scatter. The measurement of these
four-momenta relies on the reconstruction of hadronic jets, resulting
from the quark or gluon fragmentation.

At the Collider Detector at Fermilab (CDF) jets are observed as
clustered energy depositions in the calorimeters. In this article we
describe how these jets are then corrected to correspond to the energy
of the parent parton. The precision to which this can be achieved
determines the precision of many measurements, e.g. a 1\% uncertainty
on the energy scale of jets results in an uncertainty of 10 \% on the
cross section for jet production at transverse momenta of 500 GeV/$c$
\cite{cdfjet} and in a $1$ GeV/$c^2$ uncertainty on the top quark mass
\cite{cdftop}.

The original parton transverse energy can be estimated by correcting
the jet for instrumental effects and for radiation and fragmentation
effects:
\begin{equation}
\ptparton=(\ptcaljet \times C_{\eta} - C_{MI}) \times C_{Abs} - C_{UE} + C_{OOC}=\ptpartjet - C_{UE} + C_{OOC}
\end{equation}
where $\ptparton$ is the transverse momentum of the parent parton the
procedure is aimed at, $\ptcaljet$ is the transverse momentum measured
in the calorimeter jet, $\ptpartjet$ is the transverse momentum of the
particle jet, that is, a jet corrected by all instrumental effects
which corresponds to the sum of the momenta of the hadrons,
leptons, and photons within the jet cone, and
\begin{itemize}

\item $C_{\eta}$, ``$\eta$-dependent'' correction, ensures homogeneous
response over the entire angular range;

\item $C_{MI}$, ``Multiple Interaction'' correction, is the energy to
subtract from the jet due to pile-up of multiple $p\bar{p}$
interactions in the same bunch crossing;

\item $C_{Abs}$, ``Absolute'' correction, is the correction of the
calorimeter response to the momentum of the particle jet. Particle
jets can be compared directly to data from other experiments or
theoretical predictions which include parton radiation and
hadronization.

\item $C_{UE}$ and $C_{OOC}$, the ``Underlying Event'' and
``Out-Of-Cone'' corrections, correct for parton radiation and
hadronization effects due to the finite size of the jet cone algorithm
that is used. Note that these corrections are independent of the
experimental setup, i.e. the CDF detector environment.
\end{itemize}

These corrections and their systematic uncertainties will be described
below and in the following sections. All the correction factors are
determined as a function of the jet transverse momentum but they apply
to all components of the four-momentum of the jet.

The $C_{Abs}$ correction is derived using a detailed Monte Carlo (MC)
simulation of the physics processes and the detector response. The
corrections $C_{UE}$ and $C_{OOC}$ are determined using the {\tt
PYTHIA} MC generator. Thus the major task is the tuning and validation
of the detector simulation as well as of the physics modeling used in
the simulation. The other corrections are mostly derived directly from
data but are also compared to the simulation.

The following ingredients are necessary for deriving the corrections
described above:

\begin{itemize}

\item The energy scale for the electromagnetic calorimeter is set
using electrons from the decay $Z\rightarrow e^+e^-$. The energy scale
for the hadronic calorimeter is set to the test-beam scale of
$50$~GeV/$c$ charged pions.  Section \ref{sec:Detector} describes
the CDF detector and the definition of the calorimeter energy scales.

\item Jets are defined using a cone algorithm either on calorimeter
towers, on stable particles, or on partons in the MC.  Different cone
sizes are studied and all the corrections are derived for the specific
cone size. The details of the algorithm used for defining a jet are
given in Sec. \ref{sec:JetAlgo}.

\item Many datasets are used for either developing a correction
procedure, or for validating or tuning the MC simulation. These are
described in Sec. \ref{sec:Datasets}.

\item Since the simulation is used to correlate a particle jet to a
calorimeter jet a detailed understanding of the detector simulation is
needed. Therefore the simulation is tuned to model the response of the
calorimeter to single particles by comparing the calorimeter energy
measurement, $E$, to the particle momentum, $p$, measured in tracking
detectors.  Here, measurements based on both test beam and CDF data
taken during Run II are used. The details of the simulation are given
in Sec. \ref{sec:Simulation}.

\item The calorimeter simulation is most reliable in the central part
of the calorimeters since the tracking coverage in the forward regions
is limited. Therefore, the forward calorimeter jet response is
calibrated with respect to the central, to flatten out the jet
response versus the jet polar angle. This procedure also corrects for
the lower response in poorly instrumented regions of the
calorimeters. The $\eta$-dependent correction, $C_{\eta}$, is
described in Sec. \ref{sec:RelCorr}.

\item After tuning the simulation to the individual particles response
and achieving a jet response independent of the polar angle,
calorimeter jets are corrected to a particle jet, i.e. they are
corrected for the central calorimeter response. The absolute
correction, $C_{Abs}$, is derived from the simulation and described in
Sec.~\ref{sec:Absolute}.  Since the correction is derived from
simulation, it is also important to ensure that the multiplicity and
momentum spectrum of particles in the data is well reproduced by the
simulation. This is also presented in Sec.~\ref{sec:Absolute}.

\item A further correction is made for pile-up of additional
$p\bar{p}$ interactions. This pile-up can lead to an overestimate of
the jet energy if particles produced in the additional interactions
happen to overlap those produced in the hard scattering process. A
correction, $C_{MI}$, for this is derived from data and described in
Sec. \ref{sec:mi}.

\item The jet energy needs to be corrected for particles from the underlying
event, i.e.  interactions from spectator quarks and initial state QCD
radiation, since the measurement aims at estimating the parton
energy. This correction, $C_{UE}$, is described in Sec.
\ref{sec:ooc}.

\item Since the jet cone is of finite size some particles originating
from the initial parton may escape from the jet cone either in the
fragmentation process or due to parton radiation. The out-of-cone
energy, $C_{OOC}$, is measured in MC simulated events and
compared to the data. This correction is described in
Sec. \ref{sec:ooc}.

\item Various cross-checks using different physics processes are
presented to validate the universality of the procedure and verify the
systematic uncertainties. These are presented in Sec.
\ref{sec:crosschecks}.

\item A summary of the systematic uncertainties is given in
Sec. \ref{sec:sumsyst}.  They take into account any differences
observed between the data and the simulation and possible systematic
biases in the procedure used to determine the corrections.

\end{itemize}

In Sec. \ref{sec:Conclusions} we present conclusions and an outlook of
future possible improvements on this correction procedure and the
systematic uncertainties.

\clearpage

\section{CDF Detector}\label{sec:Detector}

The CDF Run II detector has been described in detail elsewhere
\cite{CDF_run2}.  In the following we use a cylindrical coordinate
system with the origin at the center of the detector where the $z$
axis points along the beam pipe in which $\theta$ is the polar angle,
$\phi$ is the azimuthal angle and $\eta=-\ln \tan(\theta/2)$ is the
pseudo-rapidity.  The transverse energy, $E_T$, is defined as
$E\sin\theta$ and the transverse momentum, $p_T$, as $p \sin \theta$,
where $E$ is the energy measured by the calorimeter and $p$ the
momentum measured in the tracking system.  The imbalance in transverse
energy, $\met$, is the magnitude of $\vec{\met}$ with
$\vec{\met}=-\sum_i E_T^i \vec{n_i}$, where $\vec{n_i}$ is a unit
vector that points from the interaction vertex to the $i$th
calorimeter tower in the transverse plane.

Transverse momenta of charged particles ($p_T$) are measured by an
eight-layer silicon strip detector~\cite{SVX} and a 96-layer drift
chamber inside a 1.4 Tesla magnetic field.  The innermost layer of the
silicon detector is located on the beam pipe at a radius of $1.5$~cm,
with the outermost layer located at $28$ cm.  The silicon detector
provides tracking in the pseudo-rapidity region $|\eta|\leq 2$, with
partial coverage up to $|\eta|<2.8$.  Outside of the silicon detector,
the Central Outer Tracker (COT)~\cite{COT} is a $3.1$ m long,
open-cell drift chamber with an active tracking region extending
radially from $41$~cm to $137$~cm. The COT's 96 layers are divided
into super-layers of 12 wires each that alternate between axial and
stereo orientation.  The COT provides coverage for $|\eta|\leq 1$. The
efficiency for finding charged particle tracks is close to 100\% for
$|\eta|<1$ and falls to about 40\% for $|\eta|\approx 2$.  The
momentum resolution is $\sigma(p_T)/p_T = 0.15\% \times p_T$ for
$|\eta|\leq 1.0$ and degrades with increasing $|\eta|$.

Located outside the solenoid, a segmented sampling calorimeter is
installed for the measurement of the electromagnetic and the hadronic
energy depositions, which is described in detail in
Sec. \ref{sec:calo}. The central and forward part of the calorimeter
have their own shower profile detector positioned at the expected
maximum of the lateral shower profile, the Central Electromagnetic
Showermax (CES \cite{ces}) and the Plug Electromagnetic Showermax (PES
\cite{pes}) detectors. Located at the inner face of the central
calorimeter, the Central Pre-Radiator (CPR \cite{cpr}) chambers use
the solenoid coil as a radiator to measure the shower development.
These three detectors are mainly used for photon and electron
identification.  Drift chambers located outside the central
calorimeters and detectors behind a $60$~cm iron shield detect energy
depositions from muons with $|\eta|<0.6$ \cite{cmu}.  Additional drift
chambers and scintillation counters detect muons in the region
$0.6<|\eta|<1.0$.  Luminosity monitoring is provided by the Cherenkov
Luminosity Counter (CLC)
\cite{CLC}.

\subsection{Calorimeters}
\label{sec:calo}

\begin{figure}[htbp]
  \begin{center} \includegraphics[width=0.7\textwidth,clip=]{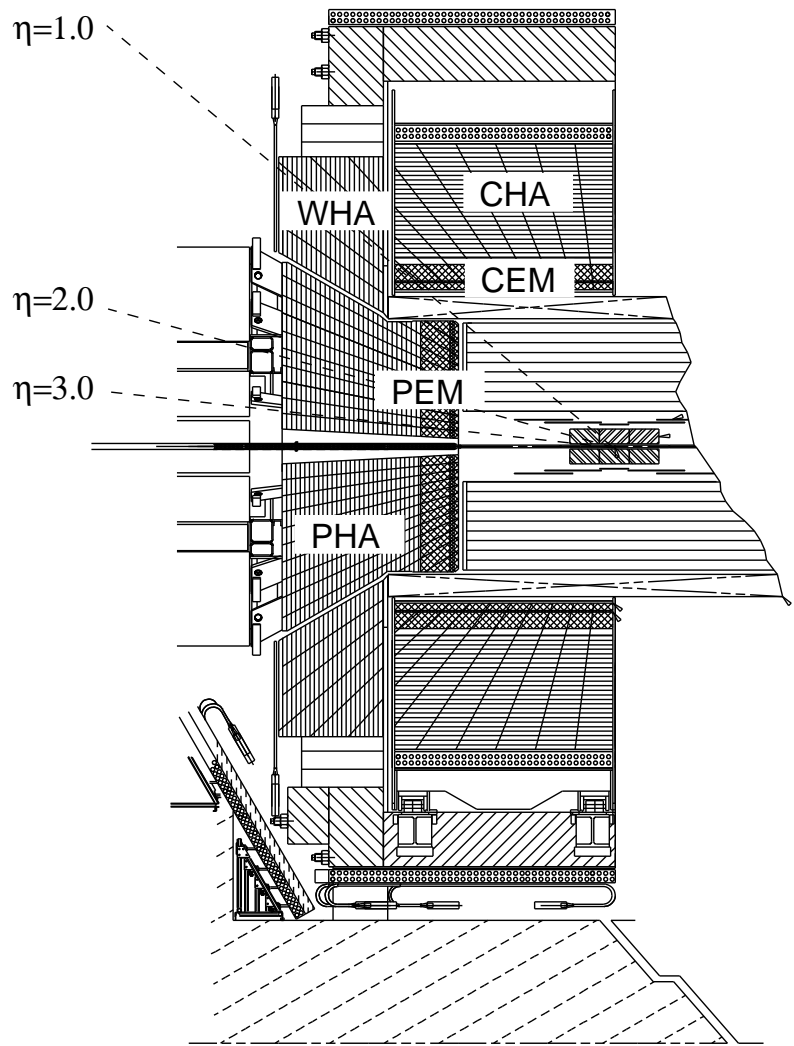}
\caption{\label{fig:cdfcalo} \sl Elevation view of one half of the CDF
detector displaying the components of the CDF calorimeter: CEM, CHA,
WHA, PEM and PHA.}
\end{center}
\end{figure}

The CDF calorimeter is divided into a central and a forward section. A
schematic view is shown in Fig. \ref{fig:cdfcalo}.

There are a total of five calorimeter compartments: the central
electromagnetic, CEM~\cite{CEM}, and central hadronic, CHA~\cite{CHA},
the plug electromagnetic, PEM~\cite{PEM}, and plug hadronic,
PHA~\cite{PEM}, and the wall hadronic WHA~\cite{CHA} in the region
between the central and the forward calorimeter at $\theta \approx
\pm30^\circ$ (see Fig. \ref{fig:cdfcalo}).

The central calorimeter covers the region $|\eta|<1.1$ and is divided
in two halves at $|\eta|=0$.  It is segmented in towers of 15$^o$ in
azimuth and $0.1$ in $\eta$ with lead-scintillator sampling for the
electromagnetic measurements and steel-scintillator sampling for
hadronic measurements. The material in the CEM has a depth of $18$
radiation lengths. The energy resolution for high energy electrons and
photons is $\frac{\sigma(E_T)}{E_T}=
\frac{13.5\%}{\sqrt{E_T}}\oplus 1.5\%$. The CHA and WHA are of similar
construction, with alternating layers of steel and scintillator and
are $4.7$ interaction lengths deep.  Three of the WHA towers are
situated behind the CEM/CHA and three are behind the plug
calorimeter. The energy resolution of the CHA is $\frac{\sigma(E_T)}{E_T}=
\frac{50\%}{\sqrt{E_T}} \oplus 3\%$ and WHA is $\frac{\sigma(E_T)}{E_T}=
\frac{75\%}{\sqrt{E_T}} \oplus 4\%$ for charged pions that do not
interact in the CEM.

The forward ``plug'' calorimeters cover the angular range
corresponding to $1.1<|\eta|<3.6$. They are segmented in $7.5^\circ$
towers for $|\eta|<2.11$ and $15^\circ$ for $|\eta|>2.11$.  The PEM
has a depth of $23.2$ radiation lengths. The energy resolution for
high energy electrons and photons is $\frac{\sigma(E)}{E}=
\frac{16\%}{\sqrt{E}}\oplus 1\%$. The PHA has alternating layers of
iron and scintillating tile, for a total of $6.8$ interaction
lengths. The energy resolution of the PHA is $\frac{\sigma(E)}{E}=
\frac{80\%}{\sqrt{E}}\oplus 5\%$ for charged pions that do not
interact in the PEM.

Each calorimeter tower is read out by two photomultipliers in the CEM,
CHA and WHA and by one photomultiplier in the PEM and PHA. The
calorimeter readout electronics was upgraded for the Run II data
taking period to accommodate the $396$~ns beam bunch spacing as well
as a possible upgrade to $132$~ns.  The ADC integration gate for the
charge collection is $120$~ns wide and collects $94$ to $98\%$ of the
signal depending on the calorimeter type. During test beam and the
previous Run I data taking period (1992-1996) the time between two
bunches was $3.5$~$\mu$s and the integration time was $600$~ns for all
calorimeter compartments. The fractional energy loss due to the
shorter ADC integration gate in Run II is measured in muon, electron
and jet data with an uncertainty of $1.5\%$.

\subsection{Definition of the Energy Scale of Calorimeter Towers}

The absolute energy scale of the CEM calorimeter is set such that for
fully corrected electrons \cite{wmassrun1} the measured mass of the
$Z$ boson in the $e^+e^-$ decay mode is consistent with the mass
measured at LEP \cite{lepz} taking into account photon radiation. The
ratio of the measured calorimeter energies and the track momenta for
electron candidates, $E/p$, is used to apply additional relative
calibrations for each tower to improve the resolution of the energy
measurement. The PEM energy scale is set using $Z\to e^+ e^-$ events
with one electron in the CEM and one electron in the PEM.

For the hadronic calorimeters, CHA, WHA, and PHA, the initial energy
scale is defined by their responses to a charged pion test beam of 50
GeV/$c$ (see Sec. \ref{sec:Datasets} and
\cite{cemtestbeam,chatestbeam,plugtestbeam}) using pions with almost
no interaction in the respective electromagnetic compartments CEM and
PEM, respectively. The total raw energy deposited in a tower is given
by the sum of CEM and CHA energies.  For charged pions that interact
in the CEM the response is lower (see Secs. \ref{sec:Simulation} and
\ref{sec:Absolute}).

\subsection{Stability of the Energy Scale}\label{subsec:stability}

The stability of the CDF calorimeter is monitored online using various
calibration methods. The energy scale of both the electromagnetic and
hadronic calorimeters in general decrease with time due to aging of
both the scintillators and the photomultipliers. The online response
is kept stable to better than $3\%$ while offline a stability better
than $0.3-1.5\%$ is achieved. The following methods are used to obtain
this stability:

\begin{itemize}

\item In the CEM, $E/p$ of electrons with transverse energies,
$E_T>8$~GeV are used to monitor the time dependence. The energy scale
decreases by $3\%$ every $6$~months, and is corrected accordingly.

\item In the CHA and WHA the energy scale is monitored using three 
independent methods: a laser system \cite{CHA}, muons from $J/\psi \to
\mu^+\mu^-$ decays and minimum bias data (these datasets are defined
in Sec. \ref{sec:Datasets}). The test beam energy scale has been
maintained since 1987 using $Cs^{137}$ source calibration runs.  The
CHA response decreases by about $1\%$ and the WHA response by about
$3\%$ per year which is corrected by adjusting the calibration.

\item The PEM and PHA calorimeters are monitored using a laser system
\cite{CDF_run2,pluglaser} and a radioactive source calibration using
$Co^{60}$. The laser is only sensitive to aging of the
photomultipliers while the source is sensitive to both the
photomultiplier and scintillator aging. The plug calorimeter scale
decreases by up to 2-10\% per year, for $|\eta|=1.2-3.6$.  The largest
decrease is observed in the region closest to the beam pipe. This
decrease is calibrated accordingly.
\end{itemize}

\begin{figure}[htbp]
\begin{center}
\includegraphics[width=0.7\hsize]{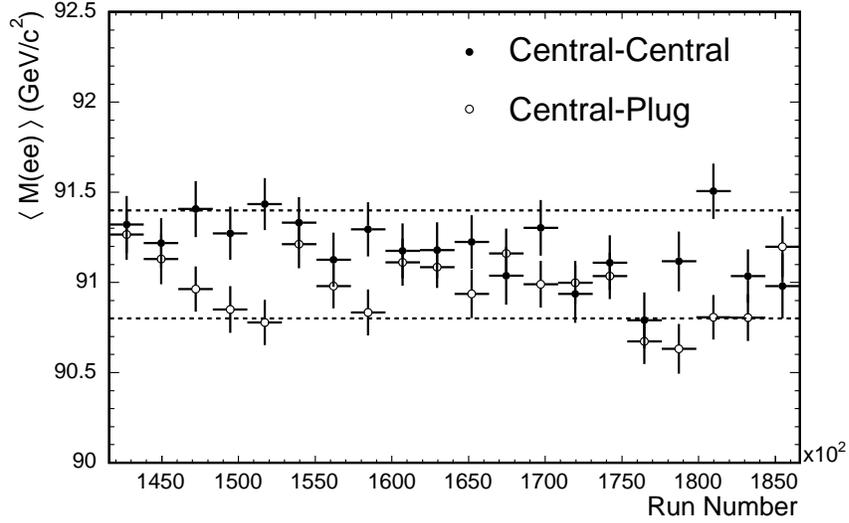}
\caption{\label{fig:zeevrun} {\sl Mean invariant mass of $Z\to e^+e^-$
candidates, $\langle M(ee) \rangle$, versus run number for events with
$86<M(ee)<98$ GeV/$c^2$. Shown are the values for events with both
electrons in the central calorimeter (full circles) and for events
with one electron in the central and one in the plug calorimeter (open
circles). The dashed lines indicate a $\pm 0.3\%$ uncertainty around
$91.1$ GeV/$c^2$. }}
\end{center}
\end{figure}

\begin{figure}[htbp]
\begin{center}
\includegraphics[width=0.7\hsize]{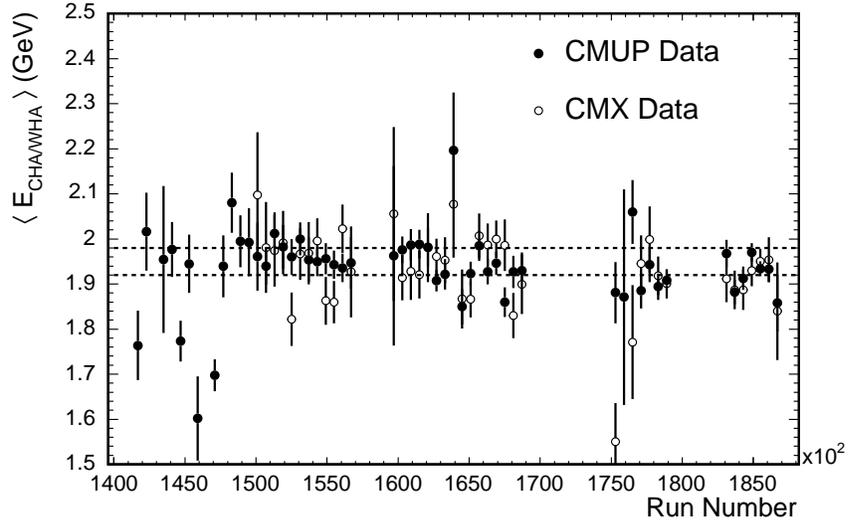}
\caption{\label{fig:wmunu} {\sl Mean energy observed in the CHA/WHA
for CMUP and CMX muons with $p_T>$20 GeV/$c$ from $W\to\mu\nu_\mu$
candidate events versus run number. The CMUP muons are confined to
$|\eta|<0.6$ and thus only sensitive to the central part of the CHA.
The CMX muons cover the region $0.6<|\eta|<1.0$ and probe the outer
part of the CHA plus the innermost part of the WHA.  The dashed lines
indicate a 1.5\% uncertainty.}}
\end{center}
\end{figure}

The calibration stability of the electromagnetic scales is verified
using the time dependence of the reconstructed invariant $Z$ boson
mass. Figure \ref{fig:zeevrun} shows the mean of the $Z \to e^+e^-$
mass distribution between 86 and 98 GeV/$c^2$ as a function of the run
number. The range of run numbers corresponds to the data taking period
from April 2002 until September 2004.  The mean energy deposited by
muons from $W\to\mu\nu_\mu$ candidate events is shown in
Fig. \ref{fig:wmunu} for muons with $|\eta|<1$ to verify the stability
of the CHA and WHA energy scales. For the PHA both muons and the jet
response are used to verify the stability.


From Figs. \ref{fig:zeevrun} and \ref{fig:wmunu} an uncertainty on the
stability of the CEM of 0.3\% and for the CHA of 1.5\% is
assigned. For a jet, which deposits typically 70\% of the energy in
the CEM and 30\% in the CHA, it is thus 0.5\%.

\clearpage

\section{Jet Clustering Algorithm}\label{sec:JetAlgo}

The energy of a jet is calculated from the energy deposited in the
calorimeter towers using different types of clustering algorithms. For
this study, jets are clustered using a cone algorithm with a fixed
cone size in which the center of the jet is defined as
($\eta^{jet},\phi^{jet}$) and the size of the jet cone as
$R=\sqrt{(\eta^{tower}-\eta^{jet})^2+(\phi^{tower}-\phi^{jet})^2}$ =
0.4, 0.7, or 1.0. The jet corrections and uncertainties are estimated
for these three cone sizes.

\subsection{Calorimeter Jets}

The jet clustering algorithm groups calorimeter towers with
$E_{Ti}>1$~GeV into jets. $E_{Ti}=E_i \sin\theta_i$ is the transverse
energy of a tower with respect to the $z$-position of the $p\bar{p}$
interaction, and the energy $E_i$ is the sum of the energies measured
in the electromagnetic and hadronic compartments of that tower.

Firstly ``seed towers'' are defined in order of decreasing
$E_{Ti}$. For each seed tower the towers within a radius of size $R$
with respect to its position are used to build ``clusters''. Once we
have an initial list of clusters, the cluster transverse energy and
the location of the cluster is calculated using the definitions:

\begin{equation}
\label{eq:jet1}
E_T^{jet}=\sum_{i=0}^{N_{tow}}E_{Ti}
\end{equation}
\begin{equation}
\phi^{jet}= \sum_{i=0}^{N_{tow}}\frac{E_{Ti}\phi_i}{E_{T}^{jet}}
\end{equation}
\begin{equation}
\eta^{jet}= \sum_{i=0}^{N_{tow}}\frac{E_{Ti}\eta_i}{E_{T}^{jet}}
\end{equation}

\noindent where $N_{tow}$ is the number of towers inside the radius $R$ with
$E_T>1$~GeV.

This procedure is repeated iteratively, a new list of towers around
the new center is determined. The jet $E_T$ and direction are
recalculated until the list of towers assigned to the clusters is
stable, that is, when the geometrical center of the tower correspond
to the cluster centroid. Overlapping jets are merged if they overlap
by more than 50$\%$. If the overlap is smaller than 50\%, each tower
in the overlap region is assigned to the nearest jet.

The final jet energy and momentum coordinates are computed from the
final list of towers:

\begin{equation}
E_{jet}=\sum_{i=0}^{N_{tow}}E_{i}
\end{equation}
\begin{equation}
p_{x}^{jet}=\sum_{i=0}^{N_{tow}}E_{i}\sin(\theta_i)\cos(\phi_i)
\end{equation}
\begin{equation}
p_{y}^{jet}=\sum_{i=0}^{N_{tow}}E_{i}\sin(\theta_i)\sin(\phi_i)
\end{equation}
\begin{equation}
\label{eq:jetlast}
p_{z}^{jet}=\sum_{i=0}^{N_{tow}}E_{i}\cos(\theta_i)
\end{equation}
\begin{equation}
p_{T}^{jet}=\sqrt{(p_{x}^{jet})^2+(p_{y}^{jet})^2}
\end{equation}

\begin{equation}
\phi_{jet}=\tan\frac{p_{y}^{jet}}{p_{x}^{jet}}
\end{equation}
\begin{equation}
\sin \theta_{jet}=\frac{p_{T}^{jet}}{\sqrt{(p_{x}^{jet})^2+(p_{y}^{jet})^2+(p_{z}^{jet})^2}}
\end{equation}
\begin{equation}
E_{T,jet}=E_{jet}\sin\theta_{jet}
\end{equation}

In general, jets with $E_T<$ 3 GeV are not used in physics analyses at
CDF.

\subsection{Particle Jets in Monte Carlo}

In Monte Carlo (MC) simulation, particle jets are obtained using the
same jet clustering algorithm on stable final state
particles\footnote{In {\tt PYTHIA} and {\tt HERWIG}, these are
particles which have a lifetime long enough to not decay within the
CDF detector volume.}, i.e. the sums in Eqns.
\ref{eq:jet1}-\ref{eq:jetlast} go over the stable particles instead of
the towers. A particle jet includes any particles produced in the
interaction, i.e. particles from the hard scattering process and also
those from the underlying event.

\clearpage

\section{Data and Monte Carlo Samples}\label{sec:Datasets}

In this section we describe the data and Monte Carlo (MC) simulation
samples used throughout this paper for studying and determining the
jet response.

The following samples are used:
\begin{itemize}
\item {\bf Test beam data}: The response of all calorimeters was first
measured in a test beam. Test beam for the CEM, CHA, and WHA were
taken in 1985 and 1988 \cite{cemtestbeam,chatestbeam} with a momentum
range $5-180$ GeV/$c$ for electrons and $7-220$ GeV/$c$ for charged
pions.  The test beam for the plug calorimeters was taken in 1996
\cite{plugtestbeam} with a momentum range $5.3-181$ GeV/$c$ for
electrons and $8.6-231$ GeV/$c$ for charged pions.

\item {\bf Minimum bias}: This sample is collected requiring at least
one $p\bar{p}$ interaction. It is triggered by activity in the
luminosity counters, i.e. the trigger requires coincident hits in both
the east and west CLC detectors. It is used for studying multiple
interactions and the underlying event (see Secs. \ref{sec:mi} and
\ref{sec:ooc}). Tracks from this sample are also used for tuning the
simulation (see Sec. \ref{sec:Simulation}).

\item {\bf Single track}: A special trigger was designed to take data
with a high momentum track at $p_T$ thresholds of $3$, $7$ and $10$
GeV/$c$. These data are used to tune the central calorimeter response
at larger $p_T$ (see Sec. \ref{sec:Simulation}). The tracks in this
dataset are confined to the region in the COT where particles can
traverse all available layers, that is $|\eta|<1$.

\item {\bf Jet}: There are four samples triggered on at least one jet
with transverse energy $E_T^{jet}>20$, $50$, $70$ and $100$ GeV
referred to as jet-20, jet-50, jet-70 and jet-100, respectively. They
are primarily used to calibrate the response as function of
pseudo-rapidity (see Sec. \ref{sec:RelCorr}). Apart from the jet-100
sample the data are taken with a fixed ``prescale'' depending on the
$p_T$ threshold. A prescale of e.g. 10 means that only 1 in 10 events
passing the trigger requirements is recorded. There is another sample
triggered on at least one trigger tower with $E_T>$5 GeV referred to
as single-tower-5.

\item {\bf\boldmath $\gamma$-jet}: This sample is triggered on an
isolated electromagnetic cluster with $E_T^\gamma>25$ GeV. It is used
to cross-check many aspects of the calibration and to determine the
systematic uncertainty on the out-of-cone correction (see
Sec. \ref{sec:ooc}). The photon selection requirements are described
in detail elsewhere \cite{promptgammaprd}.

\item {\bf\boldmath $W\to e\nu_e$ and $W\to \mu\nu_{\mu}$ }: These
data are taken by an inclusive electron (e) or muon ($\mu$) trigger
with $E_T^e>18$ GeV and $p_T^\mu>18$ GeV/$c$, respectively. The
electron and muon identification is described in detail elsewhere
\cite{wecross}. The $W\to e\nu_e$ ( $W\to \mu\nu_{\mu}$) sample is
selected by requiring one electron (muon) with $E_T^e>25$ GeV
($p_T^\mu>25$ GeV/$c$) and missing transverse energy $\met>25$ GeV.

\item {\bf\boldmath $Z\to e^+e^-$ and $Z\to \mu^+\mu^-$}: These data
are taken with the same trigger to the previous $W$ samples. They are
selected by requiring two electrons (muons) with $E_T^e>18$ GeV
($p_T^\mu>20$ GeV/$c$) and requiring the invariant mass of the
electrons and muons to be between 76 and 106 GeV/$c^2$.

\item {\bf\boldmath $J/\psi \to e^+e^-$ and $J/\psi \to \mu^+\mu^-$}:
The $J/\psi \to e^+e^-$ data are taken with a dielectron trigger with
$E^{e}_T>2$ GeV for each electron. The $J/\psi \to \mu^+ \mu^-$ data
are taken with a dimuon trigger with $p^{\mu}_T>1.5$ GeV/$c$ for each
muon.  The dilepton invariant mass is required to be between $2.5$ and
$3.5$ GeV/$c^2$ for the $e^+e^-$ decay and between $3.0$ and $3.2$
GeV/$c^2$ for the $\mu^+\mu^-$ decay.

\end{itemize}

Corresponding MC samples are used for all processes. In all cases,
Monte Carlo samples are generated using both {\tt PYTHIA} 6.216 \cite{pythia}
and {\tt HERWIG} 6.505 \cite{herwig} with CTEQ5L \cite{cteq5l} parton
distribution functions.

The following MC samples are generated:
\begin{itemize}

\item {\bf Minimum bias} and {\bf jet}: The inclusive $2 \to 2$ parton
processes ($p\bar{p} \to q\bar{q}+X$, $p\bar{p} \to gq+X$, $p\bar{p}
\to gg+X$) are generated at different thresholds for the transverse
momentum of the outgoing partons, $\hat{p}_T>$ 0, 10, 18, 40, 60, 90,
150, 200, 300, 400, 500, 600 GeV/$c$. The sample with $\hat{p}_T>0$ is
referred to as minimum bias MC.

\item {\bf Single-particle}: Single charged particles are generated
with a mixture of $60\%$ $\pi^\pm$, 30\% $K^\pm$ and 10\% $p$ and
$\bar{p}$.  

\item {\bf \boldmath $\gamma$-jet}: The processes $qg \to q\gamma+X$
and $q\bar{q} \to \gamma g+X$ are generated at different thresholds
of the transverse momentum of the outgoing partons, $\hat{p}_T>12$,
$20$ and $40$ GeV/$c$. 

\item {\bf \boldmath $W$ and $Z$}: $W$ and Drell-Yan production is
generated in both electron and muon decay channels as described in
\cite{wecross}.

\item {\bf \boldmath $J/\psi \to e^+ e^-$ and $J/\psi \to
\mu^+\mu^-$}: This sample is generated using {\tt BGEN} \cite{bgen}.

\item {\bf \boldmath $t\bar t\to$ $WbWb$}:
Top pair production is generated using {\tt HERWIG}. More details
can be found in \cite{mtop_prd_prl}.

\item {\bf \boldmath $W$+jets}: The {\tt ALPGEN 1.3} generator interfaced with {\tt HERWIG} is used to simulate these type of events \cite{mtop_prd_prl}.

\end{itemize}

The matrix element of the hard scattering process, the underlying
event, the higher order QCD correction (i.e. the parton showering
process), and the fragmentation are all handled by {\tt PYTHIA} and
{\tt HERWIG}. The matrix element is well known at leading order in
QCD. However, the other three components are not fully calculable and
thus modeled using an empirical approach. Note that {\tt PYTHIA} and
{\tt HERWIG} use different models, which is helpful to study and
understand the uncertainties associated with these effects. The
modeling of the fragmentation and parton showering is mostly tuned to
$e^+e^-$ data.  For modeling the underlying event we use a tuning
optimized to describe CDF data from Run-I, which we refer to as ``{\tt
PYTHIA} Tune A''
\cite{tuneA}. For {\tt HERWIG} the default parameters for the underlying
event are used. The recent model for the underlying event, {\tt JIMMY}
\cite{jimmy}, has not been used in this analysis.

After generation, the samples are processed through the CDF Run II
detector simulation, which is described in detail in Sec.
\ref{sec:Simulation}. They are then processed through the standard
reconstruction program as the data. The details of particle and
calorimeter jet reconstruction are in Sec. \ref{sec:JetAlgo}.

Throughout this article the data are primarily compared to {\tt
PYTHIA} MC samples but any differences between {\tt PYTHIA} and {\tt
HERWIG} are discussed. Detailed comparisons between {\tt HERWIG} and
{\tt PYTHIA} are presented in Sec. \ref{sec:crosschecks}.

\clearpage


\section{\label{sec:Simulation}Calorimeter Simulation}

In this section we describe the simulation of the CDF calorimeter.  A
parameterized shower simulation is used with the simulation parameters
tuned to describe the observed calorimeter response of single
particles. The response of single isolated charged particles is
measured using test beam data as well as minimum bias and single track
trigger data from Run II. Similarly, the tuning of electromagnetic
showers is based on electrons observed in $J/\psi \to e^+e^-$ and
$Z\to e^+e^-$ decays. In the following, we first describe the shower
parameterization, then the measurement of the calorimeter response is
discussed, and finally the systematic uncertainty is assigned based on
how well the MC simulation models the data.

\subsection{Simulation of Electromagnetic and Hadronic Showers} 

The CDF detector simulation uses {\tt GEANT}~\cite{geant} to track
generated particles through the CDF detector and to simulate secondary
physical processes such as energy loss, multiple scattering, and
inelastic interactions. After the first inelastic interaction in the
calorimeter, the particles are passed to the program {\tt
GFLASH}~\cite{gflash}, a fast simulation of electromagnetic and
hadronic particle shower. {\tt GFLASH} generates particle showers
shapes within the calorimeter and computes the energy deposited in the
calorimeter sensitive volumes, using parameterizations of the
longitudinal and lateral shower profiles. The parameterizations of
electromagnetic and hadronic showers are described in detail in
Ref.~\cite{gflash} and are briefly outlined in this section.

\subsubsection{Procedure}
\label{sec:proctuning}
The simulation of electromagnetic and hadronic showers involves two
steps. First, {\tt GFLASH} calculates the spatial distribution of
energy, $E_{dp}$, deposited by a shower within the calorimeter volume:
\begin{eqnarray}
   dE_{dp}(\vec{r}) = \frac{E_{dp}}{2\pi} \,{L}(z)
   \, {T}(r) \, \rd z \rd r \label{eqn:edep}
\end{eqnarray}
where the longitudinal energy profile, $L(z)$, depends on the shower
depth $z$, and the lateral shower profile, $T(r)$, depends on the
radial distance, $r$, from the trajectory of the particle.  This
parameterization takes dependencies on the incident particle energy
and shower fluctuations into account, and further considers the
repetitive sampling structure of the detector volume. Second, the
fraction of the deposited energy which is visible to the active
medium, $E_{vis}(\vec{r})$, is determined. $E_{vis}(\vec{r})$ is
computed depending on the relative sampling fractions of minimum
ionizing particles, electromagnetic and hadronic particles. The
relative sampling fractions for electrons, $S_{e}/S_{mip}$, and for
hadrons, $S_{had}/S_{mip}$, compared to the sampling fraction of a
particle that does not interact inelastically in the calorimeter,
$S_{mip}$, are two tunable parameters.

\subsubsection{Longitudinal shower profile}
\label{sec:simlongitudinal}

\paragraph{Electromagnetic Showers:}

For the simulation of the longitudinal profile of electromagnetic 
showers {\tt GFLASH} assumes that they follow the $\Gamma$-distribution:
\begin{eqnarray}
   L_{em}(z) = \frac{ x^{\alpha_{em}-1} e^{-x} }{ \Gamma(\alpha_{em}) }
   \;\;\hspace{3ex}\;\;
   x = \beta_{em} z(X_0)
\label{eqn:GammaFunction}
\end{eqnarray}
where $z$ is the shower depth, measured in units of radiation lengths,
$X_0$.  The parameters $\alpha_{em}$ and $\beta_{em}$ are
Gaussian-distributed, and parameterized as a
function of incident particle energy.  The longitudinal profile of
each shower is simulated by choosing a correlated 
$(\alpha_{em},\beta_{em})$ pair at random.

\paragraph{Hadronic Showers:}
{\tt GFLASH} distinguishes three classes of hadronic showers:

\begin{itemize}
\item Purely hadronic showers ($h$) whose propagation scales with the
absorption length $\lambda_0$.

\item Showers where a $\pi^0$ is produced in the first inelastic
 interaction ($f$). Their propagation scales with radiation length
 $X_0$.

\item Shower profiles from $\pi^0$'s produced at a later stage of the
hadronic showering process ($l$).
\end{itemize}

The hadronic showers in {\tt GFLASH} are calculated as a
superposition of those three shower classes:
\begin{eqnarray}
   dE_{dp} = f_{dp} \, E_{inc} \, [ c_h \, L_h(x_h) \, dx_h + c_f \,
   L_f(x_f) \, dx_f + c_l \, L_l(x_l) \, dx_l \, ]
\label{eqn:GammaFunctionHad}
\end{eqnarray}

with 
\begin{eqnarray}
   \mathcal{L_h}(x_h) & = & \frac{ x_h^{\alpha_h-1} e^{-x_h} }{ \Gamma(\alpha_h)} \hspace{3ex} x_h= \beta_h z(\lambda_0)\\
   \mathcal{L_f}(x_f) & = & \frac{ x_f^{\alpha_f-1} e^{-x_f} }{ \Gamma(\alpha_f)} \hspace{3ex} x_f= \beta_f z(X_0)\\
   \mathcal{L_l}(x_l) & = & \frac{ x_l^{\alpha_l-1} e^{-x_l} }{ \Gamma(\alpha_l)} \hspace{3ex} x_l= \beta_l z(\lambda_0)\\
\label{eqn:GammaFunctionHad2}
\end{eqnarray}

The coefficients $c_h$, $c_f$ and $c_l$ are the relative fractions of
the three contributions normalized such that $c_h+c_f+c_l=1$. The
factor $f_{dp}$ is the fraction of deposited energy with respect to
the energy of the incident particle, and takes the intrinsic losses
during the hadronic shower development into account.

The relative probabilities of the three classes $c_h$, $c_f$ and $c_l$
depend on the incident energy and are correlated through:
\begin{eqnarray}
c_h(E) = 1-f_{\pi^0}(E),\hspace{3ex} c_f(E) =f_{\pi^0}(E)(1-f^l_{\pi^0}(E)), \hspace{3ex} 
c_l(E) =f_{\pi^0}(E)f^l_{\pi^0}(E)
\end{eqnarray}
where $f_{\pi^0}(E)$ is the probability that a hadronic shower
contains any $\pi^0$ and $f^l_{\pi^0}(E)$ is the probability that a
$\pi^0$ is produced ``late'', i.e. not in the first interaction.

In total, the longitudinal hadronic shower development depends on
18 partially correlated parameters: 
the mean and $\sigma$ values of the three different $\alpha$- and
$\beta$- values for each of the three shower types, and the fractions
$f_{dp}$, $f_{\pi^0}$ and $f^l_{\pi^0}$. For the electromagnetic 
shower development there are only 4 parameters: the mean and 
$\sigma$ values of $\alpha_{em}$ and  $\beta_{em}$.

\subsubsection{Lateral shower profile}

The parameterization for the lateral energy profile of electromagnetic
and hadronic showers is taken to be
\begin{eqnarray}
   T(r) = \frac{2 r R_{50}^{2}}{(r^{2}+{R_{50}^{2}})^{2}}
   \label{eqn:latprof}
\end{eqnarray}
$R_{50}$ is given in the units of
Moli\`{e}re radius $R_M$ for electromagnetic showers, and absorption
length, $\lambda_0$, for hadronic showers. $R_{50}$ is an approximate
log-normal distribution, with a mean value $\langle R_{50}(E,z)\rangle$
and variance $\sigma_{R_{50}}$ parameterized as a function of the 
incident particle energy, $E$, and the shower depth, $z$,
\begin{eqnarray}
\langle R_{50}(E,z)\rangle  = \left [R_1+(R_2-R_3 \log E)z\right ]^n  \hspace{3ex}n=1,2 \\
\sigma_{R_{50}}(E,z) = \left [ (S_1+(S_2-S_3 \log E)z \langle R_0(E,z)\rangle \right]^2
\end{eqnarray}
The $z$-evolution of the lateral spreading is linear for hadronic
showers ($n=1$) and quadratic for electromagnetic showers ($n=2$). The
electromagnetic and hadronic profiles are determined by their own set of
adjustable $R_i$ and $S_i$ values, thus giving a total of 14 parameters.

\subsubsection{Tuning to CDF Data}

The {\tt GFLASH} longitudinal and lateral hadronic shower parameters
were tuned using single, isolated tracks selected from minimum bias
data ($0.5< p < 2.5$ GeV/$c$ in the central and $0.5<p<5$ GeV/$c$ for
the plug calorimeter) and test beam data ($7\le p \le 220$
GeV/$c$). The electromagnetic showering parameters were tuned using
test beam data and compared {\it in situ} using $Z\to e^+ e^-$ events
data and simulation. Further adjustments in addition to the above
parameters are made to the energy deposited at the boundary between
central and plug calorimeter.

Given the limited availability of CDF isolated single track and test
beam data with current statistics, not all of the 34 parameters
described above were tuned to CDF data. Furthermore, the current
tuning of various parameters based on CDF data is restricted to
relatively low particle momenta, below $2.5$~GeV/$c$.  For the
remaining parameters, we use the default setting from the H1
collaboration \cite{gflash}. At this stage the parameters that are
tuned to CDF data are:

\begin{itemize}

\item $\alpha_h$ and $\alpha_l$ for both the central and plug calorimeters.

\item $\beta_f$ for both the central and plug calorimeters.

\item $f_{dp}$ and $f_{\pi^0}$ for both the central and plug
calorimeters.

\item four of the parameters that characterize the lateral profile for
the central calorimeter: $R_1$, $R_2$, $S_1$ and $S_2$.

\item the relative sampling fractions $S_{e}/S_{mip}$ 
and $S_{had}/S_{mip}$ for the plug calorimeters.

\end{itemize}

Tuning these parameters gives a good description of the single track
data as will be seen in the following sections. However, since the tuning was
done on a limited dataset there are some discrepancies with the latest
single track data samples that extend up to momenta of $20$ GeV/$c$. 
The newly collected data that are presented in the following sections 
will be used in the near future to further tune the simulation parameters. 
This is particularly important for the plug calorimeters. 

\subsection{\label{sec:EoP}Calorimeter Response to Hadronic Particles}

Charged and stable neutral hadrons carry approximately 70\% of the jet
energy and therefore a good description of their response is pivotal
for a good simulation of the jet energy measurement. The hadronic
shower development can best be studied using isolated charged
particles.

In the rapidity range covered by the central calorimeters, CDF has an
excellent tracking coverage and efficiency. The measurement of the
response in the plug calorimeters is more difficult since there is no
track trigger available and the tracking efficiency is limited.

\subsubsection{Procedure}

Tracks are selected and extrapolated to the impact point at the
position of the CES or PES detectors, taking the magnetic field into
account. If a track is extrapolated to a calorimeter tower, this tower
is considered to be the target tower. Tracks are required to point to
the inner $0.9\times 0.9$ contour of the target tower to stay away
from the tower edges, otherwise they are not considered.

\begin{figure}[htbp]
\begin{center}
\setlength{\unitlength}{1 mm}
\begin{picture} (160,60)
\put(30,0) {\epsfig{file=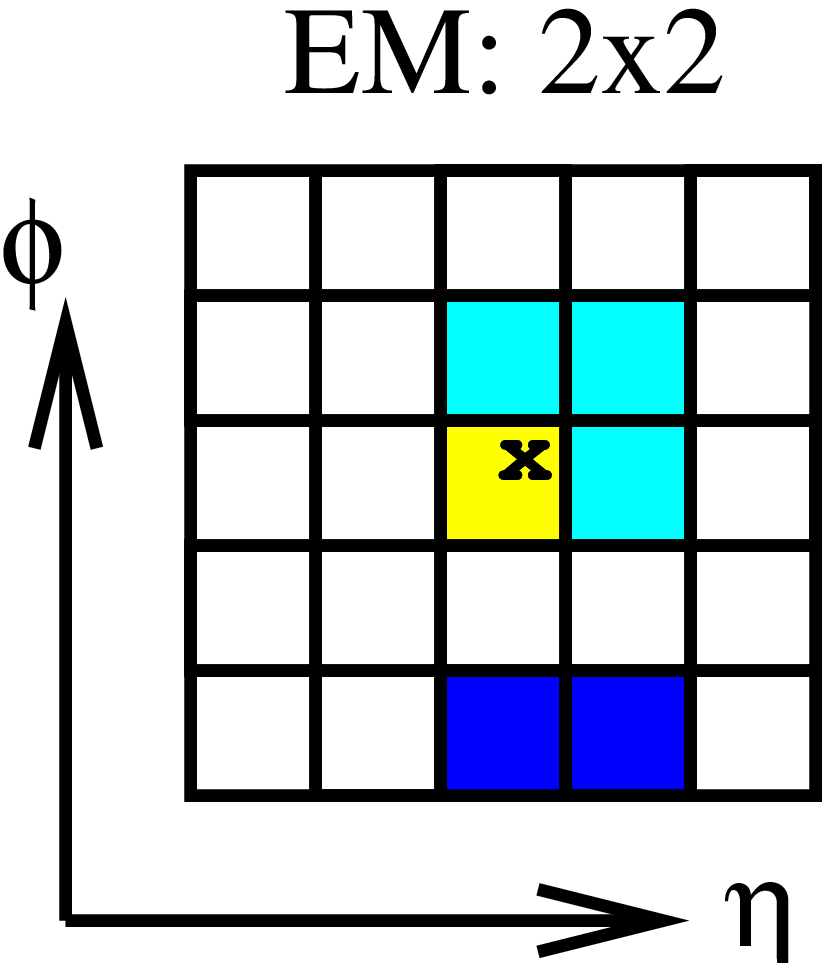,width=0.3\textwidth}}
\put(85,10) {\epsfig{file=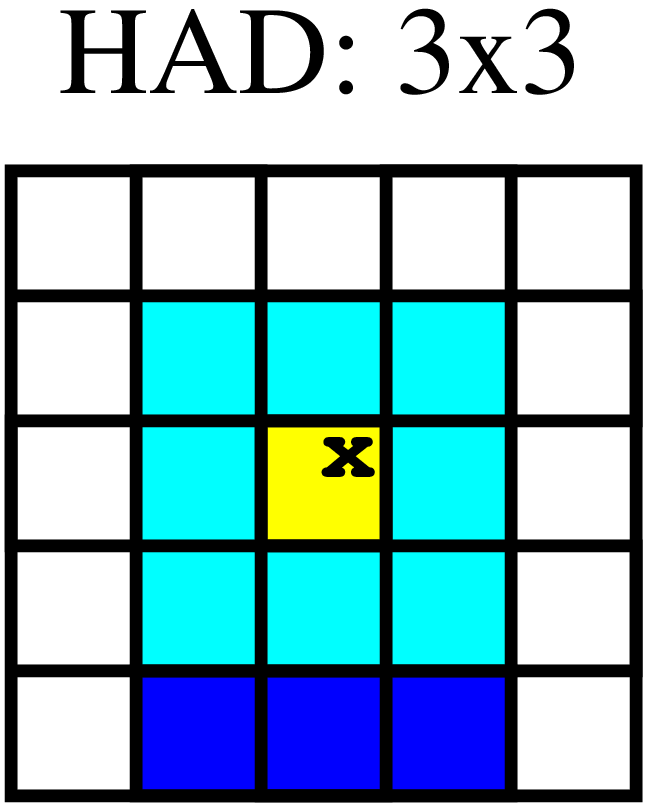,width=0.23\textwidth}}
\end{picture}
\caption{\label{fig:eopsketch} \sl Illustration of the target tower
for the electromagnetic (EM) and the hadronic (HAD) sections used in
the tuning of charged hadrons. The yellow (lightest shading) region is
the target tower. The ``x'' marks the impact point of the track. The
yellow and cyan (light and medium shading) region is the signal region
and the dark blue (darkest shading) illustrates the background
region. The horizontal axis represents the $\eta$ direction and the
vertical represents the $\phi$ direction.}
\end{center}
\end{figure}

From the eight towers surrounding the target tower, the measurement of
the CEM energy uses a $2\times 2$ subset containing the target tower
and those three adjacent towers closest to the track impact point.
For the CHA the shower spread is typically larger and all eight
adjacent towers are considered in the signal energy definition. A
sketch of the target tower and signal region definitions is shown in
Fig. \ref{fig:eopsketch} for the CEM and the CHA.

The signal tower region may contain additional particles overlapping
with the direction of the primary track that is being analyzed. This
background is reduced by requiring no additional tracks within a
$7\times 7$ block of towers around the target tower.  In addition, no
energy depositions above $1$ GeV in the CES detector are allowed
within the $7 \times 7 $ blocks of towers except within $\Delta
R=\sqrt{(\eta_{CES}-\eta_{track})^2+(\phi_{CES}-\phi_{track})^2}<0.03$
around the extrapolated track position, where $\phi_{CES}$
($\phi_{track}$) and $\eta_{CES}$ ($\eta_{track}$) denote the
azimuthal angle and the pseudo-rapidity of the CES cluster
(track). The first requirement removes charged particles and the
second requirement mostly removes photons.

Even after these isolation requirements, some overlap background
remains. The background is estimated using the energy deposited in the
same $\eta$ range but within the towers along the edge in $\phi$ of a
$5\times 5$ tower group around the target tower, using those edge
which is more distant from the track impact point, see Fig.
\ref{fig:eopsketch}. The energy measured in those background regions
is then scaled to the area covered by the signal region and
is subtracted from the energy measured in the signal
regions. For the single track MC the effect of the background
subtraction is negligible. 

\subsubsection{Central Calorimeter}

\begin{figure}[h]
  \begin{center}
  \includegraphics[width=\linewidth,clip=]{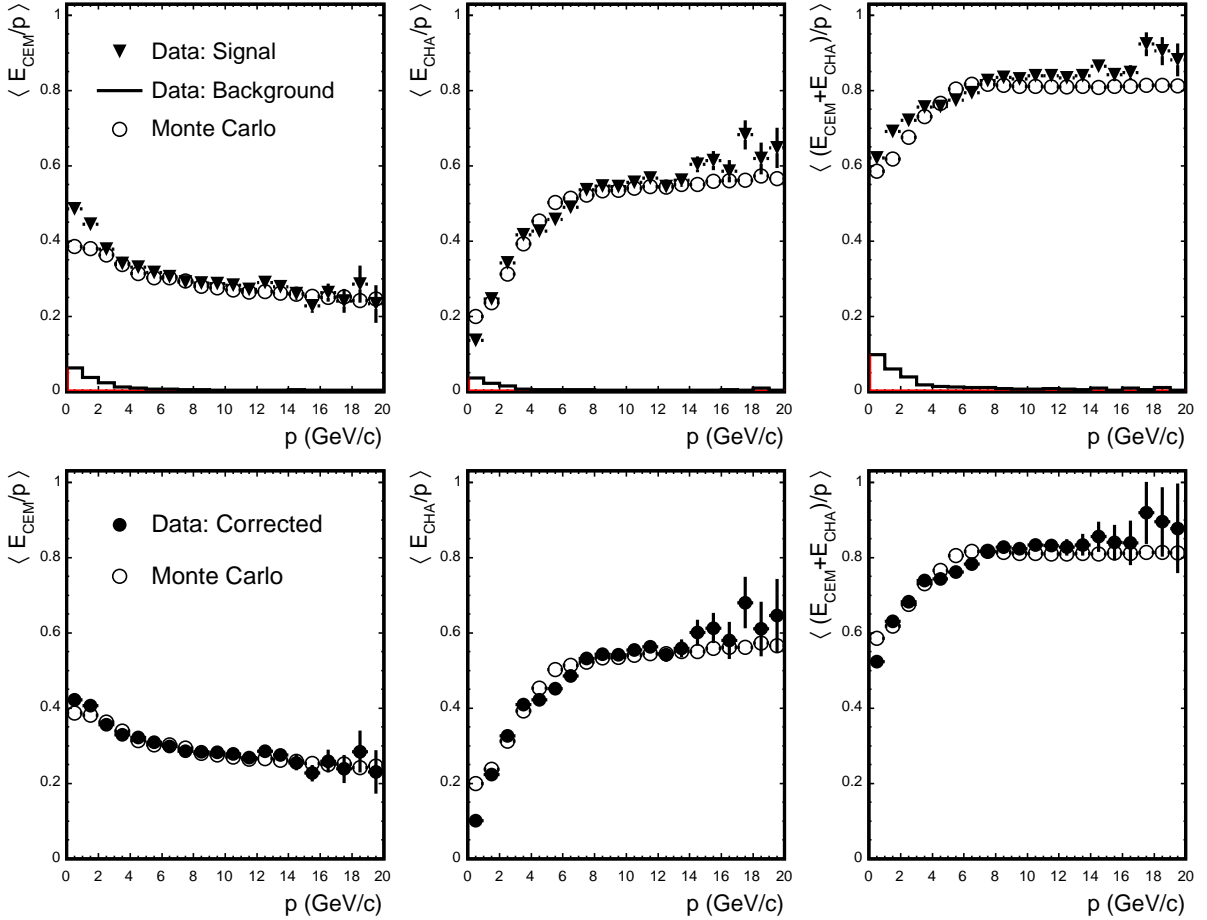}
  \caption{\sl Fractional energy observed in the central calorimeter
  as a function of incident particle momenta. The top row shows
  $\langle E_{CEM}/p \rangle$, $\langle E_{CHA}/p \rangle$ and
  $\langle (E_{CEM}+E_{CHA})/p \rangle$ for data signal (triangles)
  and background (histogram) and for single track MC simulation (open
  circles). The bottom row shows the same distributions for data after
  background subtraction (full circles) and MC simulation (open
  circles).\label{fig:eop}} \end{center}
\end{figure}

The comparison of the mean value of the $E/p$ distribution, $\langle
E/p \rangle$, between single track data and MC is shown in
Fig. \ref{fig:eop}. The data are shown before and after background
subtraction. The background contributes significantly for $p<3$
GeV/$c$ but is negligible at higher $p$. For particle momenta
increasing from $0.5$ to $20$ GeV/$c$, the fraction of energy
deposited in the CEM drops from 40\% to 25\%, whereas the fraction in
the CHA rises from 20\% to 55\%. In total $\langle E/p \rangle$ rises
from about $0.5$ at $p=0.5$ GeV/$c$ to about $0.8$ at $p\geq 5$
GeV/$c$.  For $p>5$ GeV/$c$ the response is almost independent of
$p$. In general, the mean $\langle E/p \rangle$ agrees well between
the data and the simulation separately for the CEM and the CHA
energies and for their sum. The deviations observed at very low
momenta and around $4$~GeV/$c$ are probably related to differences in
the particle spectrum between data and MC caused by momentum cutoffs
in the MC and trigger thresholds in the data.

\begin{figure}[h]
  \begin{center}
  \includegraphics[width=\linewidth,clip=]{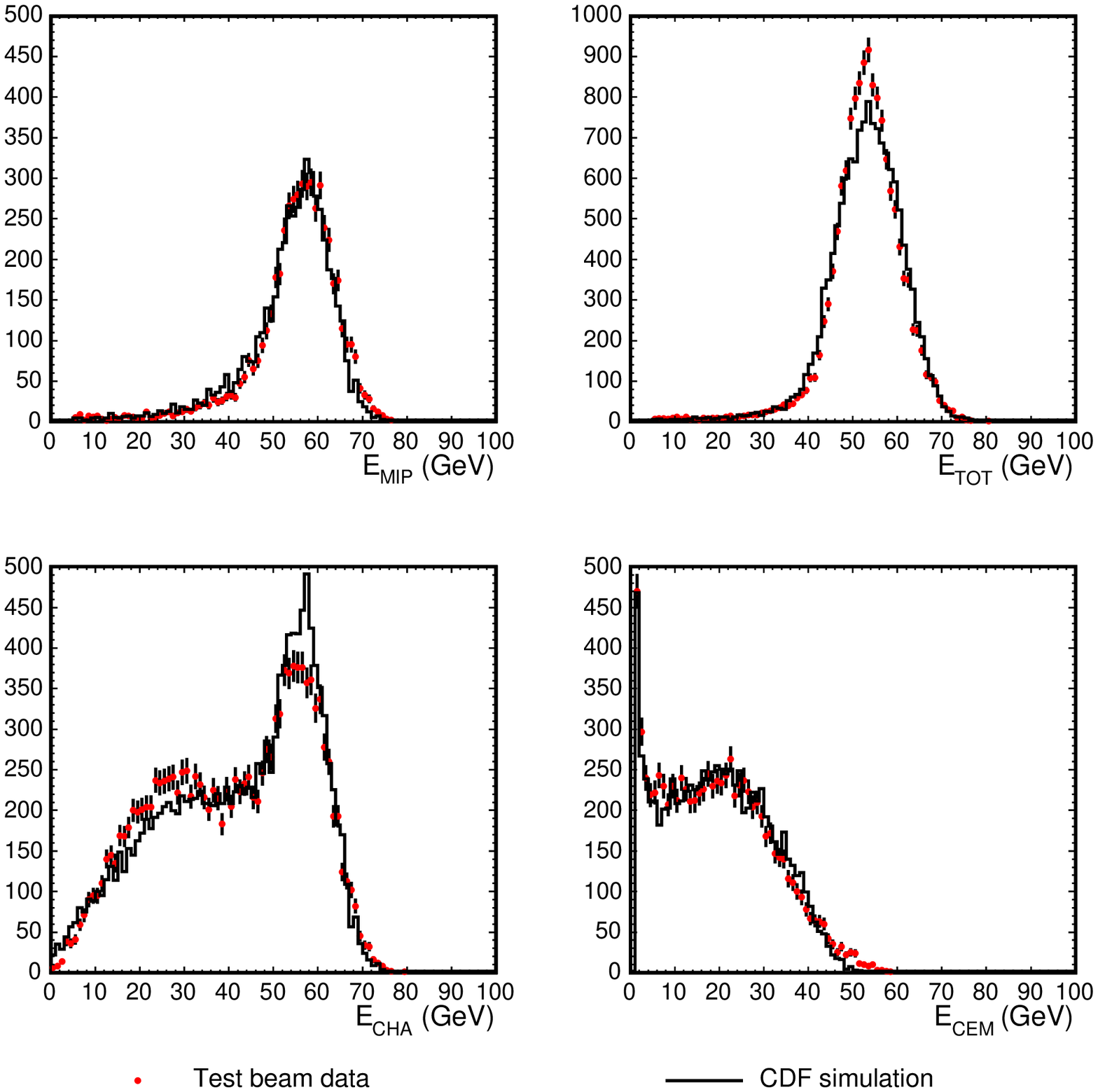}
  \caption{\sl Top left: Energy observed in CHA for particles that do
  not interact in the CEM. Total (top right), CEM (bottom right) and
  CHA (bottom left) energy for charged pions with $p=57$ GeV/$c$.  The
  test beam data (points) are compared to the CDF simulation (solid
  line).\label{fig:Pi57GeV}} \end{center}
\end{figure}

Beyond $20$~GeV/$c$ there are not enough isolated tracks to verify the
simulation and test beam data are used for studies of higher
momenta. The simulated CEM, CHA, and CEM+CHA energy is compared with
the test beam response for $57$~GeV/$c$ pions in
Fig. \ref{fig:Pi57GeV}.  For this study the charged pion sample is
divided into two categories:

\begin{itemize} 
\item pions which do not interact in the CEM are selected by requiring
the CEM energy to be less than $500$ MeV \footnote{The mean CEM
response for non-interacting particles is about $300$ MeV.}. The CHA
energy of these pions is shown in the upper left display in
Fig. \ref{fig:Pi57GeV}. This category is also used to set the absolute
energy scale of the hadronic calorimeters (see
Sec. \ref{sec:Absolute}).
\item pions which interact inelastically in the CEM: the CEM, CHA and
total energy of this category is displayed in the other three plots in
Fig. \ref{fig:Pi57GeV}.
\end{itemize}

The data are reasonably well described by the simulation.

\begin{figure}[htbp]
\begin{center}
\setlength{\unitlength}{1 mm}
\begin{picture} (160,140)
\put(0,0) {\epsfig{file=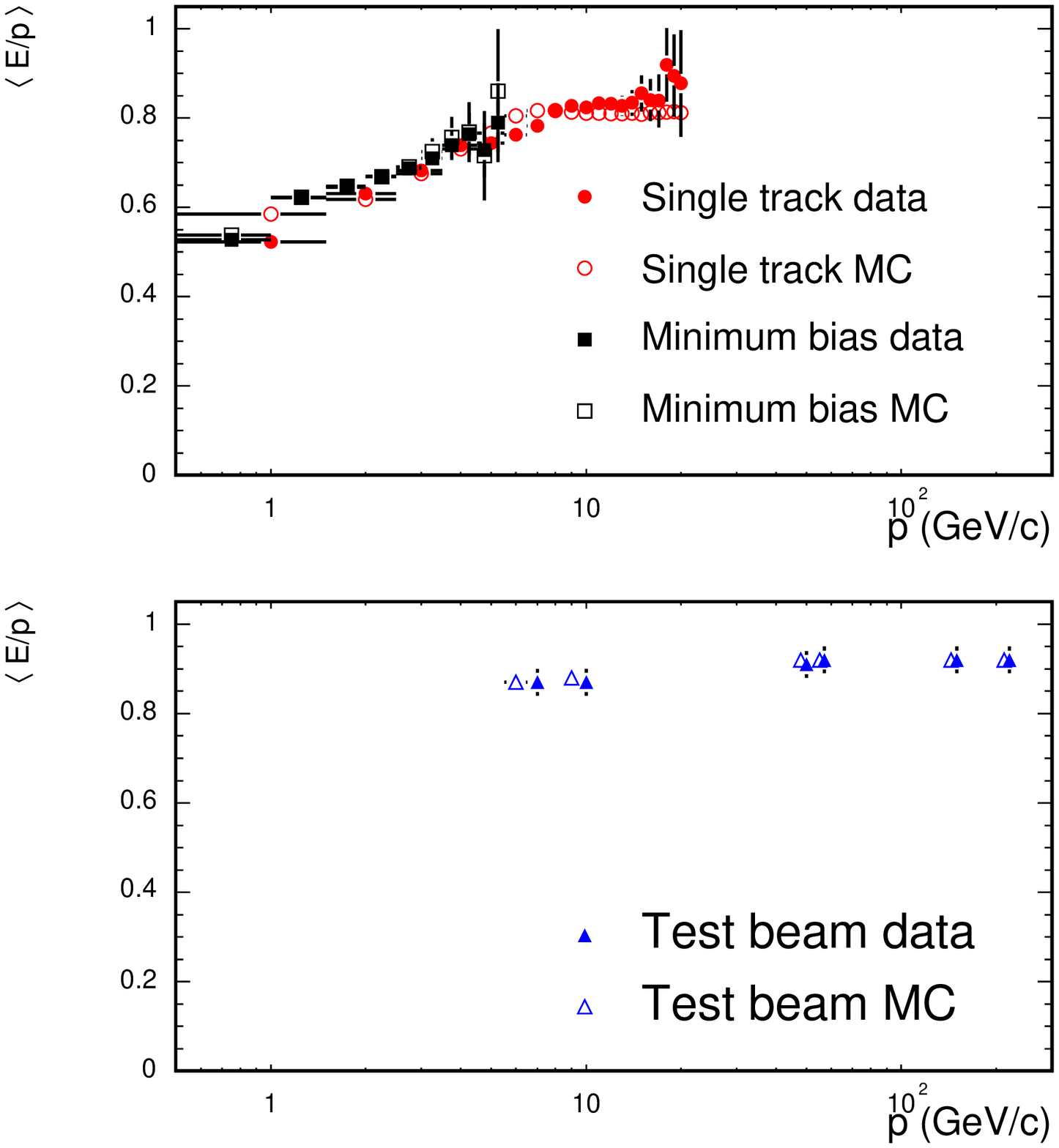,width=0.9\textwidth}}
\end{picture}
\caption{\label{fig:eopcentral} \sl $\langle E/p \rangle$ for the
central calorimeter versus particle momentum $p$. The top plot shows
the {\sl in situ} minimum bias (circles) and single track data
(squares) compared to the simulation (open symbols). The bottom plot
compares the test beam data (closed triangles) to the simulation (open
triangles).  The error bars on the minimum bias and single track data
are statistical. For the test beam data we display an uncertainty of
3.5\% due to the systematic uncertainty on the test beam momentum
scale which is correlated between all the data points.  For the test
beam comparison the MC points have been shifted slightly in $p$-value
to allow a better visual comparison. }
\end{center}
\end{figure}

Fig.~\ref{fig:eopcentral} shows a summary plot of the absolute total
response $\langle E/p\rangle$ for minimum bias, single track trigger
and test beam data together with the Monte Carlo expectation.

Note, that in test beam the particles were shot at the center of the
tower while for the single track data the inner 81\% of the towers are
used. This difference in selection is estimated to cause about a 5\%
higher response for the test beam data. In addition, for the test beam
only one central calorimeter tower (covering the region
$0.3<|\eta|<0.4$) was used. Therefore, the data cannot be compared
directly between test beam and {\sl in situ} measurements. However, in
each case the results are consistent between data and simulation.

\subsubsection{Plug Calorimeter}

In the plug calorimeter the same procedure is used as in the central
calorimeters. However, due to the smaller tower size a larger fraction
of the incident particle momentum is measured by neighboring
towers. Thus, for the $\langle E/p \rangle$ analysis, two adjacent
towers in azimuth are treated as one target tower.

Some difficulties arise in the plug calorimeter which are not present
in the central calorimeter:

\begin{itemize}

\item The tracking efficiency is rather low in the forward region of
the detector, decreasing from 70\% at $|\eta|\approx 1.2$ to 30\% at
$|\eta|=2.2$. Thus the background rejection is less efficient in the
plug than in the central region where the efficiency is nearly
100\%. Since the tracking efficiency in the very forward region
$|\eta|>2.2$ is even lower this method cannot be used to evaluate the
response.

\item The triggers dedicated to collect high momentum tracks are limited
to the COT coverage $|\eta|<1.0$.  

\item The momentum resolution for tracks is poorer than in the central
region because the track reconstruction is mostly based on the Silicon
Vertex Detector. This results in a less accurate determination of the
reference scale, $p$, particularly in the high momentum
region. However, this problem can be minimized by using tracks with
partial coverage by the COT.  Tracks with combined COT and silicon
hits are available up to $|\eta|<1.8$.

\end{itemize}

\begin{figure}[htbp]
\begin{center}
\setlength{\unitlength}{1 mm}
\begin{picture} (160,120)
\put(0,0) {\epsfig{file=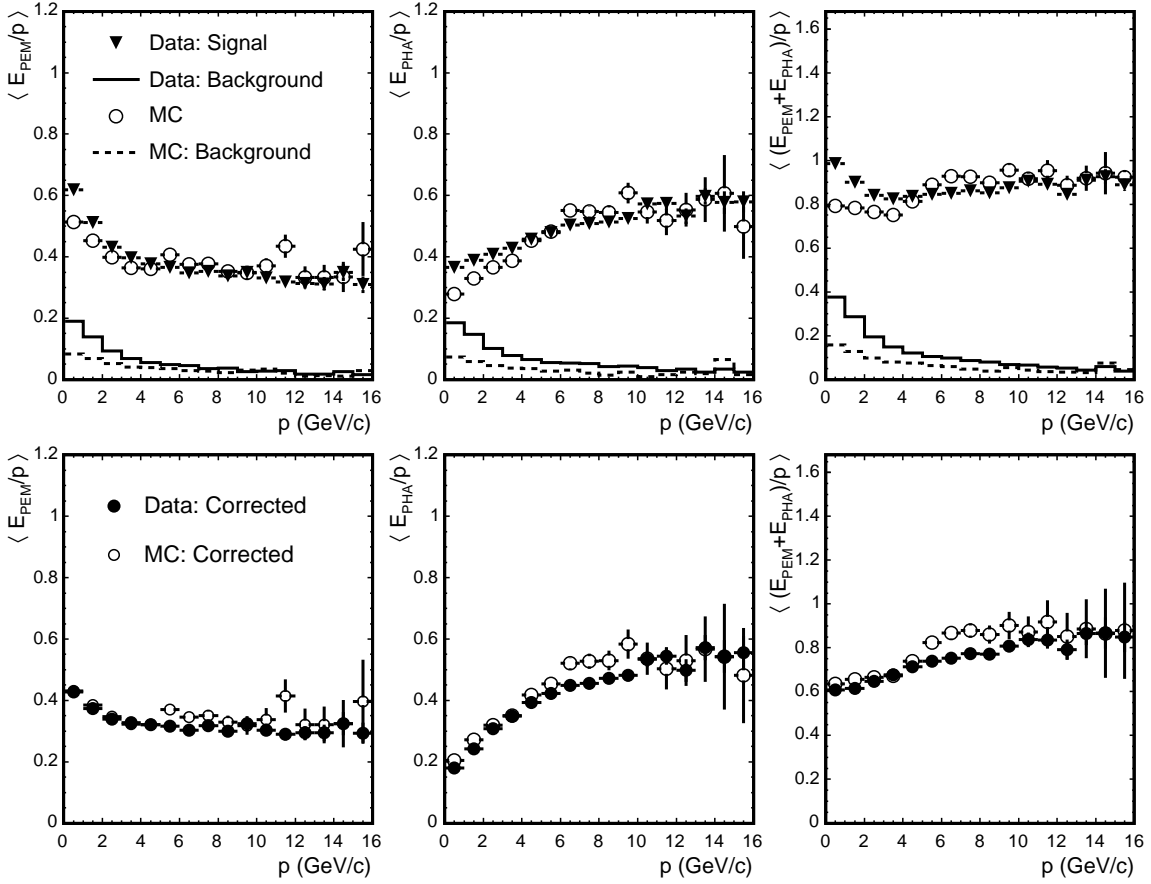,width=0.95\textwidth}}
\end{picture}
\caption{\label{fig:eopplug} \sl $\langle E/p \rangle$ observed in the 
plug calorimeter as a function of incident particle momenta. The top
rows shows $E_{PEM}/p$, $E_{PHA}/p$ and $(E_{PEM}+E_{PHA})/p$ for data
signal (closed squares) and background (solid line) and for MC
simulation signal (open squares) and background (dashed line). The
bottom row shows the same distributions for data after background
subtraction (full circles) and MC simulation (open circles). The
tuning of the calorimeter was done using tracks up to 5 GeV/$c$.}
\end{center}
\end{figure}

A measurement of the single particle response in the plug using
combined COT and silicon tracks is shown in Fig.~\ref{fig:eopplug} for
single track trigger data and simulation. The background is larger
than in the central calorimeter as expected. After background
subtraction the data and simulation agree well at low momenta but
deviations of up to 13\% are observed between $5$ and $10$
GeV/$c$. Note that for the tuning relevant for this paper the data at
the medium momentum range were not available. Using the data shown
here the plug simulation will be improved in the near future.

\begin{figure}[htbp]
\begin{center}
\setlength{\unitlength}{1 mm}
\begin{picture} (160,160)
\put(0,0) {\epsfig{file=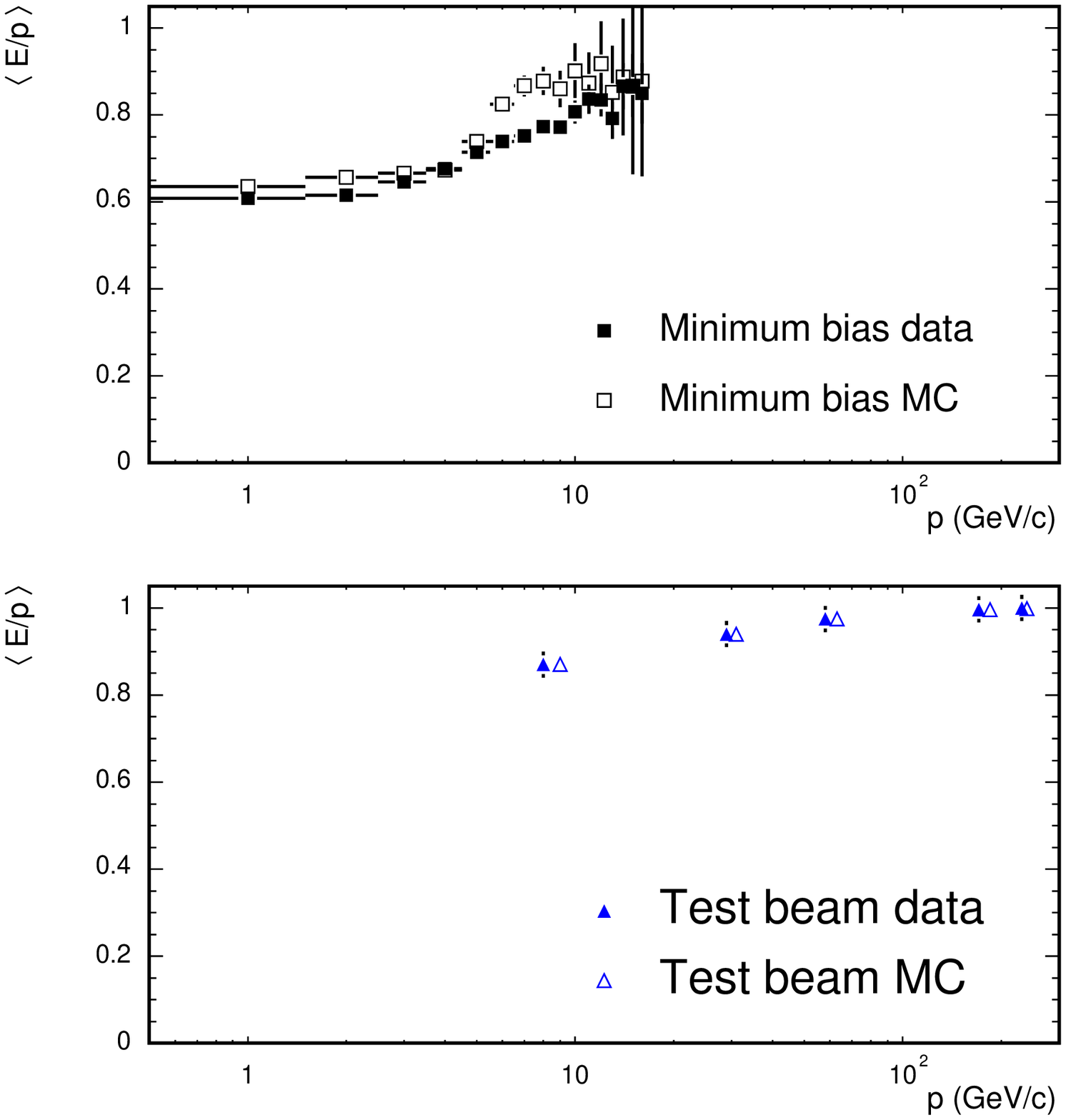,width=0.9\textwidth}}
\end{picture}
\caption{\label{fig:eopplugcen} \sl $\langle E/p \rangle $ versus $p$
for the plug calorimeters. The upper plot shows a comparison of single
track and minimum bias data (closed squares) and MC (open
squares). The bottom plot shows the comparison of test beam data
(closed triangles) and single track MC (open triangles).  The tuning
of the calorimeter was done using tracks up to 5 GeV/$c$. For the test
beam comparison the MC points have been shifted slightly in $p$-value
to allow a better visual comparison. }
\end{center}
\end{figure}

Figure \ref{fig:eopplugcen} shows $\langle E/p \rangle $ versus $p$
for the plug calorimeter for minimum bias data and test beam data
compared with the corresponding simulations. At low momenta the
response is about 60\%, increasing to nearly 100\% at high momenta.

Due to the discrepancies described above the plug calorimeter
simulation is not used for determining the absolute jet energy
scale. An additional calibration is made where the plug calorimeter
response is calibrated with respect to the response of the central
calorimeter using dijet events as will be described in Sec.
\ref{sec:RelCorr}. Therefore, no systematic uncertainty is associated
directly with the $\langle E/p \rangle$ measurements for the plug
calorimeter.

\subsection{Calorimeter Response to Electromagnetic Particles}\label{sec:emSimulation}

On average about 30\% of the particles in a jet are neutral pions
which mostly decay into two photons: $\pi^0 \to
\gamma\gamma$. Therefore a good understanding of the calorimeter
response to electromagnetic particles is important.  The
electromagnetic single particle response is studied similarly to the
hadronic response, using the track momentum as reference and EM
energies measured in the target tower and the two towers adjacent in
$\eta$ around the track impact point.

\begin{figure}[htbp]
  \begin{center} 
  \includegraphics[width=\linewidth,clip=]{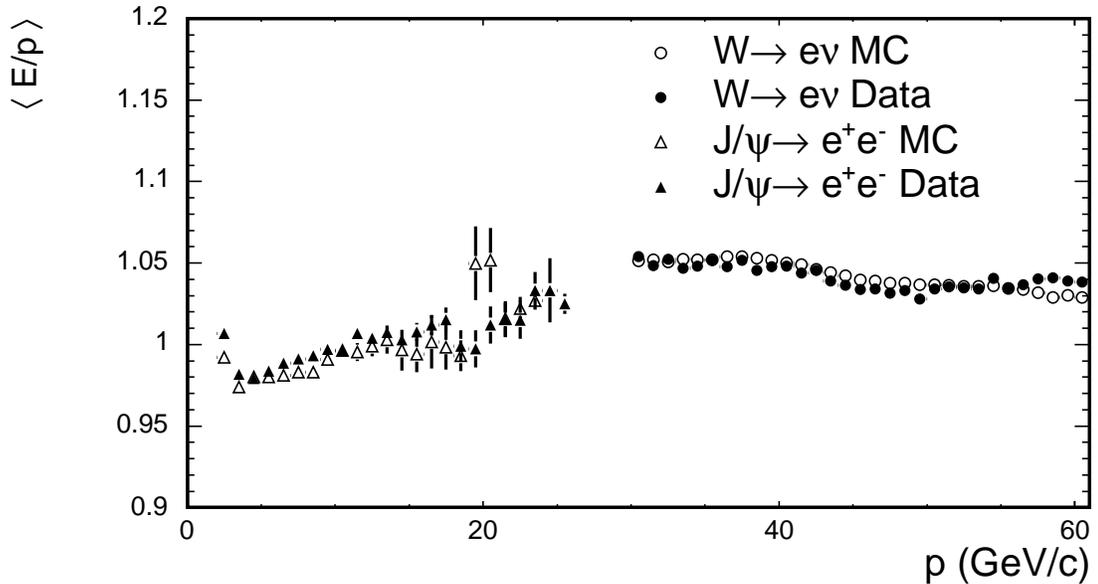}
   \caption{\sl $\langle E/p \rangle$ versus $p$ for electrons and positrons from
$J/\psi \to e^+ e^-$ and $W\to e^\pm \nu_e$ data (closed triangles and
circles) and MC samples (open triangles and circles) samples.
\label{fig:electron_eop}}
  \end{center}
\end{figure}

A comparison of the $\langle E/p \rangle$ response for electrons and positrons
is shown in Fig.~\ref{fig:electron_eop} in $W\to e^\pm \nu_e$ and
$J/\psi\rightarrow e^+e^-$ events. Overlaid is the 
simulated response from corresponding MC samples.
A certain momentum dependence of $\langle E/p \rangle$ arises from the $W$
and $J/\psi$ selection cuts and due to final state radiation of
photons. The simulation reproduces the data to better than 1\%
accuracy. 

\subsection{Uncertainties}
\label{sec:SystSim}

Figure \ref{fig:eopsyst} is used to estimate the systematic
uncertainty on the central calorimeter energy determined from the
difference between data and simulation for charged hadrons:

\begin{figure}[htbp]
  \begin{center} 
  \includegraphics[width=\linewidth,clip=]{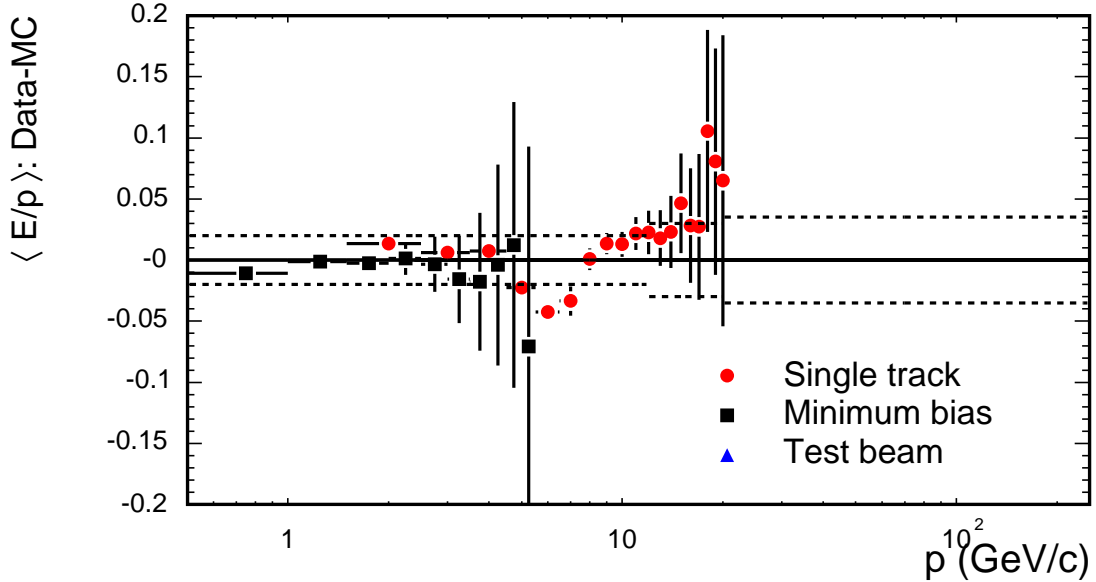}
\caption{\sl \label{fig:eopsyst} Difference between $\langle E/p
\rangle$ of data and MC versus particle momentum $p$. Shown is the
difference for minimum bias Data and MC (squares), single track data
and MC (circles) and test beam data and MC (triangles). The dashed
lines indicate the quoted systematic uncertainties. The error bars on
the minimum bias and single track data are statistical. For the test
beam data we do not display any uncertainty.}
  \end{center}
\end{figure}

We have derived the following momentum dependent estimates for the 
average differences:
\begin{itemize}
\item $1.5\%$ for $p<12$ GeV/$c$
\item $2.5\%$ for $12<p<20$ GeV/$c$
\item $3.5\%$ for $p>20$ GeV/$c$
\end{itemize}
The uncertainties at $p<20$ GeV/$c$ directly reflect the limited
performance of the calorimeter simulation as well as limited available
single isolated track statistics at medium momenta. The uncertainties
at $p>20$ GeV/c are due to uncertainties in the test beam momentum
scale ($2\%$), the shorter integration time in the CDF detector
readout compared to the test beam measurement ($1.5\%$) (see Ref.
\cite{cemtestbeam,chatestbeam} and Sec. \ref{sec:Detector}).
Adding those uncertainties linearly we obtain the systematic
uncertainty for $p>20$ GeV/$c$ of $3.5\%$.

\begin{figure}[h]
  \begin{center}
  \includegraphics[width=\linewidth,clip=]{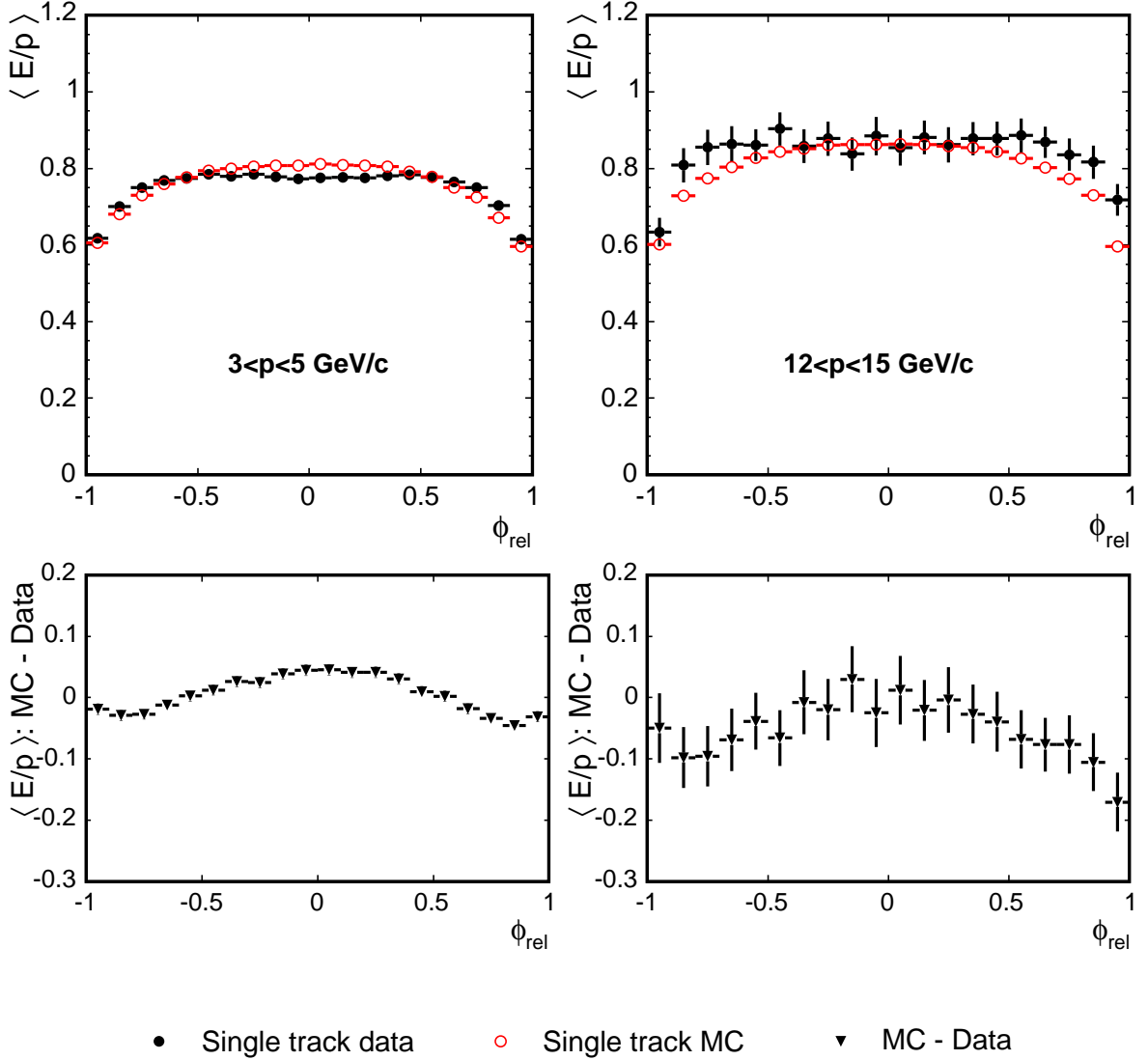}
  \caption{\sl $\langle E/p \rangle$ vs. $\phi_{rel}$ for particles
   with momenta between 3 and 5 GeV/$c$ (left) and for particles with
   momenta between $12$ and $16$ GeV/$c$ (right).  Top: data (full points) and
   MC (open squares).  Bottom: difference between data and MC.
\label{fig:eopphi}}
  \end{center}
\end{figure}

The $\langle E/p \rangle$ measurements presented so far are only
sensitive to the inner 81\% of the tower. In particular the
instrumentation between the tower $\phi$-boundaries is limited, and
the exact modeling of this region in the simulation is
difficult. Figure \ref{fig:eopphi} shows $\langle E/p \rangle$ versus
relative $\phi$ ($\phi_{rel}$), that is the azimuthal angle of the
track impact point with respect to the target tower center, normalized
to the $\phi$ of the tower edges such that $\phi_{rel}=0$ represents
the tower center and $\phi_{rel}=\pm 1$ the tower boundaries. Shown
are data and simulation for $p=3-5$~GeV/$c$ and $p=12-16$~GeV/$c$.

Data and MC differ by up to $10\%$ near the $\phi$-boundaries 
($|\phi_{rel}|>0.9$). Similar discrepancies are seen at the $\eta$ boundaries
of the towers. 
These 10\% differences are taken as systematic uncertainty. 
Since these boundaries in $\eta$ and $\phi$ correspond to 
19\% of the total tower area, the $10\%$ uncertainty at the boundaries
translates into a $1.9\%$ systematic uncertainty on the overall particle
response to charged hadrons.

\begin{figure}[htbp]
  \begin{center}
\setlength{\unitlength}{1 mm}
\begin{picture} (160,70)
\put(0,0) {\epsfig{file=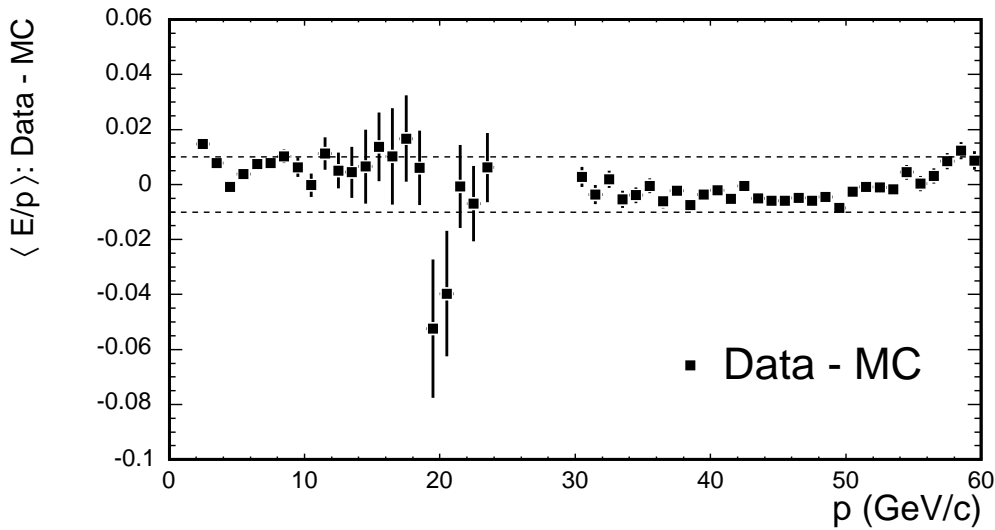,width=0.9\textwidth}}
\end{picture}
\caption{\sl Difference between $\langle E/p \rangle$ in data and
simulation versus $p$ for electrons from $J/\psi \to e^+ e^-$ and
$W\to e^\pm \nu_e$ (closed squares). The individual distributions for
data and simulation are shown in Fig. \ref{fig:electron_eop}. The
dashed line indicates the systematic uncertainty on the
electromagnetic energy scale.
\label{fig:electron_eop_rat}}
  \end{center}
\end{figure}

Figure \ref{fig:electron_eop_rat} shows the difference in $\langle E/p
\rangle$ between data and simulation for electrons. The data and
the simulation agree to within 1\% which is taken as systematic
uncertainty. This measurement only uses tracks pointing to the inner 84\% of a
tower in azimuth, $\phi_{rel}$, as a
consequence of the electron selection that involves a CES energy cluster
fiducial cut.

\begin{figure}[htbp]
  \begin{center}
\setlength{\unitlength}{1 mm}
\begin{picture} (160,140)
\put(0,0) {\epsfig{file=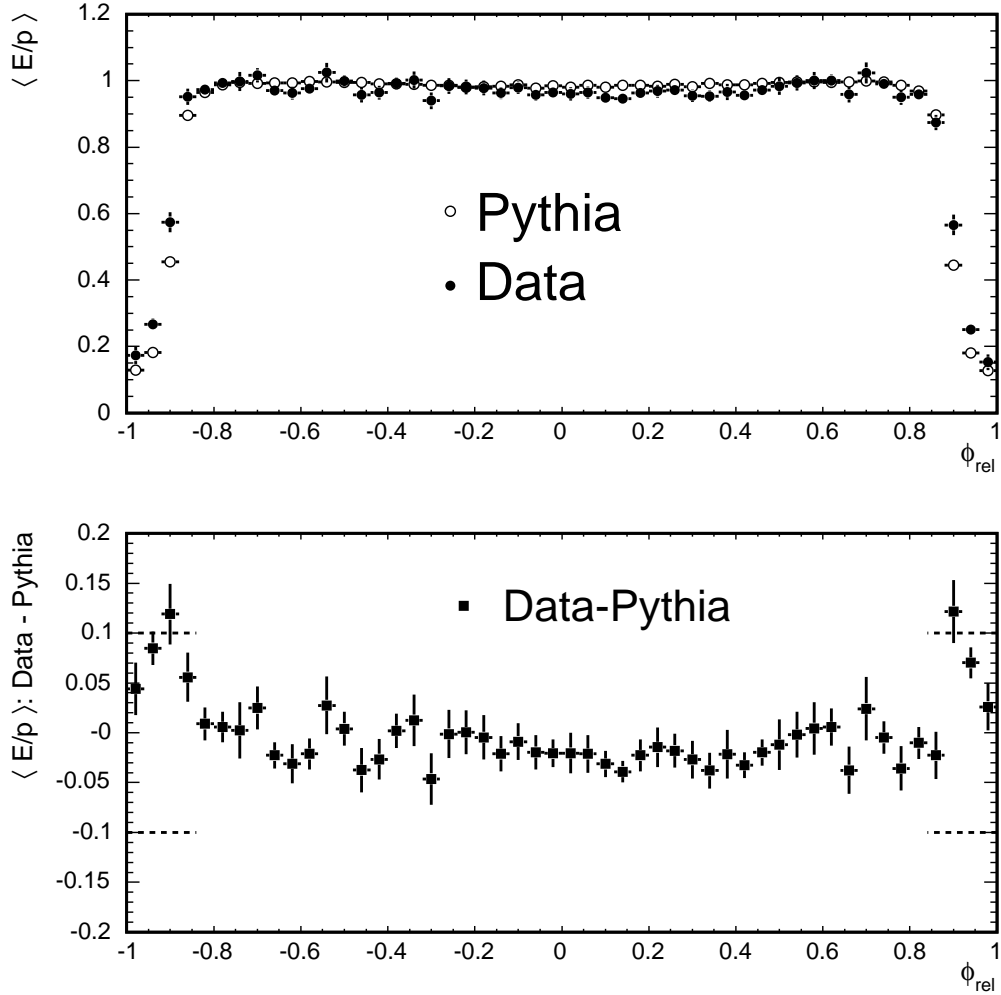,width=0.9\textwidth}}
\end{picture}
\caption{\sl Top: $\langle E/p \rangle$ versus $\phi_{rel}$ for electrons in $Z\rightarrow
e^+e^-$ events. $\langle E/p \rangle$ is shown as a function of relative
$\phi$.  Shown are data (closed points) and simulation (open circles).
Bottom: The difference in $\langle E/p \rangle$ between data and
simulation for electrons.
\label{fig:xces}}
\end{center}
\end{figure}

The remaining $16\%$ at the boundaries in $\phi$ are studied using
$Z\rightarrow e^+e^-$ events where one electron is identified using
the standard criteria~\cite{wecross} and the second electron is just
taken to be the highest momentum track with $35<p<55$~GeV/$c$ in the
angular range covered by the central calorimeter. There is no
requirement on the energy measured in the calorimeter and thus the
measurement is unbiased toward the calorimeter response. The
invariant mass between the electron and the candidate track is
required to be within 10~GeV/$c^2$ of the $Z$ boson mass. For these
track based electrons candidates, $\langle E/p \rangle$ is shown in
Fig.~\ref{fig:xces} as function of $\phi_{rel}$.

Figure \ref{fig:xces} also shows the difference between data and
simulation in $\langle E/p \rangle$ in $Z\to e^+e^-$ events as
a function of relative $\phi$. A discrepancy of about 10\% is observed
between data and simulation at $|\phi_{rel}|>0.84$. This region
corresponds to 16\% of the tower area, thus causing an overall 1.6\%
difference between data and simulation for the response of
electromagnetic particles.

A summary of the systematic uncertainties is given in Table
\ref{tab:epsys}. These uncertainties are uncorrelated and added in
quadrature. The overall systematic uncertainty on the energy of
charged hadrons is thus $2\%$ for $p<12$ GeV/$c$, $3\%$ for
$12<p<20$ GeV/$c$ and $4\%$ for $p>20$ GeV$/c^2$. The total systematic
uncertainty on the energy of electromagnetically showering
particles is $1.7\%$ for the entire momentum range studied.

\begin{table}[htbp]
\begin{center}
\begin{tabular}{|l|ccc|}
\hline
$p$ (GeV/$c$) & 0-12 & 12-20 & $>$20 \\\hline\hline
$\langle E/p \rangle$ response to hadrons &     &    &  \\ \hline
Total tower (\%)               & 1.5 &2.5 & 3.5 \\ 
Near tower $\phi$ and $\eta$-boundaries (\%) & 1.9 &1.9 & 1.9 \\\hline\hline
Total for hadrons(\%)    & 2.5 &3.0 & 4.0 \\\hline\hline
$\langle E/p \rangle$ response to EM particles    &     &    &  \\ \hline
Total tower (\%)                & 1.0 &1.0 & 1.0 \\ 
Near tower $\phi$-boundary (\%) & 1.6  &1.6  & 1.6 \\\hline\hline
Total for EM particles(\%)        & 1.7 &1.7 & 1.7 \\
\hline
\end{tabular}
\caption{Summary of the relative uncertainties due to the modeling
of the simulation of charged hadron showers and electromagnetic (EM)
showers. All the uncertainties refer to the effect on the entire
tower energy (see text). The total uncertainty is the quadratic sum of
the individual uncertainties weighted with the area covered by the
individual measurements.
\label{tab:epsys}
}
\end{center}
\end{table}

\clearpage

\section{ $\eta$-dependent Corrections}\label{sec:RelCorr}

Even after the calorimeter energy calibrations described in Sec.
\ref{sec:Detector}, the response of the CDF calorimeter is not uniform
in pseudo-rapidity. The dependencies on $\eta$ arise from the
separation of calorimeter components at $\eta = 0$ where the two
halves of the central calorimeter join and at $|\eta| \approx 1.1$
where the plug and central calorimeters join. The different responses
of the plug and central calorimeters also cause a dependence on
$\eta$. The $\eta$-dependent corrections are introduced to flatten the
$\eta$ dependence of the calorimeter response. The method implicitly
also includes corrections for both the transverse spreading of
calorimeter showers outside the jet cone and any $\eta$ dependence of
gluon radiation and multiple parton interactions.

\subsection{Correction Procedure}
\label{sec:relcor_method}

The $\eta$-dependent corrections are obtained using the ``dijet
balancing method''.  They are determined based on the assumption that
the two jets in dijet events should be balanced in $p_T$ in absence of
hard QCD radiation. To determine the corrections, we define a jet with
$0.2<|\eta|<0.6$ as a ``trigger jet'' and define the other jet as a
``probe jet''.  When both jets are in the region of
$0.2<|\eta_{jet}|<0.6$, the trigger and probe jets are assigned
randomly.

The $p_T$ balancing fraction $f_b$ is then formed:
\begin{equation}
  f_b \equiv \frac{\Delta p_T}{p_T^{ave}} =
  \frac{p_T^{probe}-p_T^{trigger}}{(p_T^{probe}+p_T^{trigger})/2}
  \label{eqn:dptf}
\end{equation}
where $p_T^{trigger}$ and $p_T^{probe}$ are the transverse momenta of
the trigger and probe jet, respectively. The correction factor
required to correct the probe jet can then be inferred through
\begin{equation}
  \beta_{dijet} \equiv
  \frac{2+\langle f_b \rangle}{2-\langle f_b \rangle}
\end{equation}

Note, that $\beta_{dijet}$ is mathematically equal to
$p_T^{probe}/p_T^{trigger}$. However, by inferring $\beta_{dijet}$
from $f_b$ we reduce the sensitivity of our measurement to
non-Gaussian tails since the $f_b$ distribution is in good
approximation a Gaussian distribution unlike the distribution of
$p_T^{probe}/p_T^{trigger}$.

The $\eta$-dependent corrections are defined as $1/\beta_{dijet}$ and
they are determined separately for data and MC and for different
$p_T^{jet}$ bins. For corrections obtained from data the samples
single-tower-5, jet-20, jet-50, jet-70, and jet-100 are used. The
corrections for the MC are obtained from the jet samples generated
with the {\tt PYTHIA} MC.

Several cuts are placed to reduce the effects of QCD radiation:

\begin{itemize}

\item The jets are required to be back to back in the $r-\phi$ plane,
i.e. the difference of their azimuthal angle,
$\Delta\phi(\mbox{jet}_{probe},\mbox{jet}_{trigger})$, should be
larger than $2.7$ radians.

\item If a 3rd jet is present in the event, the $p_T$ of this 3rd
jet should be less than $7$ GeV/$c$ for minimum bias data, less than
$8$ GeV/$c$ for the samples triggered on $20$~GeV or $50$~GeV jets,
and less than $10$ GeV/$c$ for the samples triggered on $70$~GeV or
$100$~GeV jets.

\item The average $p_T^{ave}=(p_{T}^{jet1}+p_{T}^{jet2})/2$ of the two
jets is required to be $5$ GeV/$c$ higher than the trigger threshold
of the respective sample.

\item The significance of missing $E_T$ is defined as $\met/\sqrt{\sum_i E_{T,i}}$, where the sum extends to all the calorimeter towers. It is required to be less than $2+0.018 \times p_T^{leading-jet}$ for $p_T^{leading-jet}>55$ GeV/$c$ and less than 3 $\sqrt{\mbox{GeV}}$ otherwise.

\end{itemize}

\begin{figure}[h]
  \begin{center}
  \includegraphics[width=\linewidth,clip=]{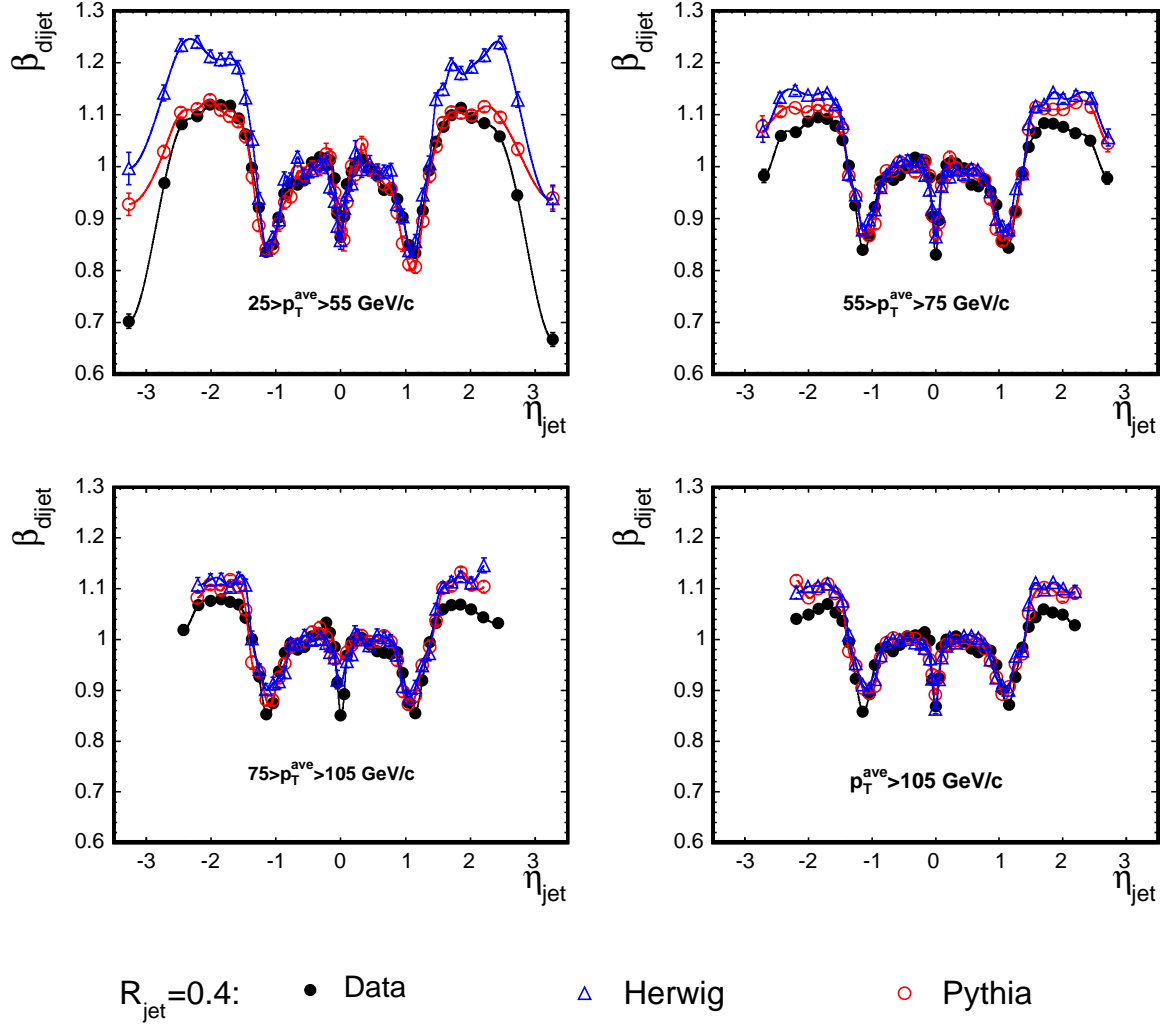} 
   \caption{\sl Dijet balance,
   $\beta_{dijet}=p_T^{probe}/p_T^{trigger}$, as a function of
   $\eta_{jet}$ in data, {\tt HERWIG} and {\tt PYTHIA} MC samples for
   $R_{jet}=0.4$ jets. Shown are the corrections for jet-$20$,
   jet-$50$, jet-$70$ and jet-$100$ jet samples, corresponding to
   $25<p_T^{ave}<55$ GeV/$c$, $55<p_T^{ave}<75$ GeV/$c$,
   $75<p_T^{ave}<105$ GeV/$c$ and $p_T^{ave}>105$ GeV/$c$,
   respectively. The lines show the interpolation between the
   individual measurements used for correcting jets.}
   \label{relplot_datamc_04}
  \end{center}
\end{figure}

\begin{figure}[h]
  \begin{center}
  \includegraphics[width=\linewidth,clip=]{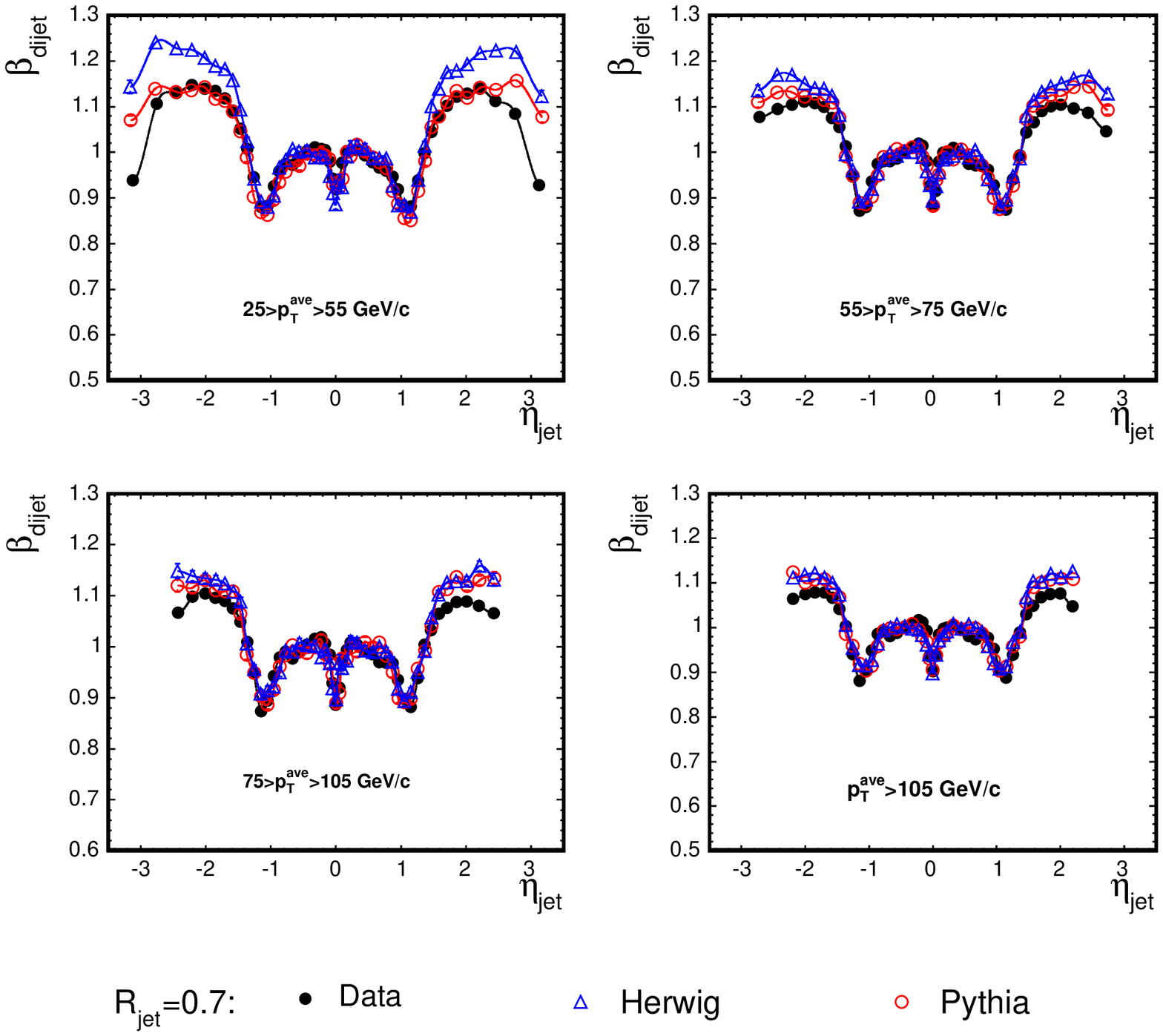} 
   \caption{\sl Dijet balance,
   $\beta_{dijet}=p_T^{probe}/p_T^{trigger}$, as a function of
   $\eta_{jet}$ in data, {\tt HERWIG} and {\tt PYTHIA} MC samples for
   $R_{jet}=0.7$ jets. Shown are the corrections for jet-$20$,
   jet-$50$, jet-$70$ and jet-$100$ jet samples, corresponding to
   $25<p_T^{ave}<55$ GeV/$c$, $55<p_T^{ave}<75$ GeV/$c$,
   $75<p_T^{ave}<105$ GeV/$c$ and $p_T^{ave}>105$ GeV/$c$,
   respectively. The lines show the interpolation between the
   individual measurements used for correcting jets.}
   \label{relplot_datamc_07}
  \end{center}
\end{figure}

\begin{figure}[h]
  \begin{center}
  \includegraphics[width=\linewidth,clip=]{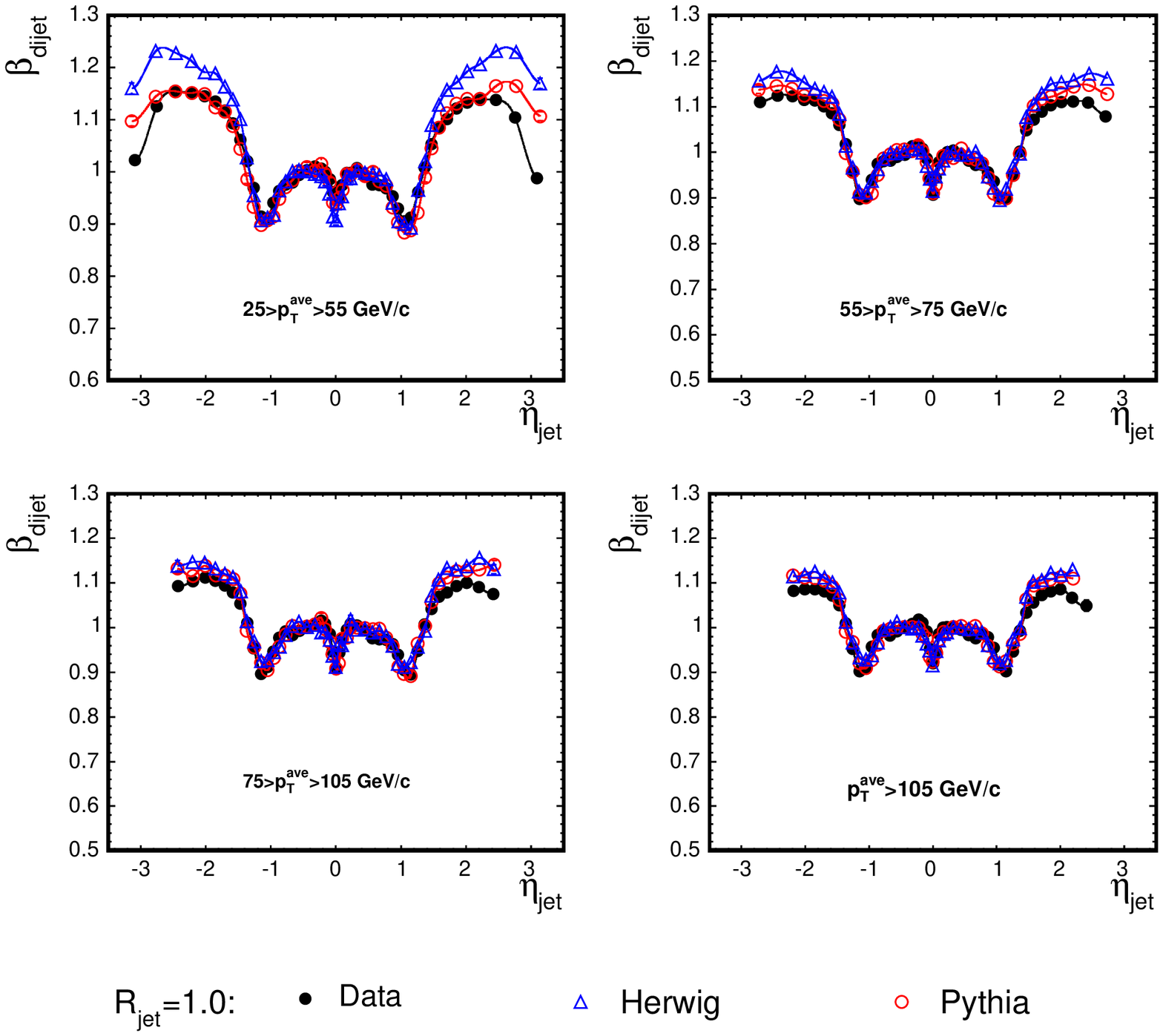} 
   \caption{\sl Dijet balance,
   $\beta_{dijet}=p_T^{probe}/p_T^{trigger}$, as a function of
   $\eta_{jet}$ in data, {\tt HERWIG} and {\tt PYTHIA} MC samples for
   $R_{jet}=1.0$ jets. Shown are the corrections for the jet-$20$,
   jet-$50$, jet-$70$ and jet-$100$ jet samples, corresponding to
   $25<p_T^{ave}<55$ GeV/$c$, $55<p_T^{ave}<75$ GeV/$c$,
   $75<p_T^{ave}<105$ GeV/$c$ and $p_T^{ave}>105$ GeV/$c$,
   respectively. The lines show the interpolation between the
   individual measurements used for correcting jets.}
   \label{relplot_datamc_10}
  \end{center}
\end{figure}

In Fig. \ref{relplot_datamc_04}, the dijet balance $\beta_{dijet}$ is
shown for data, {\tt HERWIG} and {\tt PYTHIA} MC samples for a jet
cone size of $R_{jet}=0.4$ and four transverse momentum regions
$25<p_T^{ave}<55$ GeV/$c$, $55<p_T^{ave}<75$ GeV/$c$,
$75<p_T^{ave}<105$ GeV/$c$ and $p_T^{ave}>105$ GeV/$c$. Figures
\ref{relplot_datamc_07} and
\ref{relplot_datamc_10} show the equivalent plots for $R_{jet}=0.7$
and $R_{jet}=1.0$, respectively. The lines in
Fig. \ref{relplot_datamc_04}-\ref{relplot_datamc_10} show the
interpolation between the individual measurements, and inverse of
these functions is taken to be the $\eta$-dependent correction factor.

It is seen that $\beta_{dijet}\approx 1$ in the region where the
trigger jet is selected, $0.2<|\eta|<0.6$, for both the data and the
simulation. The dips at $|\eta|\approx 0$ and $\pm 1.1$ are due to the
gap between calorimeters in these regions, resulting in a lower
average response. In the plug region, $|\eta|>1.2$, the calorimeter
response is higher than in the central region by about 10\% at low
$p_T^{ave}$ and $5\%$ at high $p_T^{ave}$. 

Both {\tt PYTHIA} and {\tt HERWIG} reproduce the data well up to
$|\eta|=1.4$ at all $p_T^{ave}$.  At larger $|\eta|$ and for
$p_T^{ave}>55$ GeV/$c$ a difference of about 4\% between data and
simulation is observed. At $p_T^{ave}<55$ GeV/$c$ {\tt HERWIG} differs
significantly from the data and {\tt PYTHIA} in the forward region.
Due to this large discrepancy we derived the MC corrections from the
{\tt PYTHIA} MC. Further discussions of these corrections and {\tt
HERWIG} MC can be found in Sec. \ref{sec:crosschecks}.

\begin{figure}[h]
  \begin{center}
  \includegraphics[width=\linewidth,clip=]{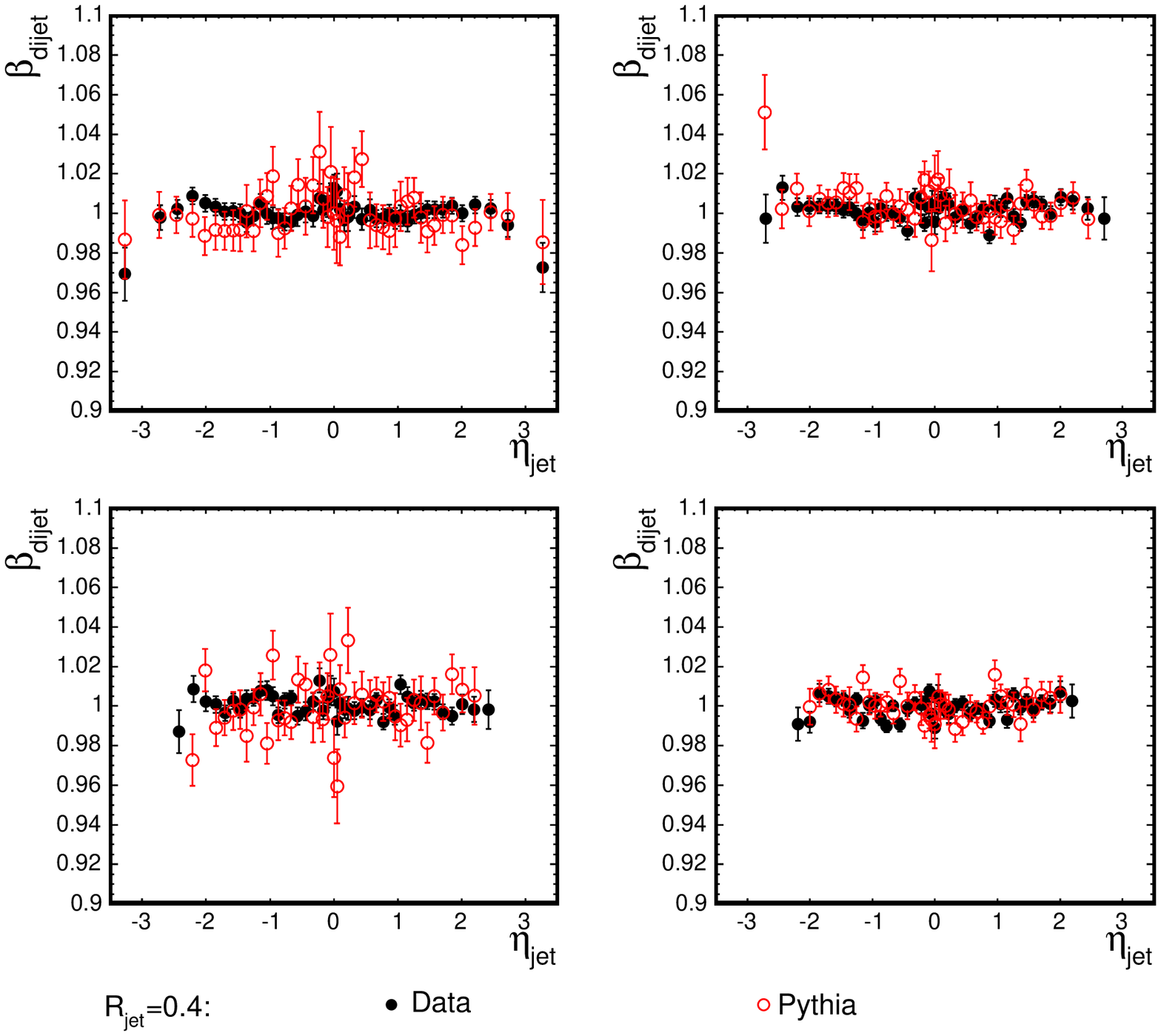} 
   \caption{\sl Dijet balance,
   $\beta_{dijet}=p_T^{probe}/p_T^{trigger}$, as a function of
   $\eta_{jet}$ in data and {\tt PYTHIA} MC samples for cone size
   $R_{jet}=0.4$ after applying the $\eta$-dependent corrections. Shown are
   the jet-$20$ (top left), jet-$50$ (top right), jet-$70$ (bottom
   left) and jet-$100$ (bottom right) jet samples, corresponding to
   $25<p_T^{ave}<55$ GeV/$c$, $55<p_T^{ave}<75$ GeV/$c$,
   $75<p_T^{ave}<105$ GeV/$c$ and $p_T^{ave}>105$ GeV/$c$,
   respectively. }  \label{relplot_datamc_after}
  \end{center}
\end{figure}

\begin{figure}[h]
  \begin{center}
  \includegraphics[width=\linewidth,clip=]{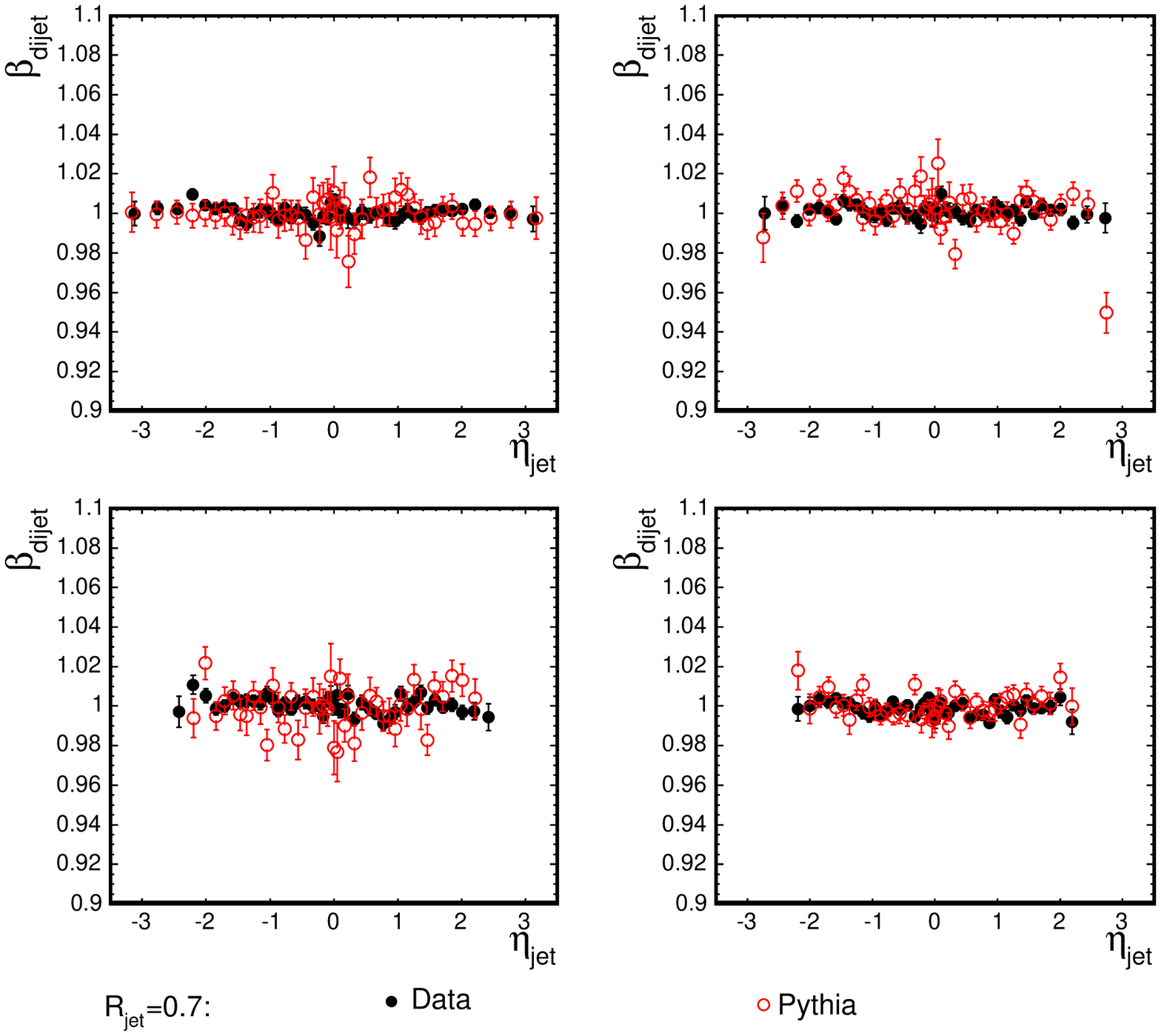} 
   \caption{\sl Dijet balance,
   $\beta_{dijet}=p_T^{probe}/p_T^{trigger}$, as a function of
   $\eta_{jet}$ in data and {\tt PYTHIA} MC samples for cone size
   $R_{jet}=0.7$ after applying the $\eta$-dependent corrections. Shown are
   the jet-$20$ (top left), jet-$50$ (top right), jet-$70$ (bottom
   left) and jet-$100$ (bottom right) jet samples, corresponding to
   $25<p_T^{ave}<55$ GeV/$c$, $55<p_T^{ave}<75$ GeV/$c$,
   $75<p_T^{ave}<105$ GeV/$c$ and $p_T^{ave}>105$ GeV/$c$,
   respectively. }  \label{relplot_datamc_after07}
  \end{center}
\end{figure}

\begin{figure}[h]
  \begin{center}
  \includegraphics[width=\linewidth,clip=]{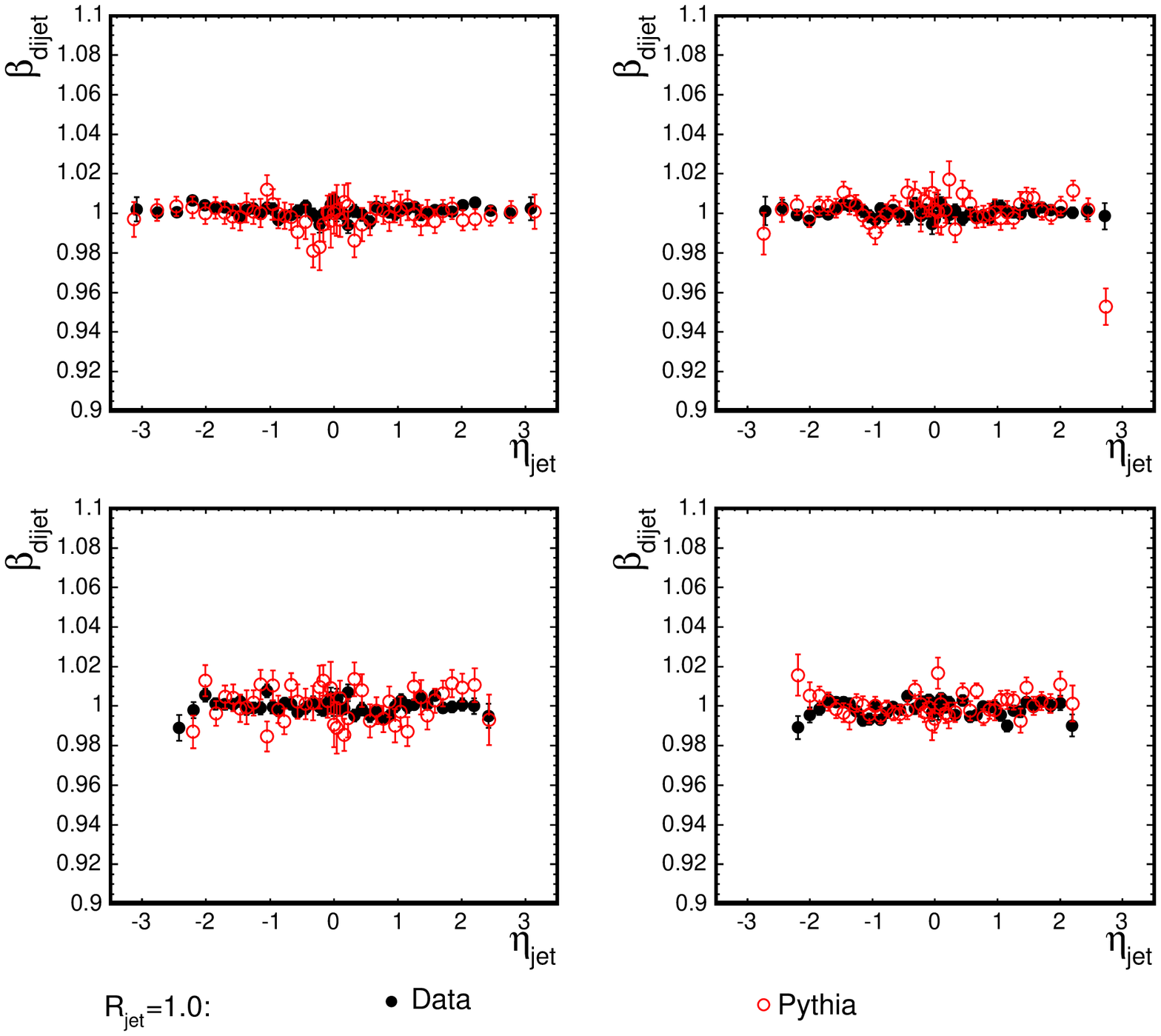} 
   \caption{\sl Dijet balance,
   $\beta_{dijet}=p_T^{probe}/p_T^{trigger}$, as a function of
   $\eta_{jet}$ in data and {\tt PYTHIA} MC samples for cone size
   $R_{jet}=1.0$ after applying the $\eta$-dependent  corrections. Shown are
   the jet-$20$ (top left), jet-$50$ (top right), jet-$70$ (bottom
   left) and jet-$100$ (bottom right) jet samples, corresponding to
   $25<p_T^{ave}<55$ GeV/$c$, $55<p_T^{ave}<75$ GeV/$c$,
   $75<p_T^{ave}<105$ GeV/$c$ and $p_T^{ave}>105$ GeV/$c$,
   respectively.}  \label{relplot_datamc_after10}
  \end{center}
\end{figure}

\subsection{Uncertainties}
\label{sec:relcor_syserr}

If this method was perfect the dijet balance applied to samples after
these $\eta$-dependent corrections would be a flat line at 1. Figures
\ref{relplot_datamc_after} to \ref{relplot_datamc_after10} show the
data and {\tt PYTHIA} MC after corrections. It is seen that the
corrections indeed flatten out the response as desired. The remaining
discrepancies from a flat distribution are due to the limitations of
the parameterization of the $\eta$- and $p_T$-dependence of the
correction and are taken as part of the systematic uncertainty of the
corrections.

The systematic uncertainties are determined by varying the event
selection requirements and the fitting procedure. Specifically, we
varied the cut on the $p_T$ of the 3rd jet and the significance of
$\met$. Any deviation of the dijet balance $\beta_{dijet}$ from unity
is taken as a systematic uncertainty.

The overall systematic uncertainty from the sources described above
are summarized in Table~\ref{tab:rel1}. In the low $p_T$ region
minimum bias data and MC are used up to $p_T<$15 GeV/$c$ and
single-tower-5 data and the MC jet sample with $\hat p_T>10$ GeV/$c$
for $15<p_T<25$ GeV/$c$. Here the systematic uncertainties are
largest.

\begin{table}[tbp]
\caption{Systematic uncertainties of the $\eta$-dependent  corrections versus $p_T^{jet}$ and $|\eta_{jet}|$.}
{\small
\setlength{\extrarowheight}{1pt}
\begin{center}
\begin{tabular}{l|ccccccc} \hline\hline
 $|\eta|$ range      & $0.0-0.2$
                      & $0.2-0.6$
                      & $0.6-0.9$
                      & $0.9-1.4$
                      & $1.4-2.0$
                      & $2.0-2.6$
                      & $2.6-3.6$  \\ \hline
  $p_T<12$  GeV/$c$        & 1.5~\% & 0.5~\% & 1.5~\% & 2.5~\% & 1.5~\% & 5.0~\% & 7.5~\%  \\ 
  $12\le p_T<25$  GeV/$c$  & 1.5~\% & 0.5~\% & 1.5~\% & 1.5~\% & 1.5~\% & 3.0~\% & 6~\%    \\
  $25\le p_T<55$  GeV/$c$  & 1.0~\% & 0.5~\% & 1.0~\% & 1.0~\% & 0.5~\% & 1.5~\% & 6~\%    \\
  $p_T\ge 55$ GeV/$c$   & 0.5~\% & 0.5~\% & 0.5~\% & 0.5~\% & 0.5~\% & 1.5~\% & 6~\%    \\

\hline\hline
\end{tabular}
\label{tab:rel1}
\end{center}
}
\end{table}

\clearpage

\section{Absolute Jet Energy Scale}\label{sec:Absolute}

The absolute correction aims to transform the jet energy measured in
the calorimeter into the energy corresponding to the underlying
particle jet. After this correction the energy scale of a jet is
independent of the CDF detector. Since the calorimeter simulation has
been optimized to reproduce the measured single particle response, we
rely on the simulation to derive corrections over a large range of jet
transverse momenta.

The accuracy of this method depends on how well jets are modeled by
the simulation.  In particular it depends on the multiplicity and
$p_T$ spectrum of the particles inside a jet and on the response of
the calorimeter to an individual particle.  These two components are
tested separately and propagated into a systematic uncertainty on the
absolute correction.

In principle, this correction depends on the initial parton type,
e.g. a jet originating from a gluon has on average a larger
multiplicity than one originating from a quark and thus may require a
different correction. However, the corrections are not derived
separately for each parton type since the parton type in the data is
$a$ $priori$ unknown.

\subsection{Correction Procedure \label{sec:absproc}}

The absolute jet energy is defined as the most probable value for a
jet transverse momentum, $p_T^{jet}$, given a particle jet with a
fixed value of $\ptpartjet$.  The corresponding probability density
function, $dP$, is parametrized as a function of $\Delta
p_T=\ptpartjet-p_T^{jet}$ according to
\begin{eqnarray}
\rd {\cal P}(\ptpartjet,p_T^{jet}) & = & f(\Delta p_T) \rd \ptpartjet \rd p_T^{jet} \nonumber \\
f(\Delta p_T) & = & \frac{1}{\sqrt{2\pi}(\sigma_1+N_2\sigma_2)}
[e^{-\frac{1}{2}(\frac{(\Delta p_T-\mu_1}{\sigma_1})^2}
	+ N_2 e^{-\frac{1}{2}(\frac{\Delta p_T-\mu_2}{\sigma_2})^2}]  
\label{eq:2gaus}
\end{eqnarray}
One Gaussian function describes the tails while the other one
reproduces the core of the distribution. Their relative contributions
are determined by the normalization of the second Gaussian, $N_2$.

\begin{figure}[h]
  \begin{center} \includegraphics[width=\linewidth,clip=]{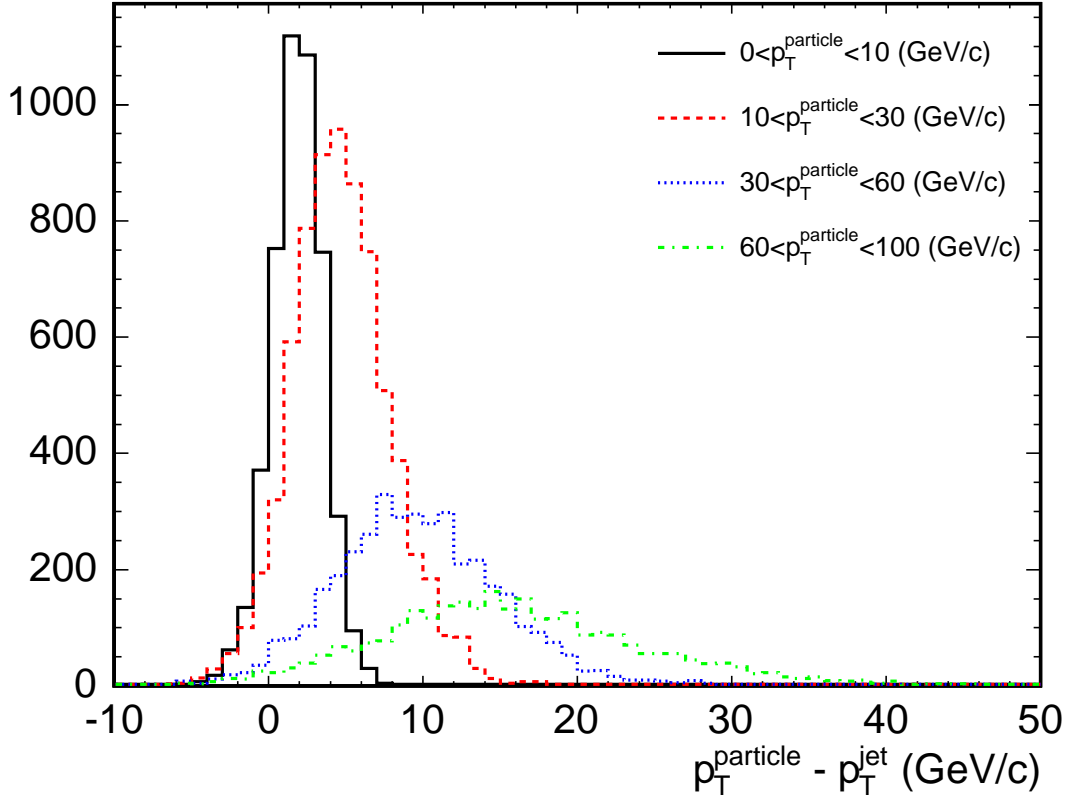}
  \caption{\sl $\Delta p_T= \ptpartjet-\ptcaljet$ for $R_{jet}$=0.4 jets
  matched using $\Delta R<$ 0.1 for different $\ptpartjet$ bins.
\label{fig:hadcal2}}
  \end{center}
\end{figure}

The parameters $\mu_1$, $\sigma_1$, $\mu_2$, $\sigma_2$ and $N_2$
depend on $\ptpartjet$ as shown by the different shapes of the
histograms in Fig. \ref{fig:hadcal2}. The dependence is modeled by a
linear parameterization of the form:
\begin{eqnarray}
\sigma_{1,2} &=&\sigma^a_{1,2}+\sigma^b_{1,2} \ptpartjet\\
\mu_{1,2} &=&\mu^a_{1,2}+\mu^b_{1,2} \ptpartjet\\
      N_2 &=& N_2^a+N_2^b \ptpartjet
\end{eqnarray}
using a total of $10$ parameters.

The number of jets $n$ with a $p_T$ between $p_T^{jet}$ and
$p_T^{jet}+dp_T^{jet}$ and given particle jet with a $p_T$ between
$p_T^{particle}$ and $p_T^{particle}+dp_T^{particle}$ is given by the
convolution of $\rd{\cal P}(p_{Ti}^{jet},\ptpartjeti)$ with the $p_T$
spectrum of the particle jets $\rd{\cal P}(\ptpartjeti)$. That
is,

\begin{equation}
n(p_T^{jet},p_T^{particle})dp_T^{jet}dp_T^{particle} = n(\ptpartjet) \times \rd {\cal P}(p_T^{jet},\ptpartjet)
\label{eq:tf1}
\end{equation}

The likelihood has the form:
\begin{equation}
{\cal L} = \Pi_{i} \rd {\cal P}(\ptpartjeti) \times \rd {\cal P}(p_T^{jet},\ptpartjeti)
\label{eq:tf11}
\end{equation}
where the product goes over all particle jets. An unbinned likelihood
fit is used to extract the parameters of $f(\Delta p_T)$ parameters,
maximizing the logarithm of the likelihood

\begin{equation}
\log {\cal L} = \sum_{i=1}^N \log \rd {\cal P}(\ptpartjeti)+ \sum_{i=1}^N \log f(\Delta p_{T,i})
\label{eq:tf2}
\end{equation}
where N is the number of particle jets. The first term is independent
of the parameters to determine and is thus ignored.

The dijet {\tt PYTHIA} MC samples, described in Sec. \ref{sec:Datasets}, are
used to calculate the likelihood. Jets are reconstructed at the
calorimeter and particle level using the standard CDF jet clustering
algorithm with cone radii of 0.4, 0.7 and 1.0. Jets are required to be
in the central region ($0.2<|\eta|<0.6$) and to be one of the two
leading jets. Each particle jet is required to match its closest
calorimeter jet within $\Delta
R=\sqrt{(\Delta\phi)^2+(\Delta\eta)^2}<0.1$. In total we selected
about 50,000 particle jets matched to calorimeter jets, with $p_T$
ranging from $0$ to $600$ GeV/$c$. The difference between the particle
jet $p_T$ and the calorimeter jet $p_T$ is shown in
Fig. \ref{fig:hadcal2} for four example $p_T$ ranges. The
distributions are not centered at zero as expected and have widths
that change with $p_T$ according to Eq. \ref{eq:2gaus}.

\begin{figure}[h]
  \begin{center} \includegraphics[width=\linewidth,clip=]{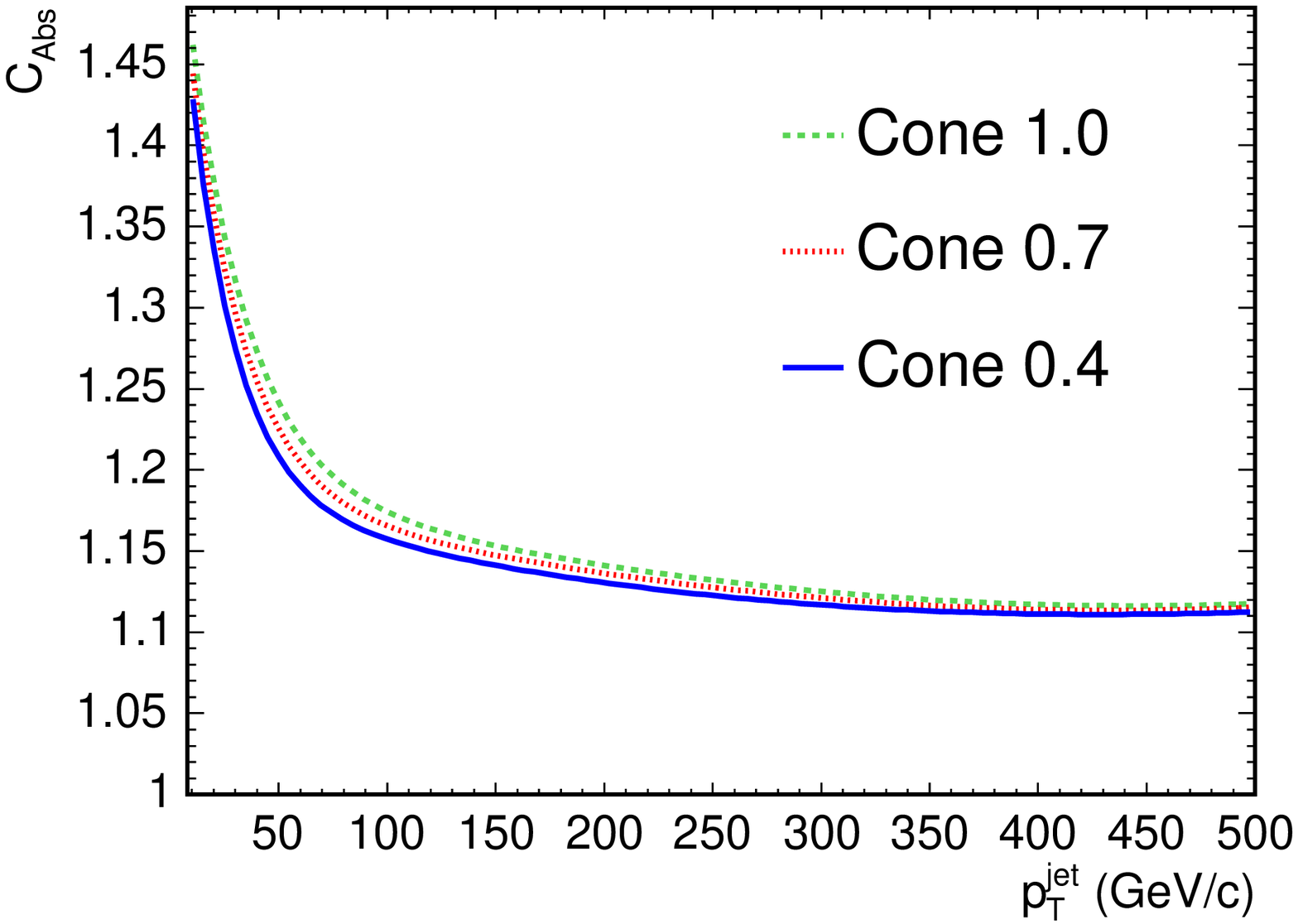}
  \caption{\sl Absolute corrections for different cone sizes as a function
  of calorimeter jet $p_T$. The solid line shows the corrections for
  cone size $0.4$, the dashed for $0.7$ and the dotted for $1.0$.
\label{fig:abs}}
  \end{center}
\end{figure}

The absolute correction is shown in Fig. \ref{fig:abs} for the three
cone radii. At $p_T^{jet} = 8$ GeV/$c$ the correction factor is about
$1.4$ and decreases toward high $p_T^{jet}$ to an asymptotic value of
about $1.12$. At high $p_T^{jet}$ the corrections are independent of
the cone size while at low $p_T^{jet}$ a slight dependence is
observed. For $p_T^{jet}<8$ GeV/$c$, a large fraction of jets are not
reconstructed since the observed single calorimeter tower energy often
falls below the $1$ GeV/$c$ seed tower threshold.  In these cases, it
is not possible to establish a mapping between calorimeter and
particle jets and no correction is derived.

\subsection{Uncertainties}
\label{sec:absolutesys}

The validity of the probability function used to determine the
absolute correction depends on how well the MC simulation models the
jet response in data. The treatment of the jet response as a
convolution of the single particle response with the $p_T$ spectrum of
the particles in a jet allows for propagation of the uncertainties of
the individual components to an uncertainty of the absolute
correction.

Given the calorimeter response $R(p)$ to charged and neutral particles
with momentum $p$, the average expected response $R_{ave}$, or jet
response, can be calculated from the generated particles using
\begin{equation}
R_{ave} = \frac{\sum_{i=1}^{N} p_i R(p_i)}{\sum_{i=1}^{N} p_i}
\end{equation}
\noindent where $N$ is the number of particles inside the jet cone
with momenta $p_i$. The uncertainties on the absolute corrections is
defined as the differences between data and simulation in the
calorimeter response to single particles, $R(p)$, the differences in
the momentum spectrum of the particles, $p$, and from the stability of
the calorimeter calibrations in data. They are addressed in the
following sections.

\subsubsection{Single Particle Response Simulation}

The measured calorimeter response $R(p)=\langle E/p\rangle$ (see
Sec. \ref{sec:Simulation}) for hadrons is parameterized as follows:
\begin{eqnarray}
\label{Eqn-Response}
p<20\mbox{ GeV}/c: & & R(p) =0.70+0.09\tanh(2.13(\log(p)-0.93)) \nonumber\\
\label{Eqn-Response2}
p>20\mbox{ GeV}/c: & & R(p) =0.70+0.14\tanh(0.49(\log(p)-1.15))
\end{eqnarray}
using the data in Fig. \ref{fig:eopcentral}. For electromagnetic
particles (electrons and photons) we set $R(p)=1.0$.

The relative uncertainty on the jet energy scale response, $\Delta
E/E$, is then
\begin{eqnarray}
\nonumber
\Delta E / E & = & (E - E^{\pm})/E \\
\nonumber
             & = & [\sum_{i=1}^{N} p_i R(p_i) - \sum_{i=1}^N p_i R(p_i) (1 \pm \Delta(E/p))]/E\\
             & = & \mp \sum_{i=1}^N p_i R(p_i) \Delta(E/p) / E
\end{eqnarray}

\noindent where $\Delta(E/p)$ is the uncertainty on $\langle E/p \rangle$ given
in Table \ref{tab:epsys}. For hadronic particles it is between 2.5\%
and 4\%.  However, since only 70\% of the jet energy arises from
hadronic particles the uncertainty on the jet energy is only
$1.8-2.8\%$. The response function is then convoluted with the
particle momentum spectra in jets to estimate the systematic
uncertainty as function of the jet energy. The uncertainty on the
simulation of the response of electromagnetic particles is
$1.7\%$. Since about 30\% of the jet energy is due to electromagnetic
particles, this results in a systematic uncertainty of 0.5\% on jet
energy scale, independent of jet $p_T$.  The uncertainties are shown
versus jet $p_T$ in Fig. \ref{Fig-abssys}.

\subsubsection{Fragmentation}

Uncertainties related to the particle momentum spectrum in a jet
originate from the modeling of hadronization effects using {\tt
PYTHIA} and {\tt HERWIG} as well as from the estimate of track
reconstruction efficiencies in data and detector simulation. The
transverse momentum spectrum of tracks in data is corrected for
inefficiencies as follows. The track reconstruction efficiency is
measured by embedding simulated tracks inside jets in data events,
after tuning the simulation of COT hits to distributions observed in
data, and then parameterized versus jet $p_T$, track momentum and
distance of track from the jet core. This parameterization is used to
correct the data for any inefficiencies \cite{simon}.

\begin{figure}[h]
  \begin{center}
  \includegraphics[width=\linewidth,clip=]{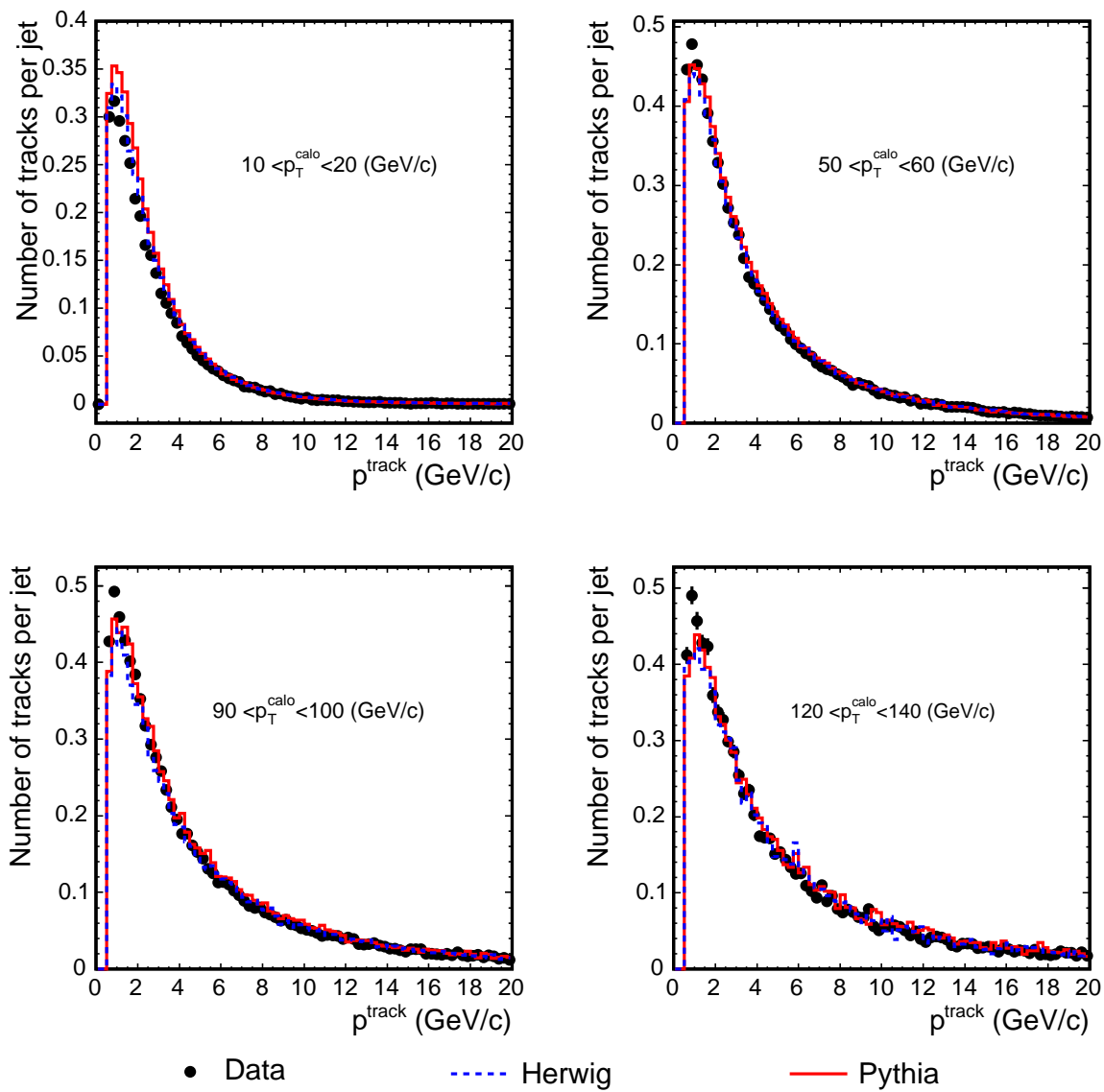}
  \caption{\sl Comparison of the particle momentum spectra in dijet
  events between data with {\tt PYTHIA} and {\tt HERWIG}. The CDF data
  have been corrected for track reconstruction
  efficiency.\label{Fig-ptfrag}} \end{center}
\end{figure}

The average number of tracks is measured as a function of the track
momentum for different values of $p_{T}^{jet}$. Only tracks that are
within the jet cone are associated with the jet. To account for
underlying event contributions, tracks from the region transverse to
the leading jet are subtracted event by event. Only events with
exactly one reconstructed $z$-vertex are used to reject events with
additional $p\bar{p}$ interactions. Figure \ref{Fig-ptfrag} shows a
comparison of the track momentum spectra in data with {\tt PYTHIA} and
{\tt HERWIG} simulation for four values of $\ptcaljet$.
The data are generally in good agreement with the MC, apart from some
discrepancies at very low track momenta.  

\begin{figure}[h]
  \begin{center} \includegraphics[width=\linewidth,clip=]{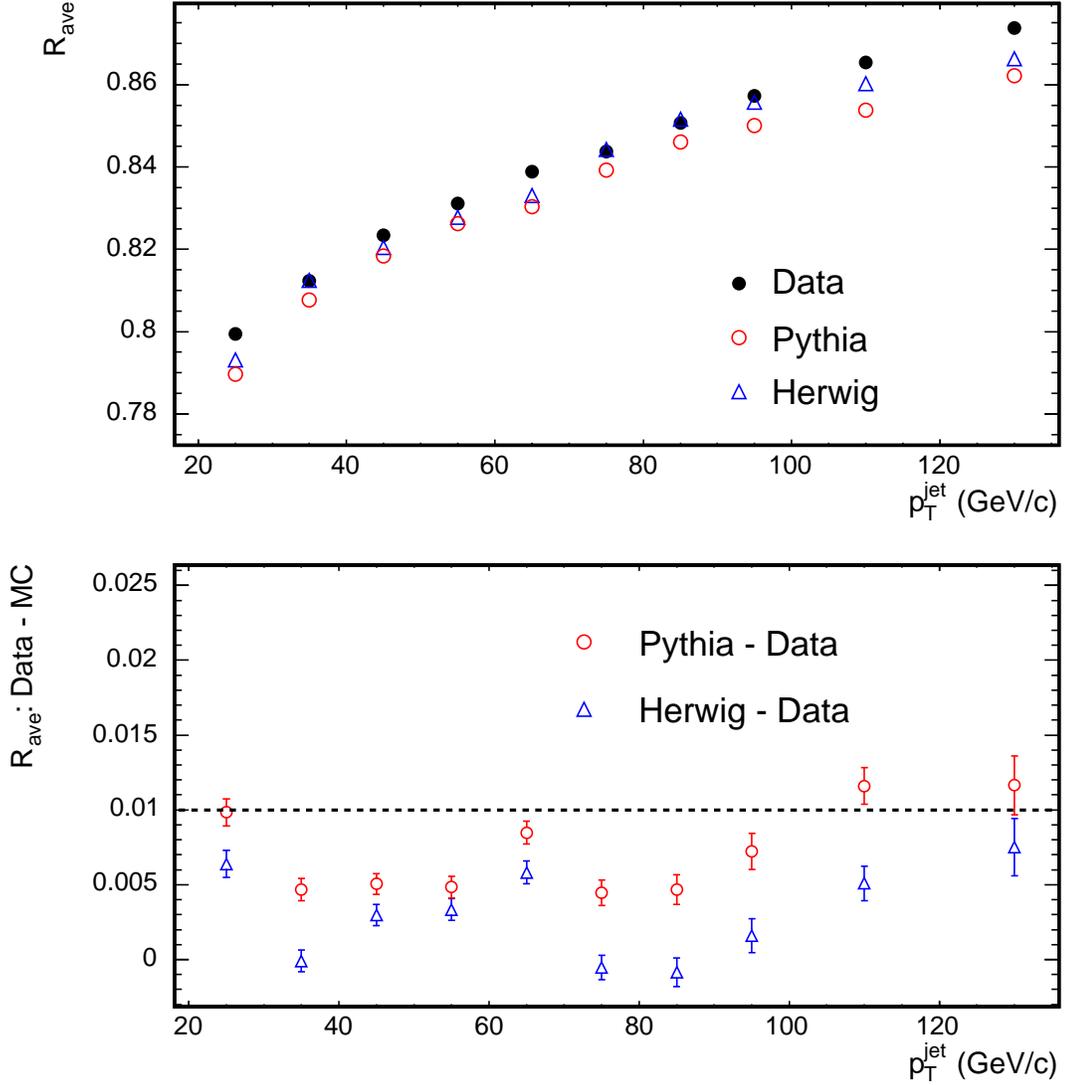}
  \caption{\sl Top: Jet response $R_{ave}$ for data (closed circles),
  {\tt PYTHIA} (open circles) and {\tt HERWIG} (open triangles) for $R_{jet}=0.4$
  jets as function of $p_{T}^{jet}$. Bottom: Difference between data
  and {\tt PYTHIA} (open circles) and data and {\tt HERWIG} (open
  triangles).\label{Fig-Loss07}}
  \end{center}
\end{figure}

The systematic uncertainty on the jet $p_T$ due to the particle
multiplicity is calculated from the differences between the measured
and simulated average calorimeter response, $R_{ave}$ for a fixed
single particle response $R(p)$. Figure \ref{Fig-Loss07} shows the jet
response for data, {\tt PYTHIA} and {\tt HERWIG}. Note that this is an
indirect measurement of the energy inferred from the track momenta.
If the response of the calorimeter, $R(p)$, was one, as is the case of
electromagnetic particles, this quantity would be unity. The deviation
from one thus quantifies the fraction of the jet energy measured due
to the low calorimeter response to charged hadrons. For a fixed $R(p)$
any difference between data and simulation can only arise from a
difference in the momentum spectrum.

The top plot of Fig. \ref{Fig-Loss07} shows $R_{ave}$ versus
$p_{T}^{jet}$ for data taken with an trigger threshold of 20 GeV/$c$
(see Sec. \ref{sec:Datasets}) compared to {\tt PYTHIA} and {\tt
HERWIG} simulation. At $p_{T}^{jet}=15$ GeV/$c$ about 20\% of the jet
energy is not measured. The response improves with increasing
$p_{T}^{jet}$ as expected: e.g., for $p_{T}^{jet}=120$ GeV/$c$ only
14\% is not measured. The data response is about 0.5\% higher than
that of the simulation. The bottom plot shows the difference between
data and simulation. The largest observed difference of 1\% is taken
as a $p_T$ independent systematic uncertainty on the jet energy scale
due to differences in the particle momentum spectrum (see
Fig. \ref{Fig-abssys}).

\subsubsection{Stability of the Calorimeter Energy Scale}

The simulation is tuned using Run II data collected during a fixed
period of time. The calorimeter calibration is kept constant to within
0.5\% as described in Sec. \ref{subsec:stability}. This value is taken
to be an additional systematic uncertainty.

\subsection{Summary}

\begin{figure}[h]
  \begin{center} \includegraphics[width=\linewidth,clip=]{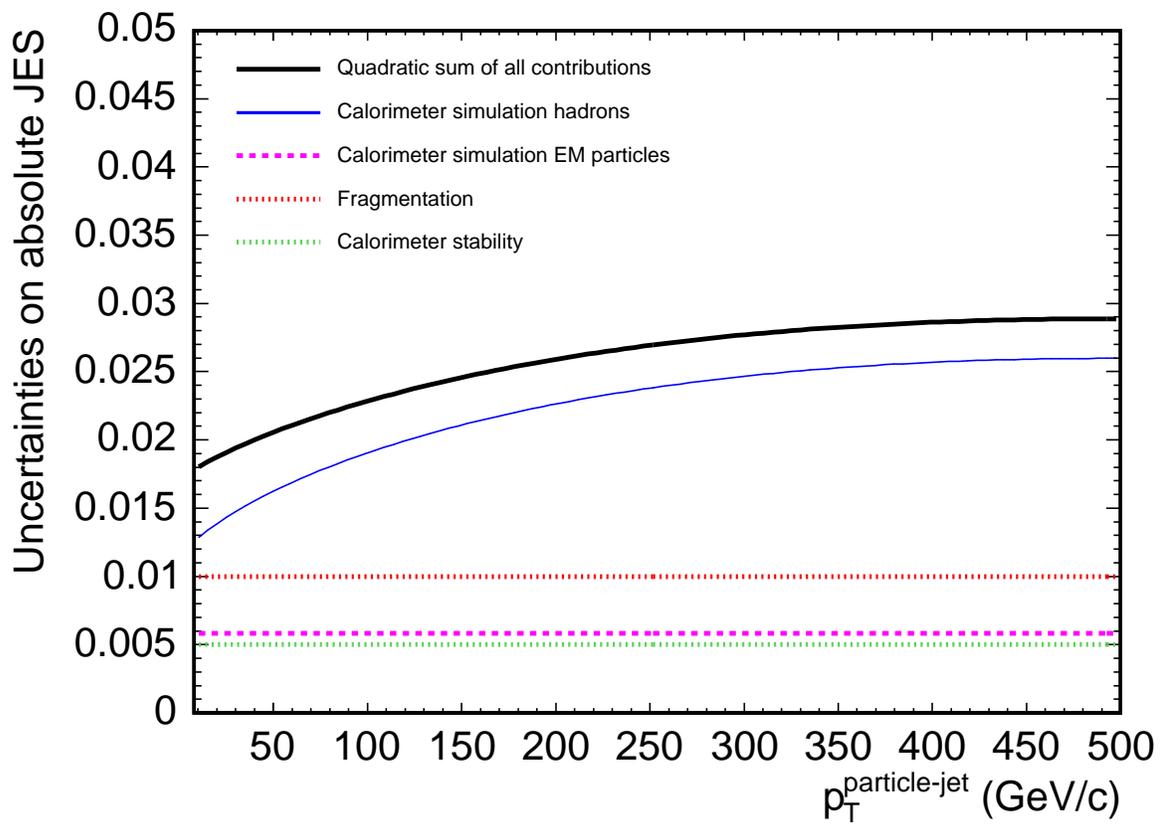}
  \caption{\sl Systematic uncertainty on the jet energy scale due to
  the calorimeter calibration and simulation. The solid line shows the
  total uncertainty and the other lines show individual
  contributions. \label{Fig-abssys}} \end{center}
\end{figure}

A summary of all the uncertainties is shown in
Fig. \ref{Fig-abssys}. It rises from 2\% at low $p_T^{jet}$ to 3\% at
high $p_T^{jet}$. The dominant uncertainty arises from the uncertainty
on the simulation of the calorimeter response to charged hadrons. The
individual uncertainties are added in quadrature to give the total
uncertainty. The uncertainties apply to all cone sizes. Further
studies of the validity of the systematic uncertainties are presented
in Sec. \ref{sec:crosschecks}.

\clearpage

\section{Multiple $p\overline p$ Interactions}
\label{sec:mi}

At high instantaneous luminosities more than one $p\bar{p}$
interaction occurs in the same bunch crossing at the Tevatron due to
the large $p\bar{p}$ cross section. The number of $p\bar{p}$
interactions per bunch crossing, $N$, follows a Poisson distribution
with a mean $\langle N \rangle$ which depends linearly on the
instantaneous luminosity. For the Run II configuration of the
Tevatron, the average number of interactions is about one at
$L=0.4\times 10^{32}$ cm$^{-2}$s$^{-1}$ and increases to three at
$L=1\times 10^{32}$ cm$^{-2}$s$^{-1}$ and eight at $L=3\times 10^{32}$
cm$^{-2}$s$^{-1}$. For the data taken up to September 2004 the
instantaneous luminosity ranges between $0.1\times 10^{32}$
cm$^{-2}$s$^{-1}$ and $1\times 10^{32}$ cm$^{-2}$s$^{-1}$.

These extra $p\bar{p}$ interactions increase the energy of the jets
from the hard scatter if their final state hadrons accidentally
overlap with the jets. This extra energy therefore needs to be
subtracted from the jet energy.

\subsection{Correction Procedure}

\begin{figure}[htbp]
\begin{center}
\setlength{\unitlength}{1 mm}
\begin{picture} (120,100)
\put(-15,0) {\epsfig{file=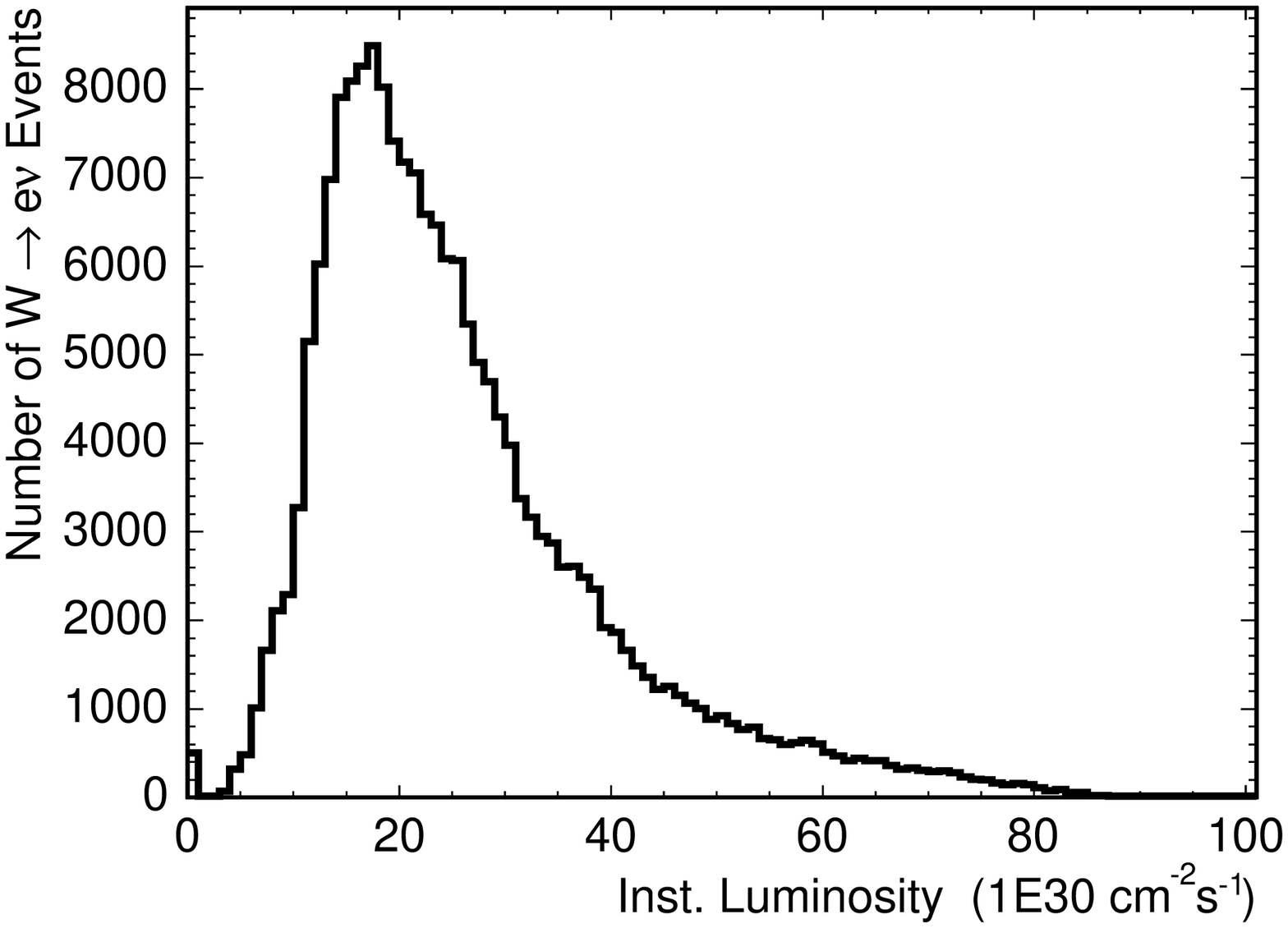,width=0.5\textwidth}}
\put(60,0) {\epsfig{file=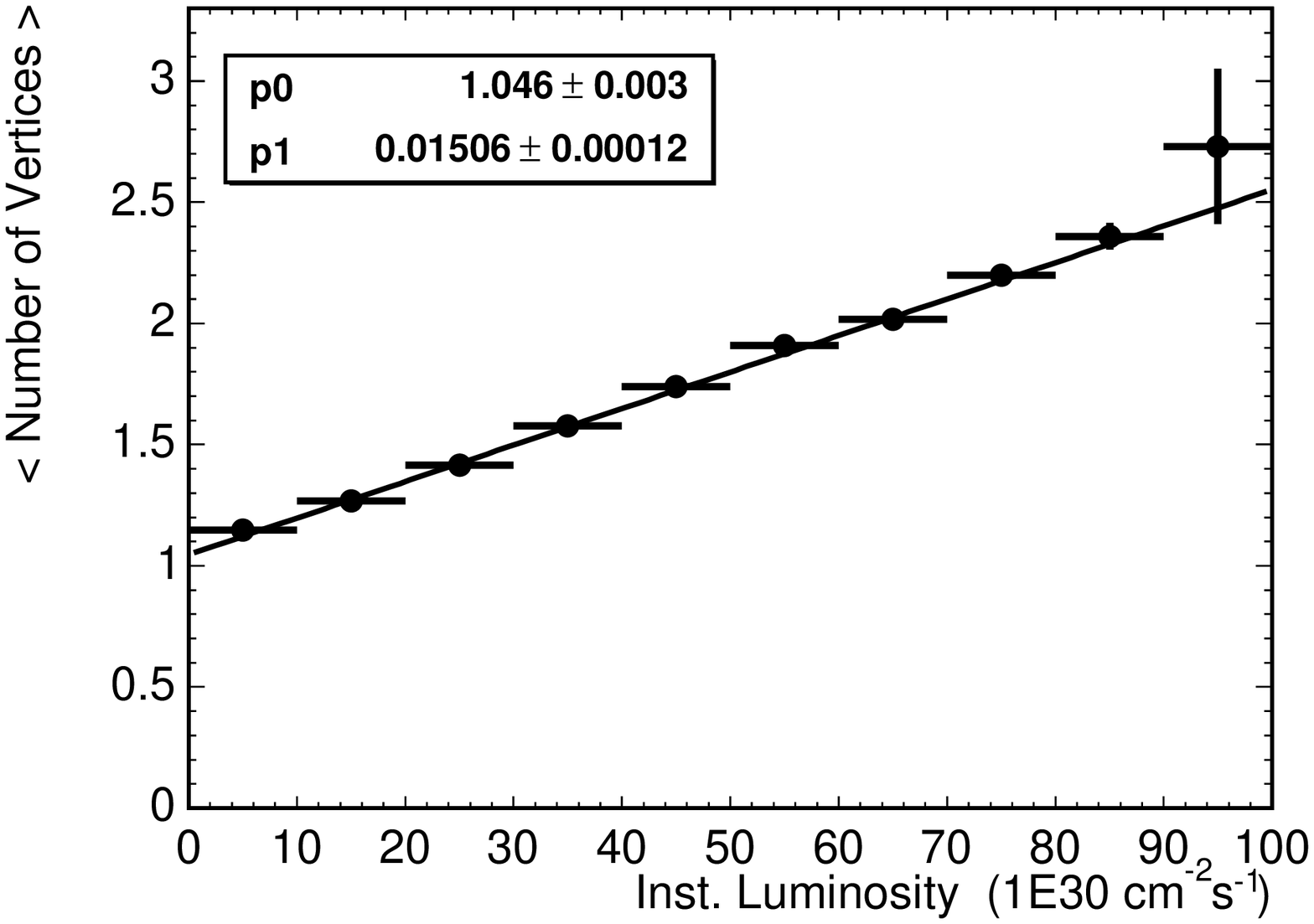,width=0.5\textwidth}}

\end{picture}
\caption{\label{fig:nvtxlum} \sl Left: Instantaneous luminosity for
$W\to e\nu$ events measured up to September 2004. Right: Mean number
of reconstructed vertices in $W\to e\nu$ events versus the
instantaneous luminosity. Also shown is a straight line fit.}
\end{center}
\end{figure}

The number of reconstructed $z$-vertices, $N_{vtx}$, is the best
estimate of the number of interactions in a bunch crossing. Vertices
are reconstructed using the intersections of the tracks with the beam
line. Figure \ref{fig:nvtxlum} shows the number of vertices as a
function of the instantaneous luminosity in the first $350$ pb$^{-1}$
of CDF data in $W\to e\nu_e$ candidate events. The mean number of
$z$-vertices, $\langle N_{vtx} \rangle$, is also shown. At low vertex
multiplicity the expected linear correlation is observed.

The efficiency of the vertex finding algorithm depends on the track
multiplicity. It is about $80\%$ for minimum bias events, $98\%$ for
$W\to e\nu_e$ events and greater than $99.9\%$ for $t\bar{t}$ events.  These
efficiencies have been determined in MC samples and verified using the
fraction of $W\to e\nu$ data events whose vertex is within $5$~cm of
the $z$-position of the electron track at the beam line.

\begin{figure}[htbp]
  \begin{center} 
  \includegraphics[width=0.49\linewidth,clip=]{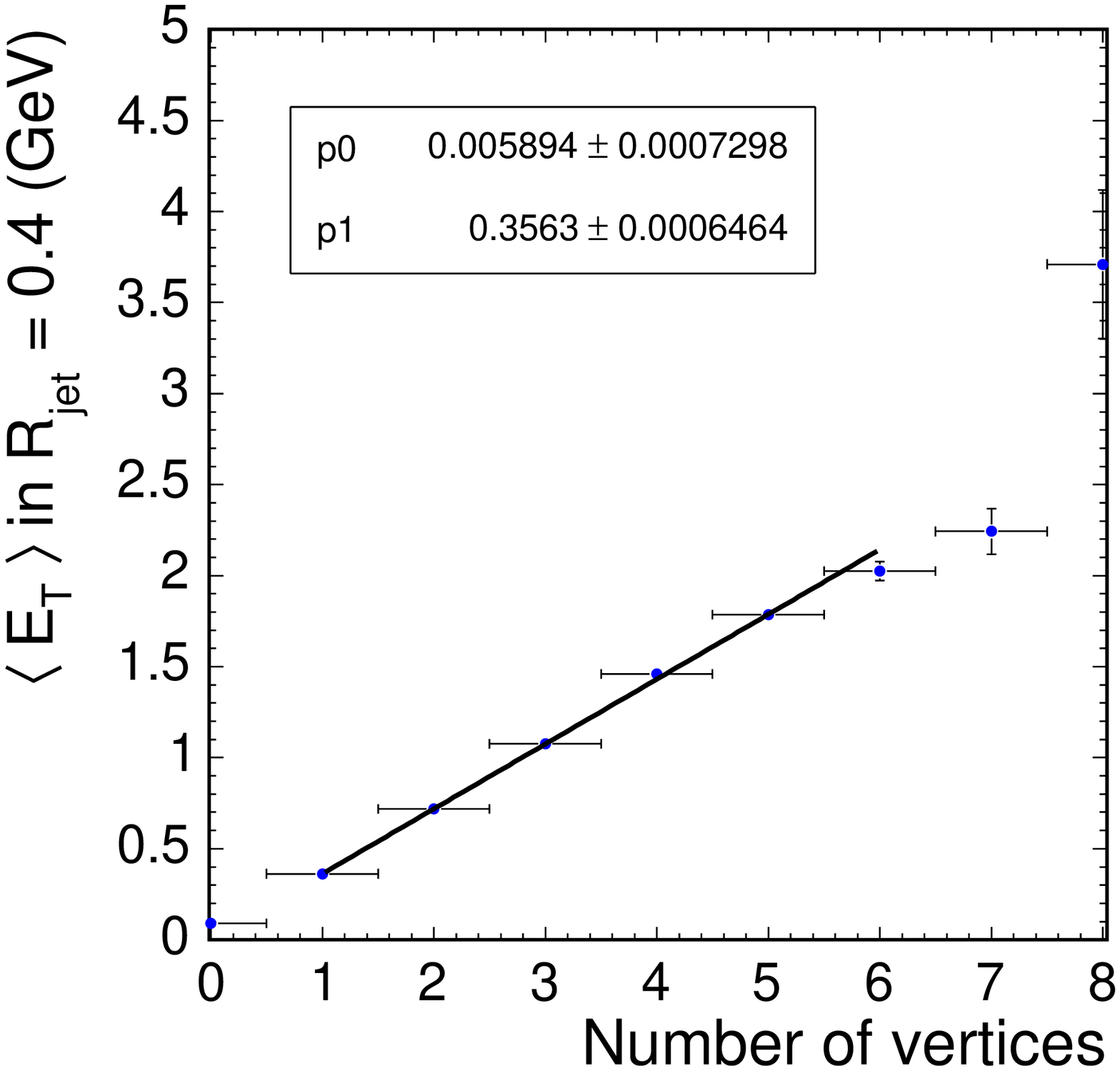}
  \includegraphics[width=0.49\linewidth,clip=]{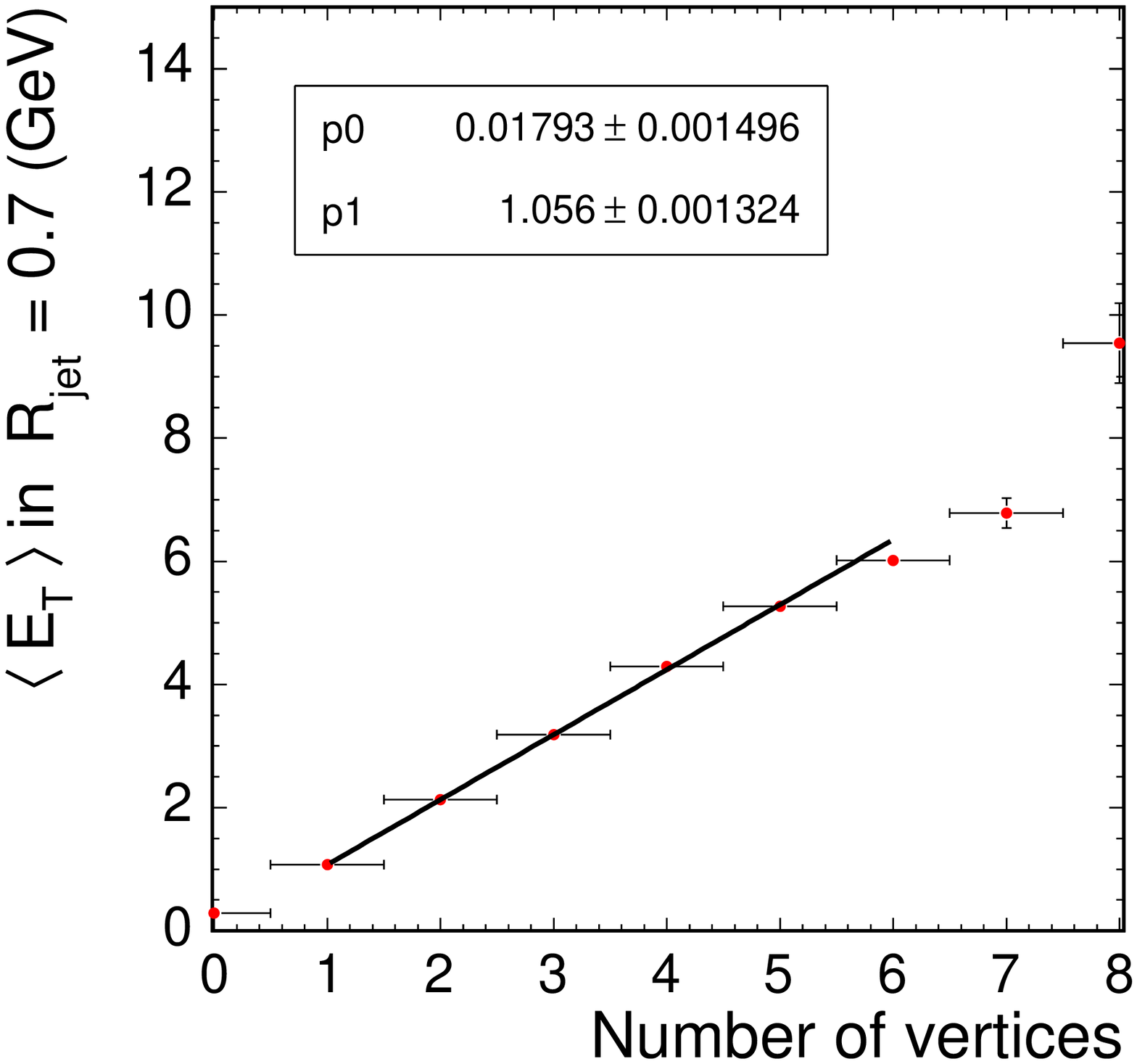}
  \includegraphics[width=0.49\linewidth,clip=]{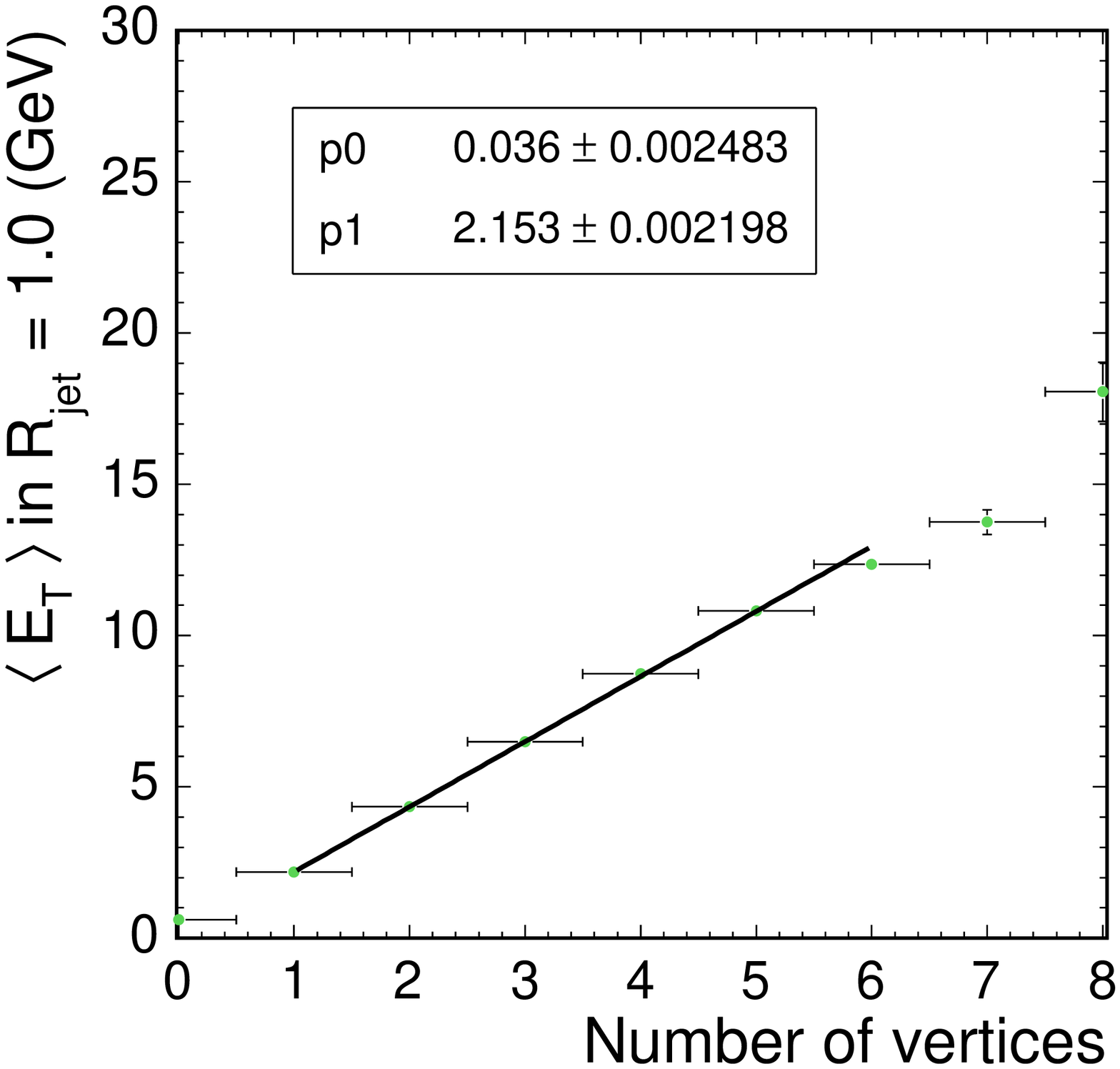}
  \caption{\label{fig:mimeasurement} \sl $\langle E_T \rangle$ versus
the number of $z$-vertices for $R_{jet}=0.4$ (top left), $R_{jet}=0.7$
(top right) and $R_{jet}=1.0$ (bottom left). A linear fit is also
shown. }
  \end{center}
\end{figure}

The average transverse energy in a cone is measured using the minimum bias
data sample. The cone is defined using a seed tower randomly selected
in the central calorimeter region $0.2<|\eta|<0.6$. The transverse
energy, $E_T^{R}$, in this cone is measured as a function of the
number of vertices for three cone sizes. Figure
\ref{fig:mimeasurement} shows $E_T$ versus $N_{vtx}$. The data show
the expected linear behavior.

The data are parameterized using a fitted straight line with
coefficients given in Table \ref{tab:mi}. The slope parameters give
the extra transverse energy per interaction as a function of
$N_{vtx}$. Ideally, the intercept should be zero.  The intercept
values are in all cases close but are consistently larger than zero as
a consequence of the vertex finding inefficiency. Note that the slope
and intercept values measured for the three cone sizes are the same
when taking into account the cone areas.

\begin{table}[h]
\caption{Intercepts and slopes of the multiple interaction correction for the three cone sizes. }
\begin{center}
\renewcommand{\arraystretch}{1.4} \setlength\tabcolsep{5pt}
\begin{tabular}{|l|l|l|l|}
\hline
Fit parameter  & Cone 0.4 (GeV)          &  Cone 0.7 (GeV)      & Cone 1.0 (GeV)      \\
\hline
\hline
intercept     &  $0.006 \pm 0.001$ &  $0.018 \pm 0.002$ & $0.036 \pm 0.002$ \\\hline
slope      &  $0.356 \pm 0.001$   &  $1.056 \pm 0.001$   & $2.153 \pm 0.002$  \\
\hline
\end{tabular}
\end{center}
\label{tab:mi}
\end{table}

\subsection{Uncertainty}

The validity of this method depends on two aspects of the vertex
finding algorithm:

\begin{itemize}

\item {\bf Vertex reconstruction efficiency}: The efficiency of
finding vertices from additional interactions may depend on the
topology of the hard interaction. Any inefficiency will result in a
steeper slope parameter and a larger intercept. 

\item {\bf Vertex fake rate}: In events with high occupancy it may
happen that fake vertices are found, i.e. vertices are reconstructed
in $z$-positions where no interaction took place.  This fake rate also
depends on the event topology since the probability of confusion in
both the tracking and the vertex finding increases with increasing
number of tracks.

\end{itemize}

\begin{figure}[htbp]
  \begin{center} 
  \includegraphics[width=0.8\linewidth,clip=]{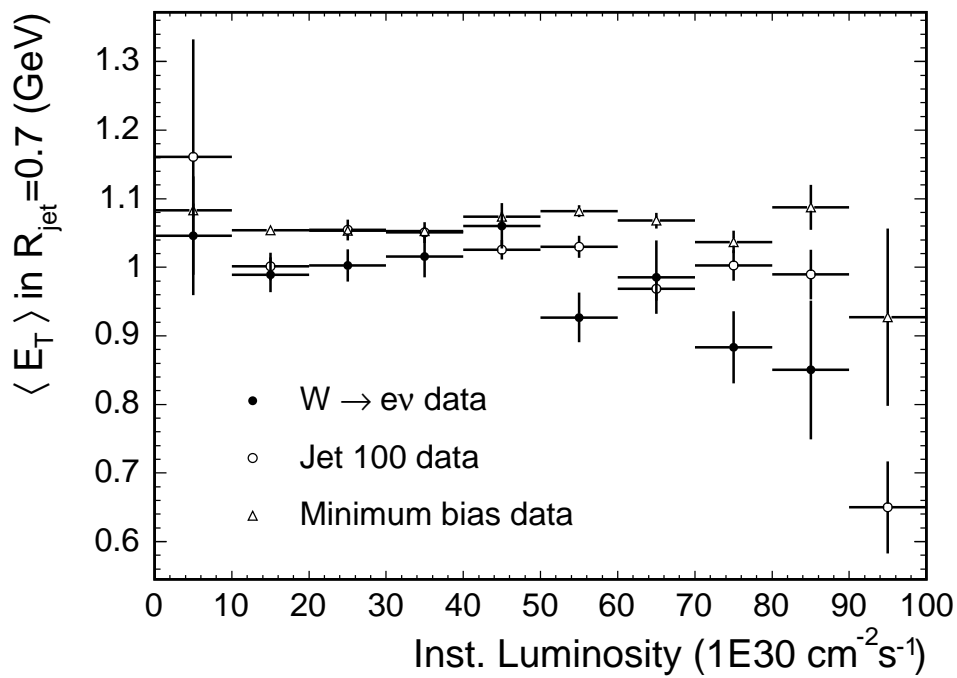}
  \caption{\sl Slope parameter of the multiple interaction correction
for $R_{jet}$=0.7 versus instantaneous luminosity in $W\rightarrow e
\nu$, minimum bias, as well as in a jet sample with $E_T$ threshold of
100 GeV. \label{fig:systlum}}
  \end{center}
\end{figure}

The impact of these effects is tested by repeating the multiple
interaction measurement using several samples: $W\to e\nu_e$, minimum
bias and a jet sample with $E_T$ thresholds of $100$ GeV. Figure
\ref{fig:systlum} shows no indication for any dependence on the
instantaneous luminosity or on the topology of these samples. However,
with the current statistical precision a 15\% effect cannot be
excluded, and it is taken as systematic uncertainty. This value of
this uncertainty corresponds to $50$~MeV for $R_{jet}=0.4$, $150$~MeV
for $R_{jet}=0.7$ and $300$~MeV for $R_{jet}=1.0$ per additional
interaction.

\clearpage

\section{Out-of-Cone Energy and Underlying Event}
\label{sec:ooc}

It is often desirable to reconstruct the energy of the original parton
rather than the energy of the jet, e.g. for the measurement of the top
quark mass or the search for the Higgs boson, where parton energies
are used to compute the invariant mass of the decaying products.

The reconstruction of the parton energy from the particle jet energy
is subject to several difficulties. A fraction of the parton energy
can be lost from the jet cone due to final state gluon radiation (FSR)
at large angles with respect to the parent parton or due to particles
exiting the cone either in the fragmentation process or due to low
$p_T$ particles bending in the magnetic field. This energy is called
``Out-of-Cone'' (OOC) energy. On the other hand the particle jet can
also have contributions not related to the actual mother parton of the
hard interaction of interest defining the jet, such as particles from
the initial state gluon radiation (ISR), or particles from spectator
partons with color connection to the other partons of the proton
(``beam-beam-remnant'', BBR). These two contributions are called
``Underlying Event'' (UE).

Final state radiation and hadronization effects are correlated with
the primary jet direction and the jet energy and are expected to
decrease with increasing distance from the jet core. The UE is thought
to be uncorrelated with the direction of the outgoing parton and thus
independent of the distance from the jet in $\eta-\phi$ space and
almost independent of the jet energy.

In this section, we derive corrections for the OOC energy and the UE
simultaneously using {\tt PYTHIA} dijet MC samples. As in the case of
the absolute corrections, the corrections are obtained using jets with
$0.2<|\eta_{jet}|<0.6$ since any $\eta$-dependence of the OOC energy
is taken into account by the relative corrections.  The corrections
are solely determined from MC simulation at particle generator level
independent of the CDF detector. The systematic uncertainties of the
OOC and UE corrections are derived from comparisons of the energy
measured in calorimeter towers in certain annuli around the jet cone
with the simulation based on {\tt PYTHIA} and {\tt HERWIG}.

\subsection{Correction Procedure}

\begin{figure}[htbp]
  \begin{center}
  \includegraphics[width=0.6\linewidth,clip=]{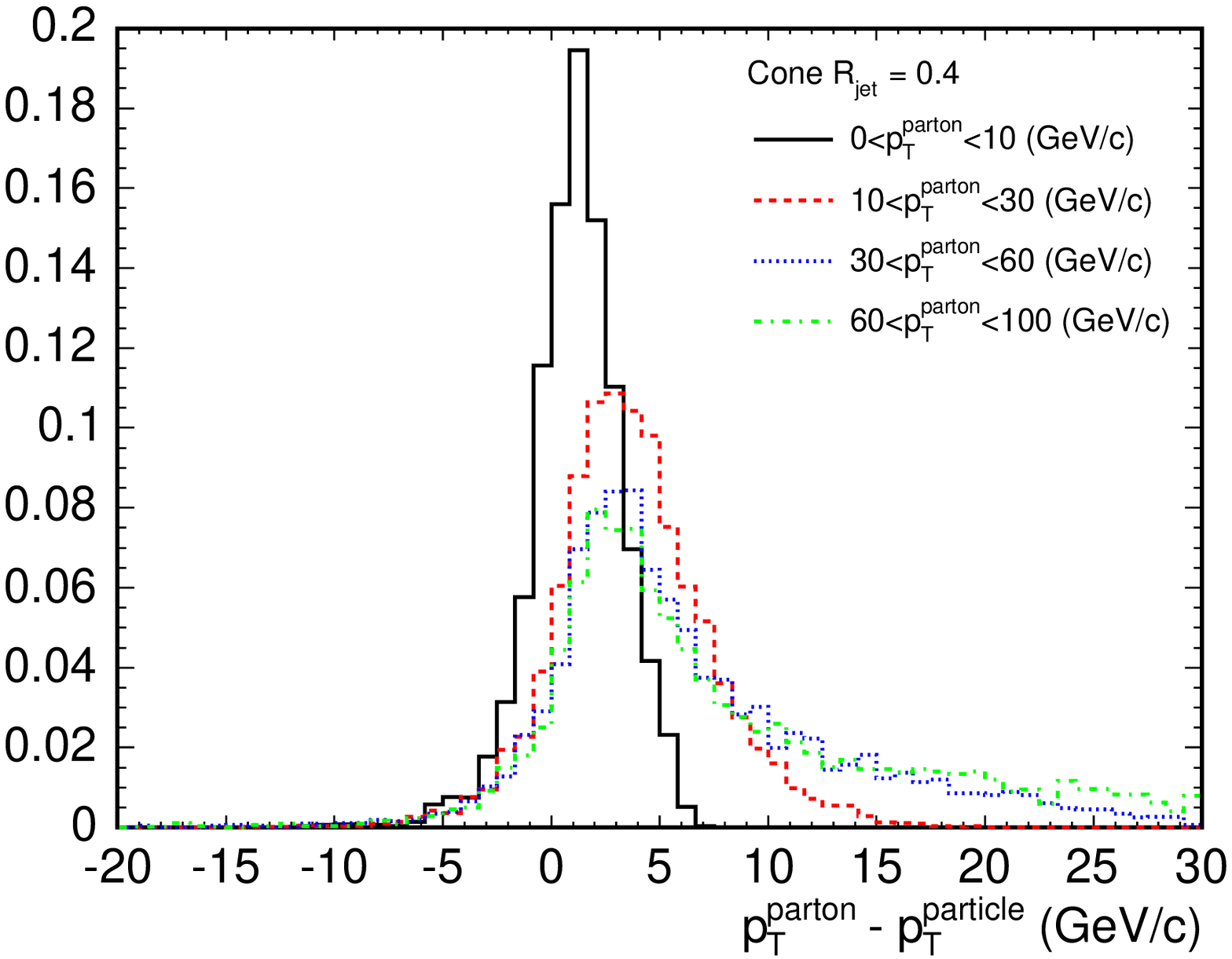}
  \includegraphics[width=0.6\linewidth,clip=]{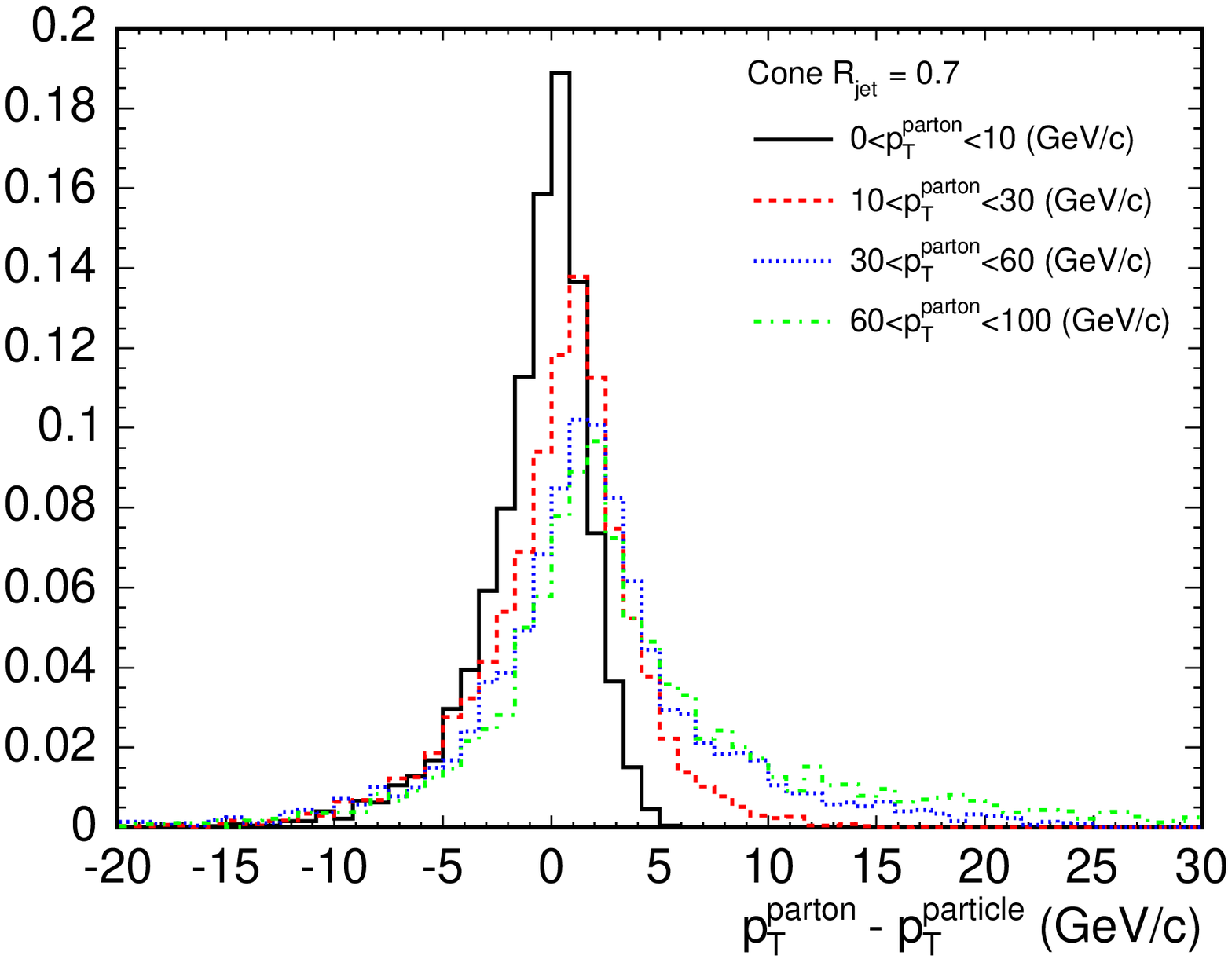}
  \includegraphics[width=0.6\linewidth,clip=]{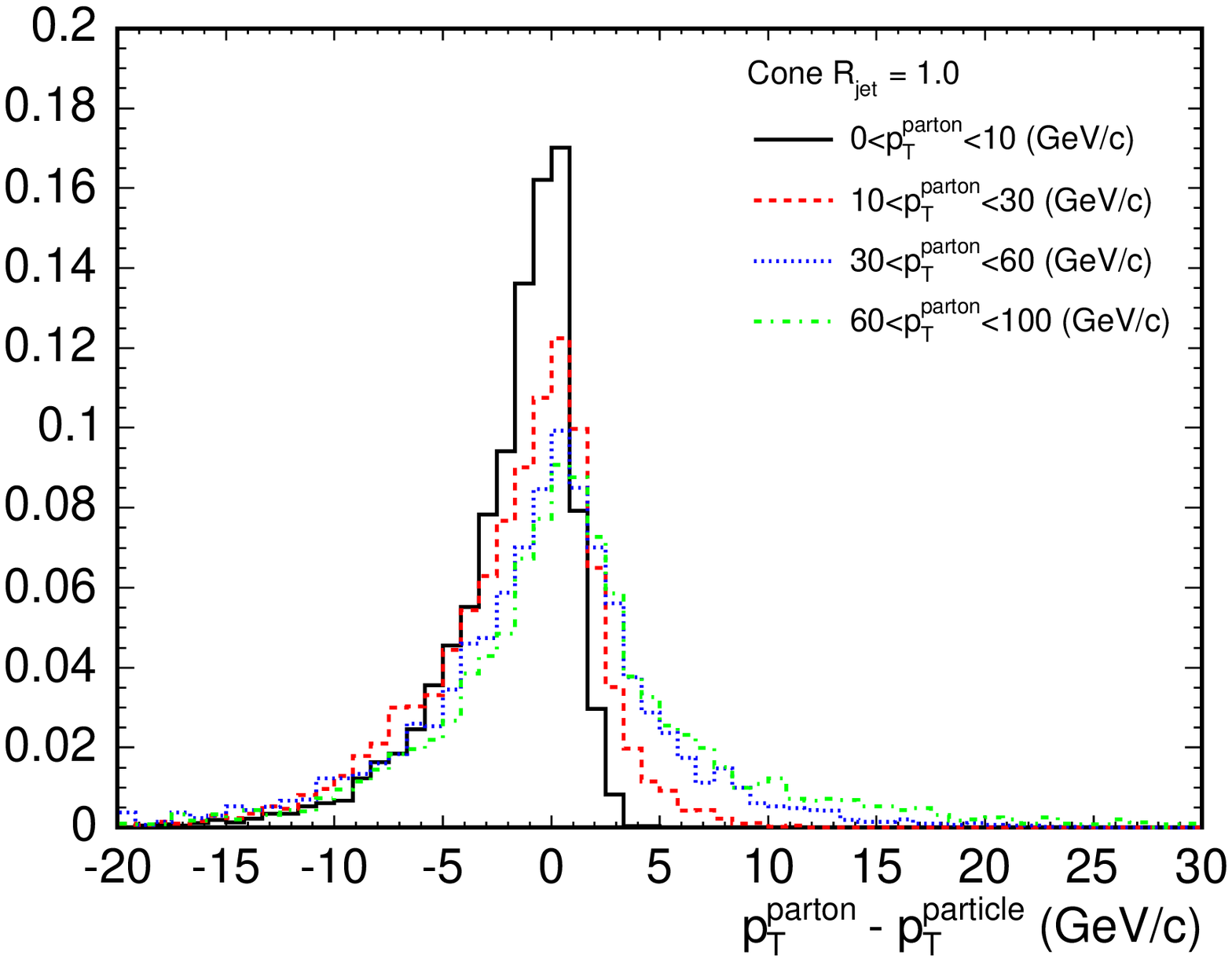}
  \caption{\sl Normalized distributions of $p_{T}^{parton} -
  p_{T}^{particle}$ for different $p_{T}^{parton}$ for cone sizes
  $0.4$ (top), $0.7$ (middle) and $1.0$ (bottom).\label{ooc_dr}}
  \end{center}
\end{figure}

\begin{figure}[htbp]
  \begin{center}
  \includegraphics[width=\linewidth,clip=]{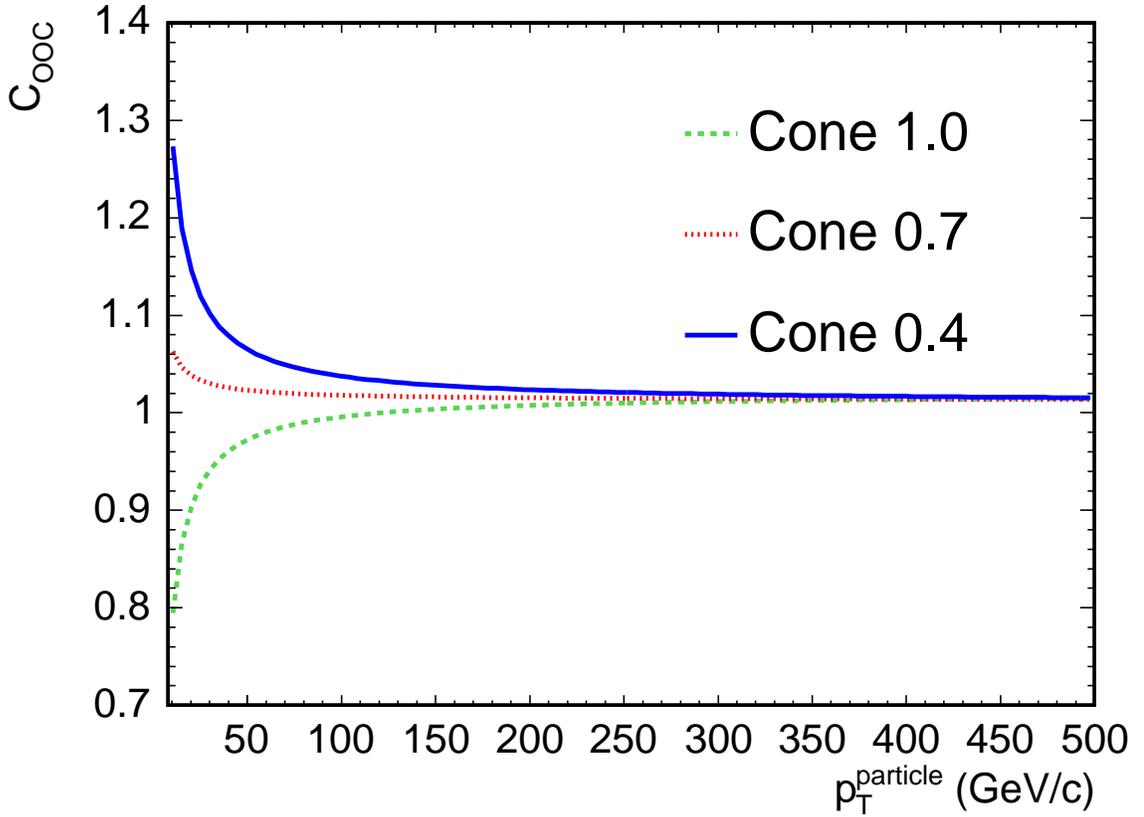}
  \caption{\sl OOC correction, $C_{OOC}$ versus
  $p_{T}^{parton}/p_{T}^{particle}$, versus $p_{T}^{particle}$ for
  cone sizes $0.4$ (solid line), $0.7$ (dashed line) and $1.0$ (dotted
  line).\label{oocc2}} \end{center}
\end{figure}

The OOC and UE corrections are obtained from {\tt PYTHIA} dijet
samples using particle jets which match a primary parton within
$\Delta R<0.4$. We parameterize the difference of the energy between
the particle jet and the parton using the same method as for the
absolute corrections (see Sec. \ref{sec:absproc}). Figure \ref{ooc_dr}
shows $p_{T}^{parton} - p_{T}^{particle}$ for different parton momenta
and $R_{jet}=0.4,0.7$ and $1.0$. The energy outside the jet cone
depends strongly on the cone size.
The OOC corrections $p_{T}^{parton}/p_{T}^{particle}$ are shown in
Fig. \ref{oocc2}. For the smallest cone size, $R_{jet}=0.4$, it is
about $+18\%$ at $p_{T}^{particle}=20$ GeV/$c$. For the largest cone
size, the correction is negative: $-6\%$ at $p_{T}^{particle}=20$
GeV/$c$, corresponding to $1.2$ GeV/$c$.
This shows that at small cone sizes the OOC losses dominate over the
energy increase due to the UE, and at large cone sizes the extra
energy from the UE is larger than the OOC losses. We have estimated
that the UE transverse energy is about $0.4$~GeV, $1.1$~GeV and
$2.2$~GeV for cone sizes of $0.4$, $0.7$ and $1.0$, respectively (see
Sec. \ref{sec:uesys}).

\subsection{Uncertainties}

\subsubsection{Out-of Cone Energy}

We determine the uncertainty on the OOC energy using $\gamma$+jets
samples. The reference energy scale is the photon $p_T$ which serves
as an estimator of the corrected jet $p_T$, i.e. $p_T^{\gamma}\equiv
p_T^{corr}$.

\begin{figure}[htbp]
  \begin{center}
  \includegraphics[width=\linewidth,clip=]{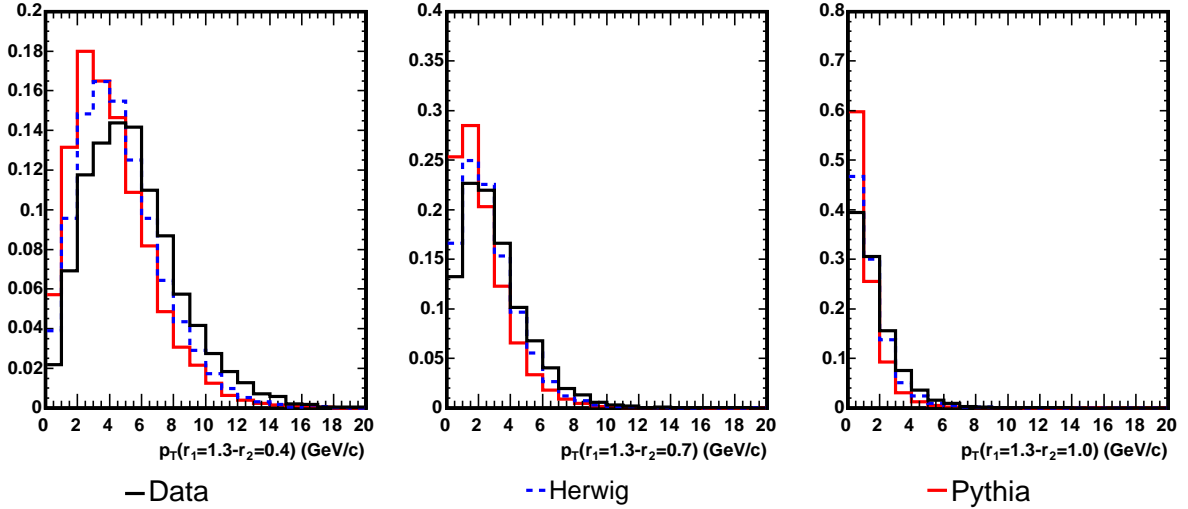}
  \caption{\sl Normalized distributions of the momentum in different
  annuli outside the jet cone for data, {\tt PYTHIA} and {\tt HERWIG}
  $\gamma$+jets events.\label{fig:oocsys_r1r2}} \end{center}
\end{figure}

The transverse energy around a jet of cone size $R_{jet}$ is measured
by adding the transverse energy in towers within the annulus defined
by radii $r_1$ and $r_2$ around the jet axis, that is
\begin{equation}
p_T(r_1-r_2) =\sqrt{(\sum_{i=1}^{N} E_x^i)^2 + (\sum_{i=1}^{N}
E_y^i)^2}
\end{equation}
where N is the number of towers for which
$r_1>\sqrt{(\eta_{jet}-\eta_i)^2+(\phi_{jet}-\phi_i)^2}>r_2$.  Figure
\ref{fig:oocsys_r1r2} shows $p_T(r_1-r_2)$ in data, {\tt PYTHIA} and
{\tt HERWIG} for different jet annuli. The shapes of the data and
simulation distributions agree rather well, and in the following the
mean value is used to quantify any disagreement between them.

\begin{figure}[h]
  \begin{center} 
  \includegraphics[width=\linewidth,clip=]{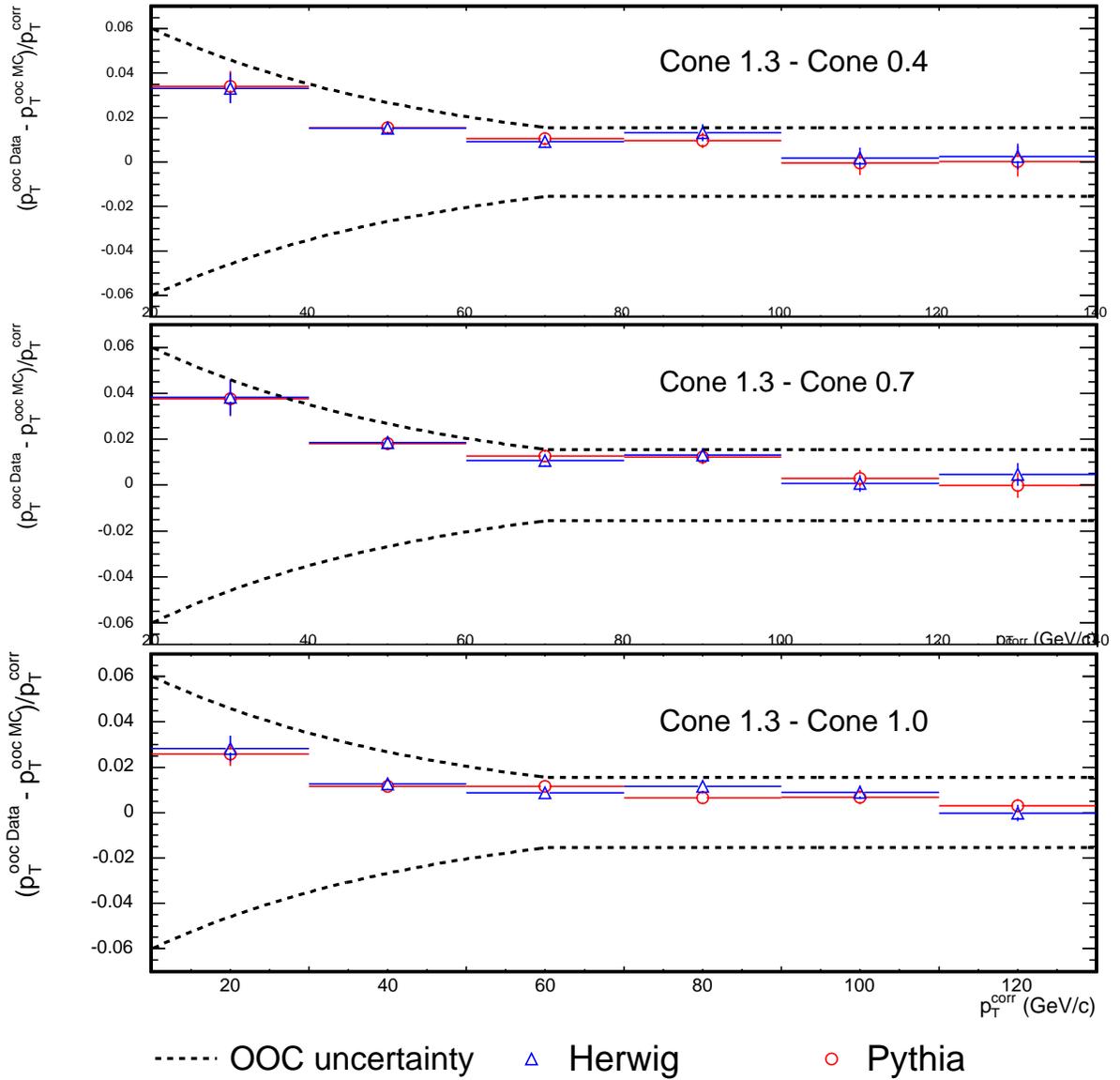}
  \caption{\sl Systematic uncertainty of OOC corrections for different
  cone sizes. The systematic uncertainty is taken as the largest
  difference between data and either {\tt PYTHIA} or {\tt HERWIG} and multiplied
  by the particle-jet/calorimeter-jet factors.\label{fig:oocsys}}
  \end{center}
\end{figure}

Figure \ref{fig:oocsys} shows the difference between data and
simulation for the mean values of $p_T(r_1-r_2)$ as a function of
$p_T^{corr}$.  The largest difference is observed at low $p_T^{corr}$
and is about 4\%. The systematic uncertainty is defined as the largest
difference between data and either {\tt PYTHIA} or {\tt HERWIG}, and
is parameterized as a function of $p_T^{corr}$ but independent of the
jet cone size. Since this measurement is made at the calorimeter tower
level and we apply this correction to a jet after absolute correction,
the uncertainties shown in Fig. \ref{fig:oocsys} are multiplied by the
factors $1.1, 1.35$ and $1.6$, for $R_{jet}$=0.4, 0.7 and 1.0,
respectively. These factors were determined from {\tt PYTHIA} by
comparing the particle and the calorimeter energy inside the annuli
around the jet cone.

By considering alternative generators various modeling uncertainties
contributing to the systematic error are taken into account. {\tt
HERWIG} and {\tt PYTHIA} have very different beam-beam remnant
contributions. Furthermore, they differ in the modeling of QCD
radiation and fragmentation.

\subsubsection{Underlying Event}
\label{sec:uesys}

\begin{figure}[h]
  \begin{center} 
  \includegraphics[width=\linewidth,clip=]{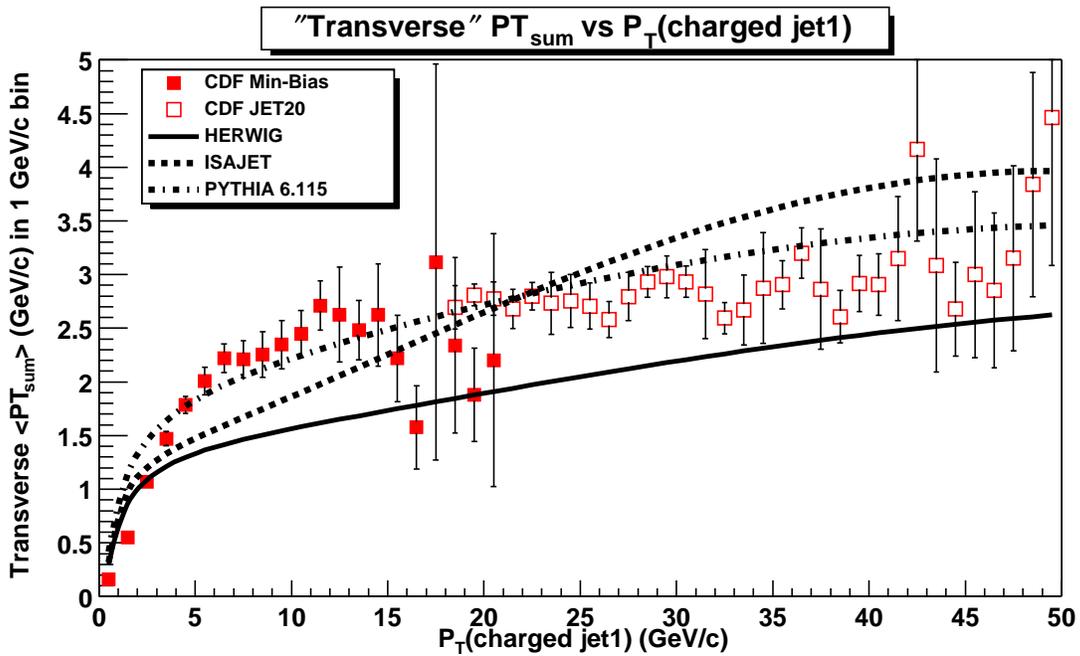}
  \caption{\label{fig:uesys} \sl The average transverse momentum of
    charged particles in the transverse region as described in the
    text versus leading jet $p_T$ \cite{cdfue} calculated using tracks
    in $R_{jet}$=0.7. The data are shown as points and are compared to
    the predictions from {\tt PYTHIA} (dashed-dotted line), {\tt HERWIG} (solid
    line) and {\tt ISAJET} (dashed line).}
  \end{center}
\end{figure}

Another source of systematic uncertainty comes from the varying UE in
different physics processes. To first order these dependencies are
taken into account by the MC generators, e.g. {\tt PYTHIA} has been
tuned to describe the UE in the data ({\tt PYTHIA} Tune A,
\cite{tuneA}). The UE uncertainties are derived from comparisons of
the UE in data, {\tt PYTHIA} Tune A and {\tt HERWIG}
\cite{cdfue}. This comparison is done using tracks with $p_T>0.5$
GeV$/c$ that are separated from the leading jet in azimuth by
$60^\circ <\Delta\Phi\mbox{(jet,track)} <120^\circ$, which is referred
to as the ``transverse region''. It is mostly sensitive to ISR and
multiple parton interactions.  Figure \ref{fig:uesys} shows the
average momenta of the tracks in the transverse region versus the
leading jet $p_T$. The data agree well with {\tt PYTHIA} but differ by
up to 30\% from {\tt HERWIG}. This value is taken as the relative
systematic uncertainty. As a further cross check, Fig. \ref{fig:uesys}
shows also the corresponding transverse momentum spectrum simulated by
{\tt ISAJET}
\cite{isajet} which has an alternate hadronization model. To get an
estimate of the absolute UE uncertainty we use the energy measured in
minimum bias data for $N_{vtx}=1$ as shown in
Fig. \ref{fig:mimeasurement}. The numbers are $0.4$ GeV, 1.1 GeV, and
2.2 GeV for $R_{jet}=0.4$, 0.7 and 1.0, respectively, which translate
to UE uncertainties of $0.11$ GeV, $0.32$ GeV and $0.66$ GeV. We have
also compared the average transverse momenta between data, {\tt PYTHIA} and
{\tt HERWIG} in $\gamma$+jet and $Z$-jet events and find a similar
agreement.

The resulting contribution to the systematic uncertainty of the jet
energy scale is about 10\% at $p_T=10$ GeV/$c$ and decreases to about
2\% at $p_T=70$ GeV/$c$. It can be further improved by a more detailed
comparison of data and simulation, thus leading to a better
understanding of the physics effects, or using improved versions of
the MC generators as {\tt JIMMY} or {\tt PYTHIA 6.3}.

\subsubsection{Splash-Out}

The OOC energy refers only to the energy lost outside the jet cone up
to $R_{jet}$=1.3. In {\tt PYTHIA} MC samples we measured that an
additional energy of 0.5 GeV falls outside a cone of $1.3$.  We take
half of this energy as systematic uncertainty, i.e. 0.25 GeV, and
refer to it as ``splash-out'' uncertainty.

\clearpage

\section{Validation of the Jet Energy Scale Determination}
\label{sec:crosschecks}

Several consistency checks and further studies are presented in this
section. The jet energy corrections, which are mostly derived from
dijet samples are applied to $\gamma$-jet, and $Z$-jet, and $t\bar t$
events to verify the validity of the corrections and systematic
uncertainties. Furthermore, we present additional studies on the
$\eta$-dependent corrections.

\subsection{Test of the Jet Corrections}

\subsubsection{Using $\gamma$-jet Events}
\label{sec:testgamjet}

The $\gamma$-jet data sample is ideal for studying the jet energy
scale. The photon energy $p_T^{\gamma}$ is measured accurately in the
CEM calorimeter and thus provides a perfect reference for the jet
energy. At tree-level the jet energy should always balance the photon
energy: $\frac{p_T^{jet}}{p_T^{\gamma}}=1$. Even in the presence of
higher order QCD corrections, which spoil this exact balancing, the
comparison of the jet energy measurement in these events provides an
excellent testing ground for the uncertainty on the jet energy
scale. In particular, the extent to which the data agree with the MC
simulation tests the systematic uncertainties in the jet energy
measurement.

Photons are selected with $p_T^\gamma>27$ GeV/$c$ and
$|\eta^\gamma|<0.9$. Jets fragmenting into a $\pi^0$ or $\eta$ can
decay into photons which constitute a significant background. We
require less than $1$ GeV of extra transverse energy in a cone of
radius 0.4 around the photon in the calorimeter and less than
$2$~GeV/$c$ for the scalar sum of the track $p_T$ values inside the
cone. We also apply cuts on the shower shape and the number of
clusters in the CES detector. However, even after these cuts there is
still a residual background of about 30\% at $p_T^\gamma =27$ GeV/$c$.
The background estimate is based on the number of hits in the CPR
detector \cite {promptgammaprd}. We estimate the $\gamma$-jet balance
separately for the signal and background and find the results to be
consistent for to within 1\%. However, the agreement strongly depends
on the cuts used in the analysis, and for looser cuts we observe
differences of up to 5\%.

Further cuts are applied to reduce the effects from QCD radiation:

\begin{itemize}
\item The photon and the jet are required to be back-to-back in
azimuthal angle: $\Delta \phi(\gamma,jet)>3$ radians.

\item The event has no more than one jet with $p_T>3$ GeV/$c$ and
$|\eta|<2.4$.
\end{itemize}

Furthermore, only events with one reconstructed vertex are used and
thus no correction for multiple $p\bar p$ interactions is necessary.

\begin{figure}[htbp]
  \begin{center}
  \includegraphics[width=\linewidth,clip=]{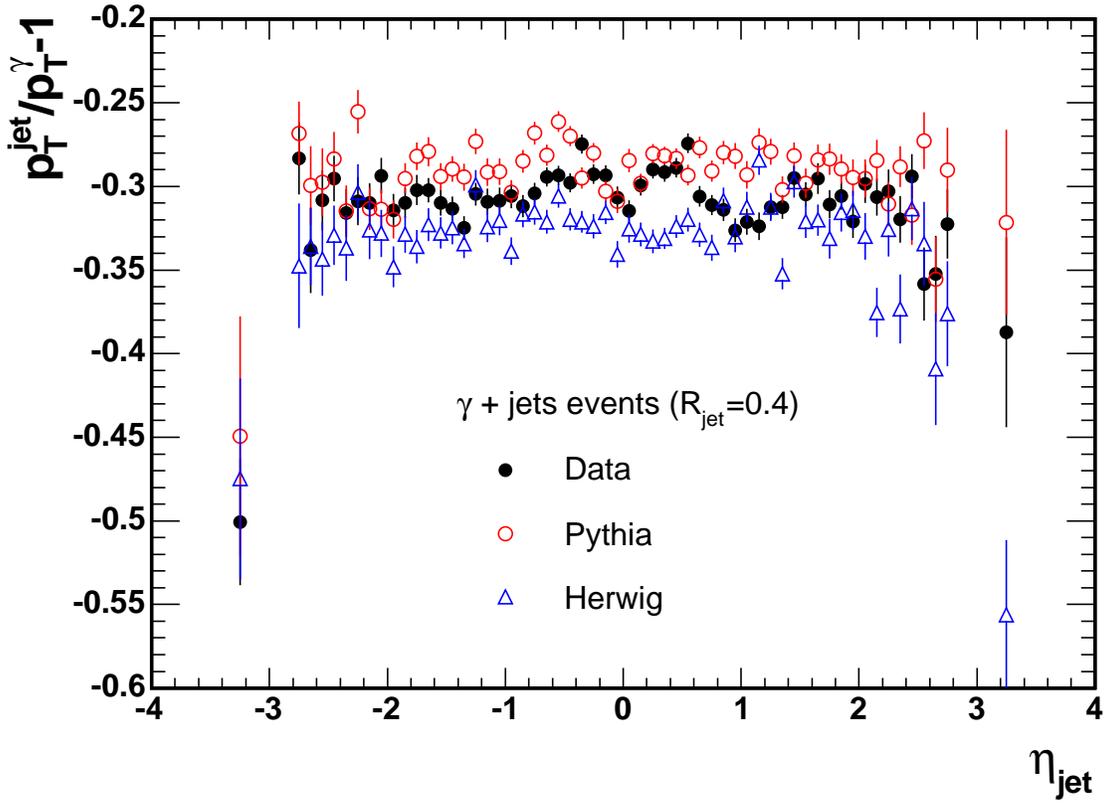}
  \caption{\sl $p_T$ balance, $\frac{p_T^{jet}}{p_T^\gamma}-1$, in
  data (full circles), {\tt PYTHIA} (open circles) and {\tt HERWIG}
  (open triangles) as function of $\eta_{jet}$ for $R_{jet}=0.4$
\label{fig:gamjeteta}}
\end{center}
\end{figure}

The $\gamma$-jet balance is shown versus jet $\eta$ after applying the
$\eta$-dependent corrections in Fig. \ref{fig:gamjeteta} for jet with
$R_{jet}=0.4$. The data, {\tt PYTHIA} and {\tt HERWIG} show no
residual dependence on $\eta_{jet}$ as desired. Note that the $p_T$
balance between the jet and the photon is not expected to be zero at
this stage of the correction procedure. The overall scale difference
between data, {\tt PYTHIA} and {\tt HERWIG} will be discussed in
Sec. \ref{sec:systest}.

\begin{figure}[htbp]
  \begin{center}
  \includegraphics[width=\linewidth,clip=]{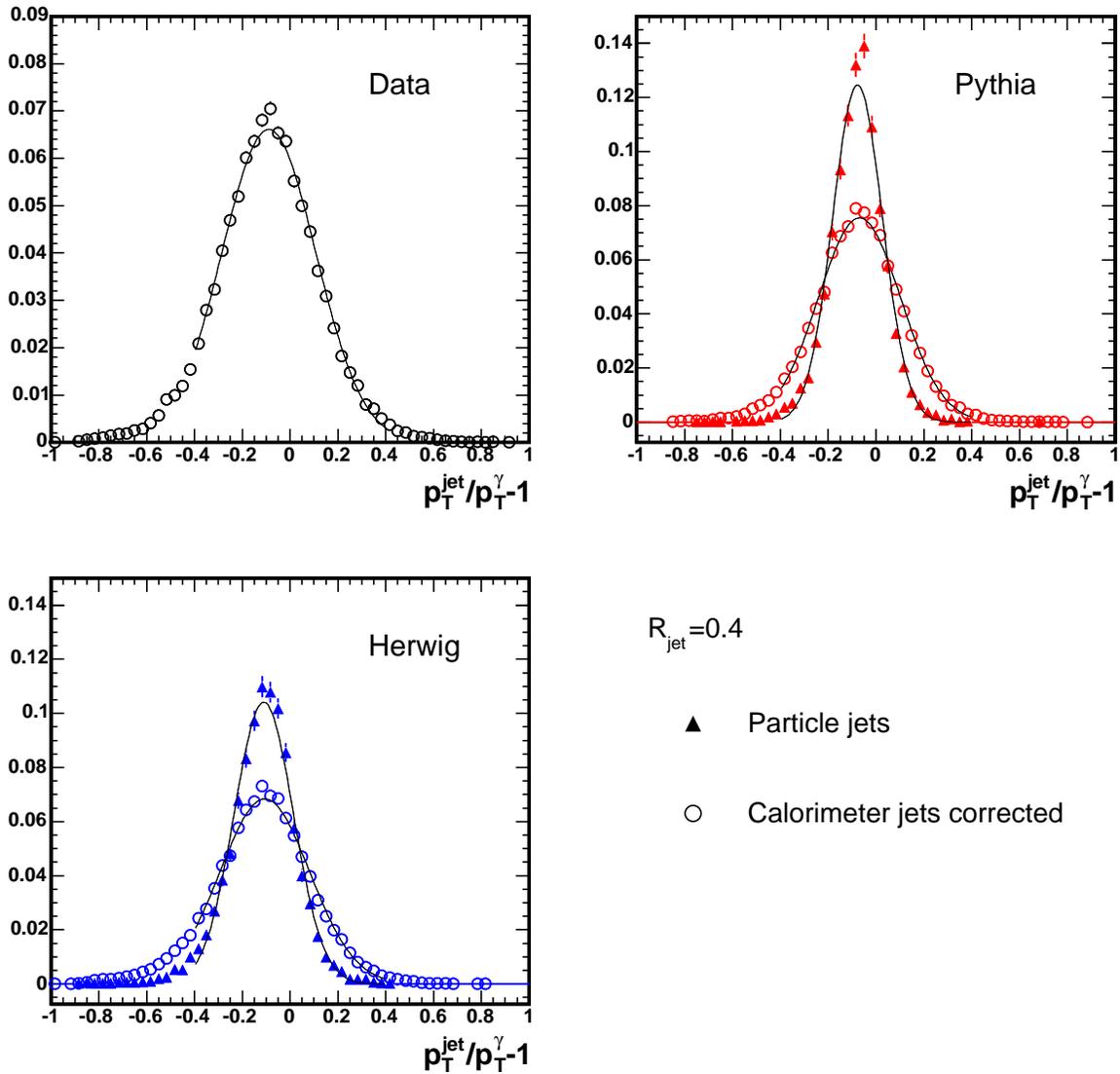}
  \caption{\sl $\gamma$-jet balance in data, {\tt PYTHIA} and {\tt
  HERWIG} for $R_{jet}$=0.4. Overlaid is the corresponding
  $\gamma$-jet balance on particle level jets (triangles) calculated
  using particles at generator level without detector simulation. The
  distributions are normalized to 1. \label{fig:ptbal04_l5}}
  \end{center}
\end{figure}

Next, we apply also the absolute correction to the jet momenta making
the jet $p_T$ independent of the calorimeter response. Figure
\ref{fig:ptbal04_l5} shows the resulting $p_T$ balance for data,
{\tt PYTHIA} and {\tt HERWIG} for $R_{jet}$=0.4. For comparison, also
the $p_T$ balance calculated using particles at generator level
without detector simulation is overlaid. Tables \ref{tab:gamjetmean}
and
\ref{tab:gamjetrms} summarize the mean and the width obtained from fits
of Gaussians to these distributions within the range -0.4 to 0.4 for
all jet cone sizes. For jets from data with $R_{jet}=0.4$ the mean is
measured to be about $9\%$ lower than the photon. Compared to the
data, the mean found in {\tt PYTHIA} is about 2\% higher and for {\tt
HERWIG} it is 2\% lower.  Generally, the $p_T$ of the jet is smaller
than the $p_T$ of the photon due to the energy lost outside the
cone. In fact, one observes that for the larger cone sizes, in
particular $R_{jet}=1.0$, the mean is much closer to $0$ since there
is nearly no energy lost outside the cone. For particle jets, the mean
values of {\tt PYTHIA} and {\tt HERWIG} agree with the respective
calorimeter jets to within 1\%, which proves the validity of the
absolute correction procedure. The observed differences between the
generators reflect the different modeling of the underlying physics
process. This difference is largest for $R_{jet}=0.4$. However, the
data generally lie between {\tt PYTHIA} and {\tt HERWIG} and agree to
within 2\% with both.

\begin{table}[htbp]
\caption{Mean value of $p^{jet}_{T}/p_{T}^{\gamma}-1$ after
$\eta$-dependent  and absolute energy correction, for data, {\tt PYTHIA}, and {\tt HERWIG}
for $R_{jet}$=0.4, 0.7 and 1.0. For {\tt PYTHIA} and {\tt HERWIG}, the values are
given also for particle jets.}
\begin{center}
\renewcommand{\arraystretch}{1.4}
\setlength\tabcolsep{5pt}
\begin{tabular}{|l||r|r|r|}
\hline
Sample & $R_{jet}$=0.4  & $R_{jet}$=0.7  &  $R_{jet}$=1.0 \\
\hline\hline
\multicolumn{4}{|c|} {Calorimeter jets}  \\\hline
Data   &  $-0.088 \pm 0.001$   &  $-0.016 \pm 0.001$      & $ 0.022 \pm 0.001$  \\
{\tt PYTHIA} &  $-0.070 \pm 0.001$   &  $-0.015 \pm 0.001$      & $-0.002 \pm 0.001$  \\
{\tt HERWIG} &  $-0.108 \pm 0.001$   &  $-0.043 \pm 0.001$      & $-0.024 \pm 0.001$  \\\hline\hline
\multicolumn{4}{|c|} {Particle jets}  \\\hline
{\tt PYTHIA} &  $-0.078 \pm 0.001$   &  $-0.037 \pm 0.001$      & $-0.009 \pm 0.001$  \\
{\tt HERWIG} &  $-0.113 \pm 0.002$   &  $-0.061 \pm 0.002$      & $-0.019 \pm 0.002$  \\
\hline
\end{tabular}
\end{center}
\label{tab:gamjetmean}
\end{table}

\begin{table}[htbp]
\caption{Width of $p^{jet}_{T}/p_T^\gamma -1$ after the
$\eta$-dependent  and absolute energy correction, for data, {\tt PYTHIA} and {\tt HERWIG}
for $R_{jet}$=0.4, 0.7 and 1.0. For {\tt PYTHIA} and {\tt HERWIG}, the values are
given also for particle jets.}
\begin{center}
\renewcommand{\arraystretch}{1.4}
\setlength\tabcolsep{5pt}
\begin{tabular}{|l||r|r|r|}
\hline
Sample & $R_{jet}$=0.4  & $R_{jet}$=0.7  &  $R_{jet}$=1.0 \\
\hline\hline
\multicolumn{4}{|c|} {Calorimeter jets}  \\\hline
Data   &  $0.199 \pm 0.001$   &  $0.191 \pm 0.001$      & $0.191 \pm 0.001$  \\
{\tt PYTHIA} &  $0.176 \pm 0.001$   &  $0.171 \pm 0.001$      & $0.169 \pm 0.001$  \\
{\tt HERWIG} &  $0.192 \pm 0.001$   &  $0.181 \pm 0.001$      & $0.178 \pm 0.001$  \\\hline\hline
\multicolumn{4}{|c|} {Particle jets}  \\\hline
{\tt PYTHIA} &  $0.105 \pm 0.001$   &  $0.095 \pm 0.001$      & $0.090 \pm 0.001$  \\
{\tt HERWIG} &  $0.127 \pm 0.002$   &  $0.116 \pm 0.002$      & $0.111 \pm 0.002$  \\
\hline
\end{tabular}
\end{center}
\label{tab:gamjetrms}
\end{table}

From Table \ref{tab:gamjetrms} we note that the data resolution is
around 4-7\% worse than {\tt HERWIG} and 12$\%$ worse than {\tt PYTHIA}. We
observe that {\tt HERWIG} has a wider resolution than {\tt PYTHIA} for both
calorimeter and for particle jets.

\begin{figure}[htbp]
  \begin{center} \includegraphics[width=\linewidth,clip=]{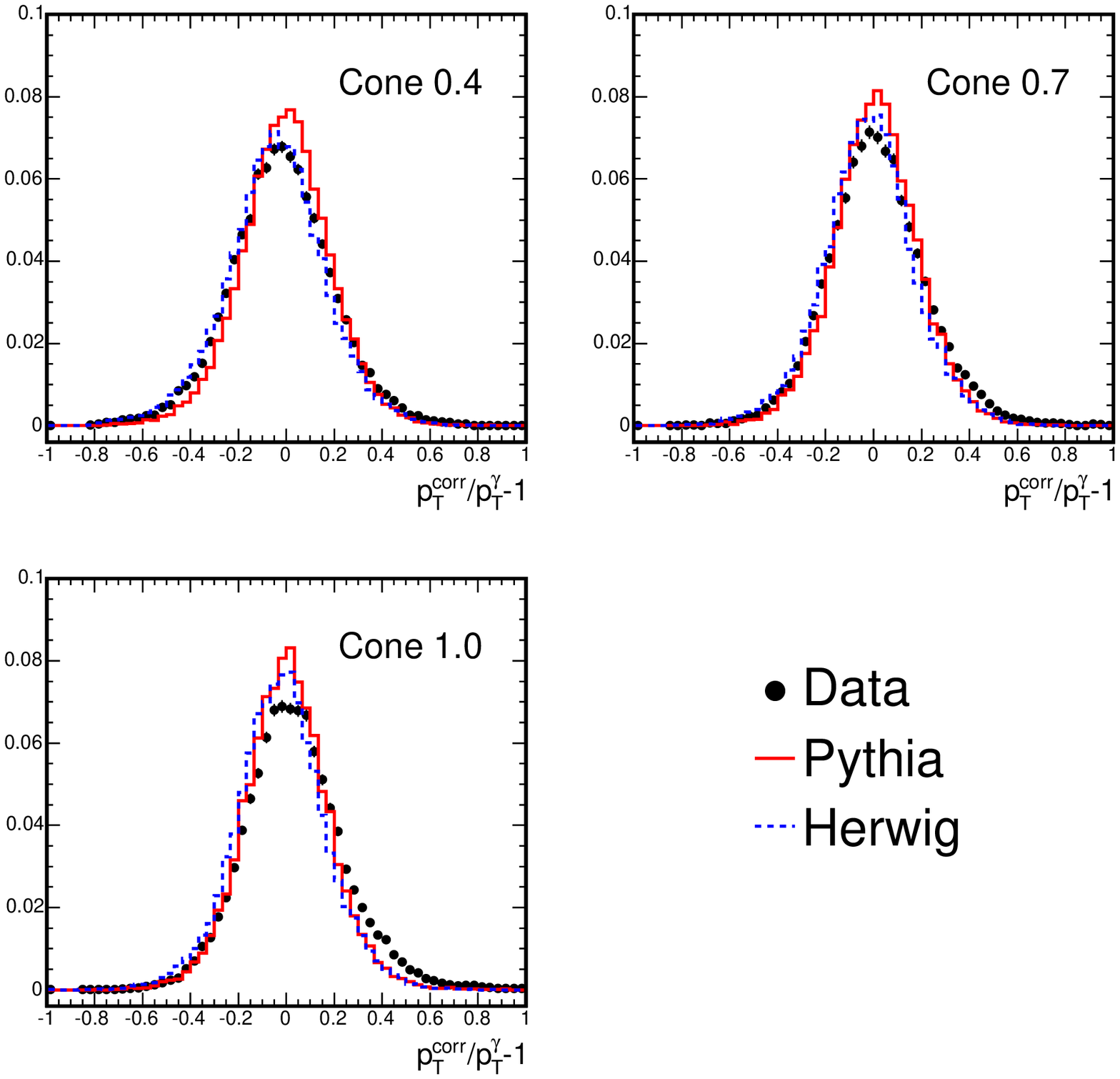}
 \caption{\sl $\gamma$-jet balance in data, {\tt PYTHIA} and {\tt HERWIG} using
  $R_{jet}$=0.4, 0.7 and 1.0 after $\eta$-dependent, absolute and OOC+UE
  corrections. \label{fig:ptbal04}}
  \end{center}
\end{figure}

After applying all corrections ($\eta$-dependent , absolute, OOC+UE)
we obtain the $\gamma$-jet balance as shown in Fig. \ref{fig:ptbal04}.
Table \ref{tab:ooccheck} contains the corresponding mean values
derived using a fit of a Gaussian to data and MC distributions for all
three cone sizes. Data and MC agree with zero to within 2\% except for
{\tt HERWIG} for a cone size of $R_{jet}$=0.4. The differences between
data and simulation are equal to those observed in Table
\ref{tab:gamjetmean}, since the OOC and UE
correction were derived from {\tt PYTHIA} and uniformly applied to all
samples.

\begin{table}[h]
\caption{Mean value of $p_T^{jet}/p_T^{\gamma} -1$ after all
corrections, including the out-of-cone energy correction for data,
{\tt PYTHIA}, and {\tt HERWIG} for jet cones of $R_{jet}=0.4$, $0.7$ and
$1.0$.}
\begin{center}
\renewcommand{\arraystretch}{1.4}
\setlength\tabcolsep{5pt}
\begin{tabular}{|l|r|r|r|}
\hline
\noalign{\smallskip}
Sample & $R_{jet}$=0.4  & $R_{jet}$=0.7  & $R_{jet}$=1.0  \\
\hline
Data        &  $-0.019 \pm 0.001$   &  $0.010 \pm 0.001$      & $0.024 \pm 0.001$  \\
{\tt PYTHIA}  &  $-0.001 \pm 0.001$   &  $0.011 \pm 0.001$      & $0.000 \pm 0.001$  \\
{\tt HERWIG}  &  $-0.040 \pm 0.001$   &  $-0.018 \pm 0.001$      & $-0.023 \pm 0.001$  \\

\hline
\end{tabular}
\end{center}
\label{tab:ooccheck}
\end{table}

\subsubsection{Using $Z$-jet Events}
\label{sec:zjet}

Another excellent calibration sample are $Z\to l^+l^-$ events where
the $p_T$ of the $Z$ boson provides a reference scale for the jet. The
advantage compared to the $\gamma$-jet sample is that it is nearly
free from background contamination, at the expense of smaller
statistics. For this study we require the jet and the $Z$ boson to be
back-to-back, $\Delta\phi({\rm jet},Z)>3$ radians, and no extra jets with
$p_T>3$ GeV/$c$ and $|\eta|<2.4$.

\begin{figure}[htbp]
  \begin{center}
   \includegraphics[width=\linewidth,clip=]{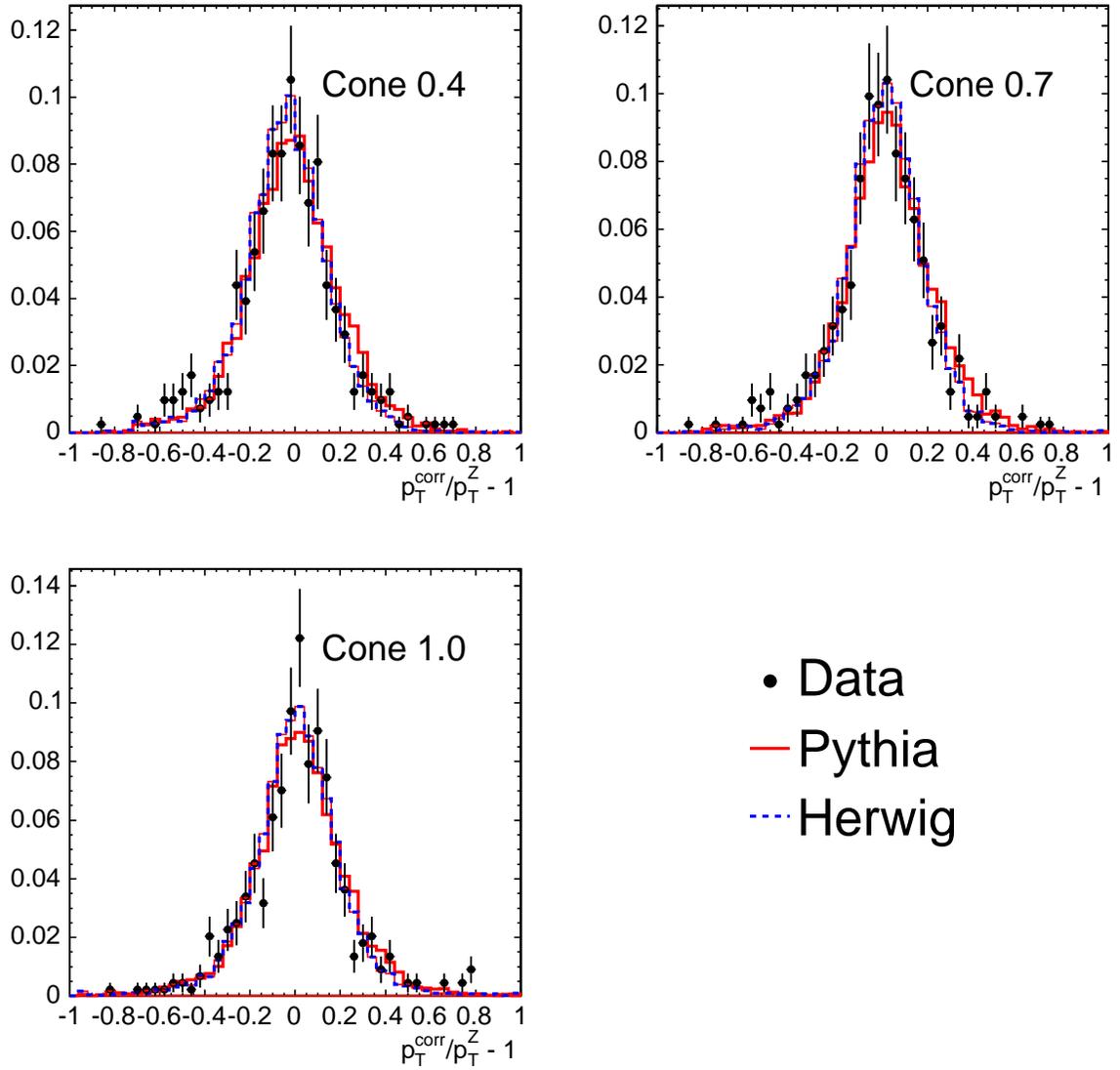}
  \caption{\sl $Z$-jet balance in data (closed
  black circles), {\tt PYTHIA} (solid red line) and {\tt HERWIG} (dashed blue
  line) for $R_{jet}=0.4$ (top left), $R_{jet}=0.7$ (top right) and
  $R_{jet}=1.0$ (bottom left) after all corrections. \label{fig:zjetbal}}
  \end{center}
\end{figure}

In Fig. \ref{fig:zjetbal} we compare the $Z$-jet balance,
$p^{jet}_T/p^{Z}_T-1$ in $Z\to e^+e^-$ and $Z\to \mu^+\mu^-$ events
for data, {\tt PYTHIA} and {\tt HERWIG} after all corrections.

The mean values derived from fits of Gaussians to the distributions
between $-0.3$ and $+0.3$ are given in Table~\ref{tab:zgauss}.  They
are reasonably close to $0$ for all cone sizes, and the data agree
with the MC to within the statistical uncertainties of 1$\%$.

\begin{table}[h]
\caption{Measured mean value of $p_T^{jet}/p_T^Z -1$ after all
corrections, including the out-of-cone energy correction for data and
{\tt PYTHIA} calorimeter jets. The result is given for $R_{jet}= 0.4$,
$0.7$ and $1.0$.}
\begin{center}
\renewcommand{\arraystretch}{1.4}
\setlength\tabcolsep{5pt}
\begin{tabular}{|l|r|r|r|}
\hline
\noalign{\smallskip}
Sample & $R_{jet}=$0.4 & $R_{jet}=$0.7 & $R_{jet}=$1.0 \\
\hline
Data              &  $-0.026 \pm 0.009$   &  $0.007 \pm 0.009$      & $0.013 \pm 0.009$  \\
{\tt PYTHIA}      &  $-0.016 \pm 0.003$   &  $0.019 \pm 0.003$      & $0.015 \pm 0.003$  \\
{\tt HERWIG}      &  $-0.032 \pm 0.003$   &  $-0.011 \pm 0.002$      & $-0.009 \pm 0.003$  \\
\hline
\end{tabular}
\end{center}
\label{tab:zgauss}
\end{table}

\subsubsection{Using Dijet Events}
\label{sec:dijet}

The $\eta$-dependent corrections have been obtained from the {\tt PYTHIA}
dijet and jet data samples. As explained in Sec. \ref{sec:RelCorr},
the data and MC have different $\eta$-dependent corrections. This
section describes the choice of the dijet balancing technique as the
$\eta$-dependent correction method and the use of only {\tt PYTHIA} MC to
correct all MC samples.

A different approach to address the $\eta$-dependence of jet response
is the so-called ``Missing $E_T$ Projection Fraction'' (MPF)
method. This approach was used in Run I by the CDF \cite{run1jets} and
DZero \cite{d0nim} experiments. The MPF is defined as
\begin{equation}
MPF =\frac{\overrightarrow{\met}\cdot\overrightarrow{p_T^{probe}}}
{(p_T^{probe}+p_T^{trigger})/2}.
\label{eqn:mpf}
\end{equation}
where the vector of the missing transverse energy, $\met$, is used to
quantify the difference between $p_T^{probe}$ and $p_T^{trigger}$
rather than using $p_T^{probe}-p_T^{trigger}$ as is done in the dijet
balancing method (see Sec.~\ref{sec:RelCorr}). In an ideal dijet
production process with no gluon radiation and fragmentation effects,
the MPF and the dijet balance methods are equivalent. That is,
\begin{equation}
  \beta(MPF) \equiv
  \frac{2-\langle MPF \rangle}{2+\langle MPF \rangle}
  =  \frac{2+\langle f_b \rangle}{2-\langle f_b \rangle}
  \equiv\beta_{dijet}.
\end{equation}
However, due to QCD radiation and out-of-cone energy losses and
underlying event contributions this does not exactly hold. In
contrast to the dijet balancing method, the MPF method does not
correct for OOC energy. The reason is that $\met$ used in the MPF
method is only affected by energy mismeasurement and has no
sensitivity to the energy flow between inside and outside the jet
cone. On the other hand, the dijet balancing method is sensitive
to the energy inside the jet cone and will thus implicitly correct for
an $\eta$-dependence of the OOC and UE effects. Since we
estimate these corrections and systematic uncertainties only in the central 
region, we choose to use the dijet balancing method as the primary correction method.

\begin{figure}[htbp]
 \begin{center} 
   \includegraphics[width=7cm]{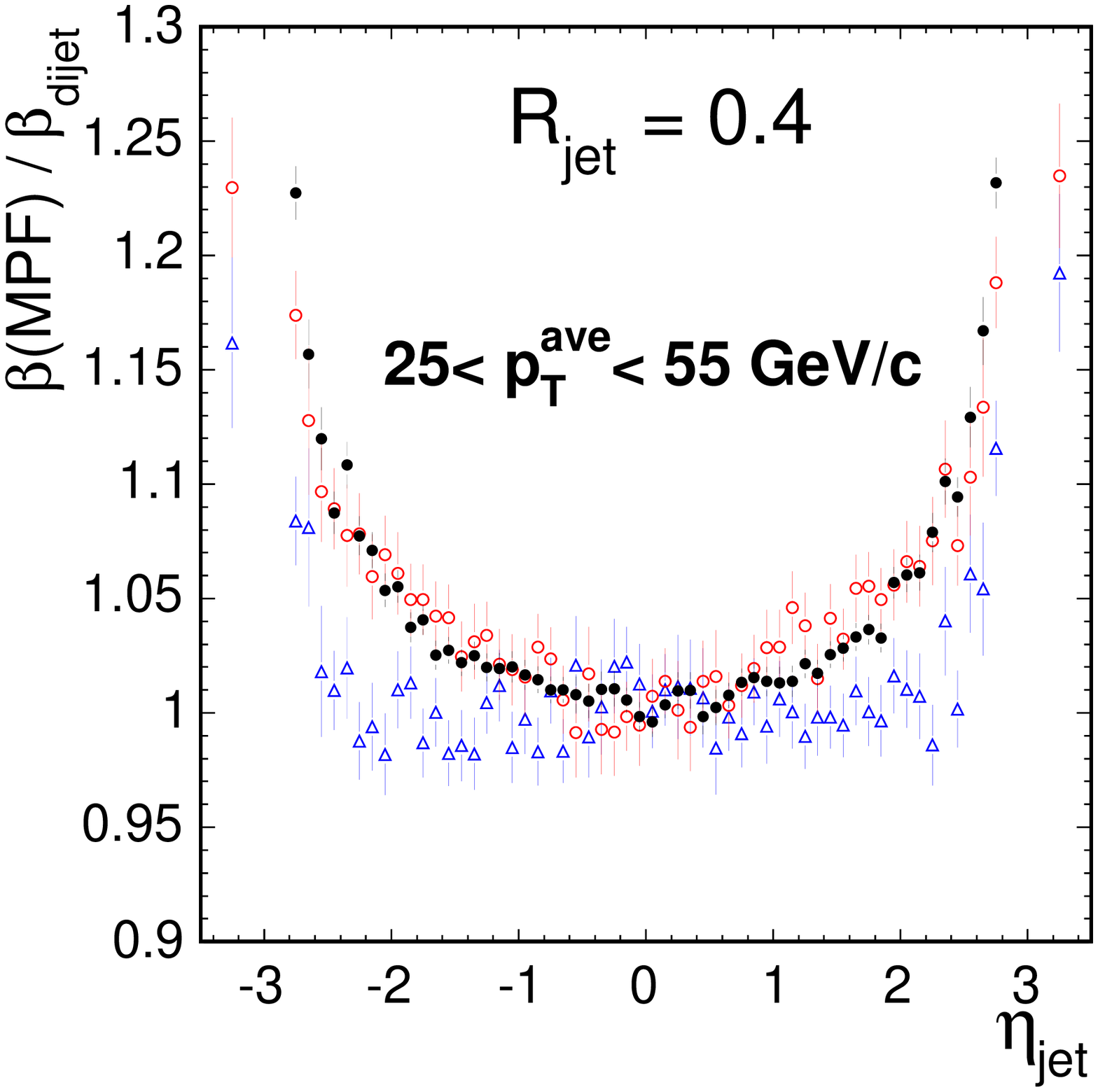} 
   \includegraphics[width=7cm]{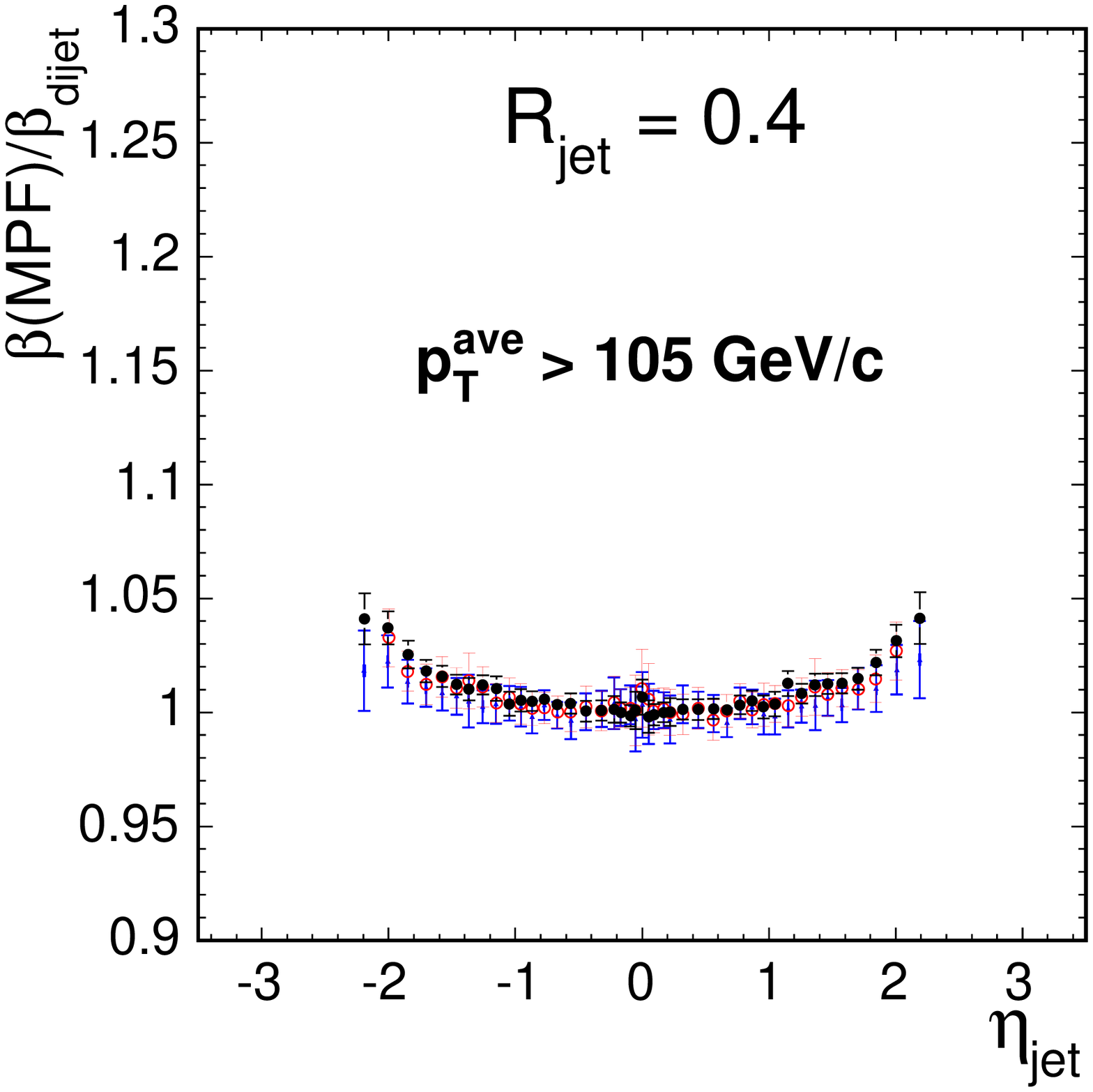}
   \caption{$\beta(MPF)/\beta_{dijet}$ as a function of $\eta_{jet}$
     for $R_{jet}=0.4$ jets in data (circles), {\tt PYTHIA} (upward
     triangles) and {\tt HERWIG} (downward triangles) for
     $25<p_T^{ave}<55$ GeV/$c$ and $p_T^{ave}>105$ GeV/$c$.}
   \label{mpfdpt}
 \end{center}
\end{figure}

However, we use the MPF method to further investigate the discrepancy
between {\tt PYTHIA} and {\tt HERWIG} in Figures
\ref{relplot_datamc_04}-\ref{relplot_datamc_10} in
Sec. \ref{sec:RelCorr}. The {\tt HERWIG} measurements are
systematically higher than {\tt PYTHIA} and data by about 10\% at
$25<p_T^{ave}<55$ GeV/$c$ and $|\eta|>0.6$ and agree very well at
higher $p_T^{ave}$ GeV/$c$. The discrepancy is larger for
$R_{jet}$=0.4 jets than for $R_{jet}$=0.7 and 1.0. To shed more light
on the origin of the discrepancy we compare the ratio
$\beta(MPF)/\beta_{dijet}$ as a function of $\eta_{jet}$ and for two
different values of $p_T^{ave}$ in Fig. \ref{mpfdpt}. In the central
region $\beta(MPF)/\beta_{dijet}$ is consistent with unity at both
values of $p_T^{ave}$. In the forward region $\beta(MPF)$ increases
with respect to $\beta_{dijet}$.
For $25<p_T^{ave}<55$ GeV/$c$ the data are well modeled by {\tt PYTHIA} MC
but large discrepancies are observed in comparison to {\tt HERWIG} MC. 
At high $p_T^{ave}>105$ GeV/$c$ the data are in good agreement with both
MC generators. 
Since the ratio $\beta(MPF)/\beta_{dijet}$ is largely independent of the
CDF jet energy scale we conclude that the observed disagreement is not due to
any residual problems of the CDF simulation but must originate from
a difference in the underlying physics between {\tt HERWIG}, {\tt PYTHIA}
and data for low $p_T$ dijet production.

\begin{figure}[htbp]
 \begin{center}
 \includegraphics[width=0.6\hsize]{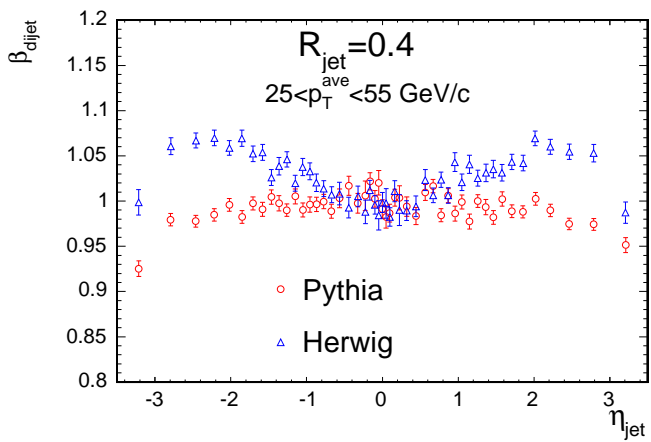}
 \caption{\sl $\beta_{dijet}$ for particle jets as a function of jet
 $\eta_{jet}$ for {\tt PYTHIA} (open circles) and {\tt HERWIG} (open
 triangles) for $\hat{p}_T>18$ GeV/$c$ and for $R_{jet}=0.4$.
 \label{fig:jetbal_pyhw_hadron}}
 \end{center}
\end{figure}

We also compare the $\beta_{dijet}$ for particle-jets with
$R_{jet}=0.4$ between {\tt PYTHIA} and {\tt HERWIG} in Figure
\ref{fig:jetbal_pyhw_hadron} for $\hat p_T>18$ GeV/$c$ and find that
for {\tt PYTHIA} it is independent of $\eta_{jet}$ while for {\tt
HERWIG} it rises with increasing $\eta_{jet}$.

Since this behavior is only found in the dijet samples, we do not
consider {\tt HERWIG} dijet samples for the determination of the
$\eta$-dependent corrections or their systematic uncertainties. In
$\gamma$-jet (see Fig. \ref{fig:gamjeteta}), $Z$-jet or $t\bar t$
events no such problems are seen. At this moment we do not have any
explanation for the differences. It could be due to initial of final
state radiation, due to the underlying event modeling or many other
effects, and it will be studied again in future versions of the
generators.

\subsubsection{Using $\wjj$ Decays in $\ttb$ Events}

The jet energy scale can be studied using the hadronic decay of
resonances with well-measured masses such as the $W$ and $Z$
bosons. Unfortunately, the decays of $W$ and $Z$ bosons to jets are
swamped by multijets QCD background in hadron collider
environments. One solution is to study hadronic $W$ boson decays
inside $\ttb$ events which have relatively small background
contamination.  This section summarizes the application of this
technique. For complete details, see \cite{jf_thesis}.

At the Tevatron, top quarks are produced primarily as top pairs and
decay to $W$ bosons and $b$ quarks nearly 100\% of the time. The $W$
bosons in turn decay into lepton-neutrino ($l \nu$) or quark pairs
($q\bar{q^\prime}$). This measurement uses the ``lepton+jets'' channel
of $\ttb$ candidates in which only one of two $W$ bosons decays to $l
\nu$ while the other decays to quark pairs.  The lepton+jets events
are selected by requiring one well-identified electron or muon, large
($\met$) due to the neutrino from the $W$ decay and at least four jets
in the final state.  The missing transverse energy, $\met$ is measured
by the imbalance in the calorimeter transverse energy and is required
to be greater than 20 GeV.  Jets are reconstructed with a radius
$R_{jet}=0.4$. The sample is divided into four subsamples with various
sensitivities for better performance. First, the events are separated
based on the number of jets that are $b$-tagged in the event. The
SECVTX \cite{secvtx} algorithm based on the identification of
secondary vertices inside jets is used to tag $b$-jets. Events with
2-,1- and 0-tag are considered separately.  Furthermore, events with
1-tag are separated based on the fourth jet $E_T$ threshold. Events in
the 1-tag(T) category have 4 jets with $E_T>$ 15 GeV, while events in
the 1-tag(L) category have 3 jets with $E_T>$ 15 GeV and the 4th jet
with $8<E_T<15$ GeV.

\begin{figure}[htbp]
\center
      \leavevmode
    \resizebox{14.0cm}{10.0cm}{\includegraphics{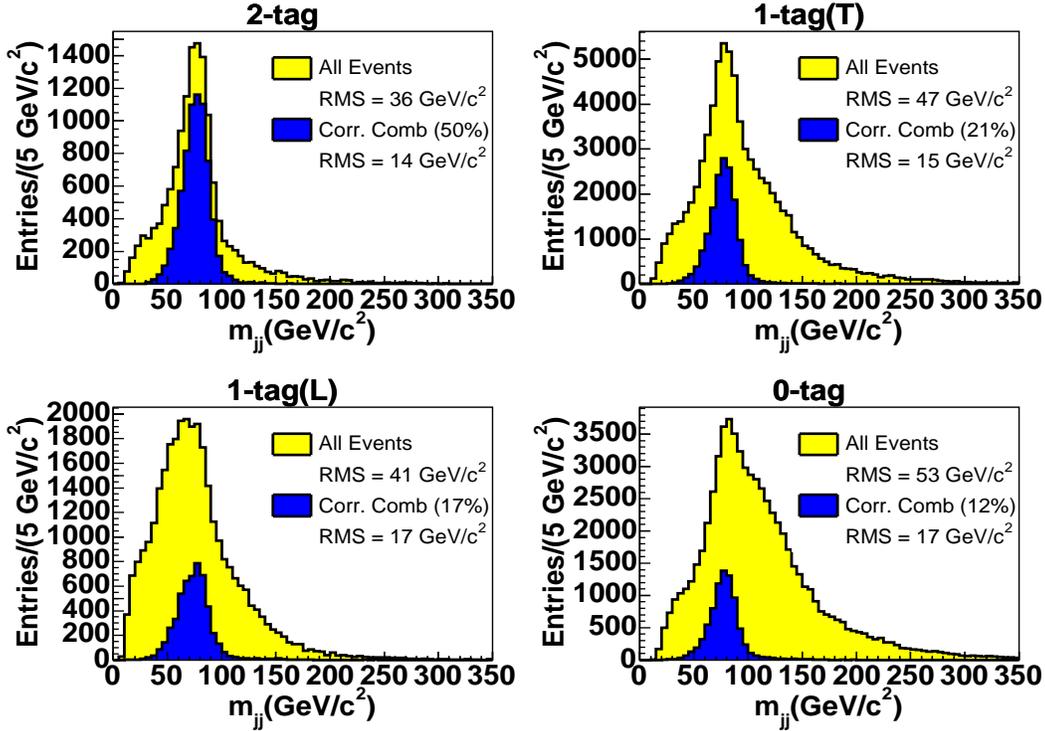}}
\caption{\label{f_defaultwtemplate} Reconstructed hadronic $W$ boson
mass from {\tt HERWIG} $\ttb$ events with $\mtop$ = 178 GeV/c$^2$ for \dt\
events (upper-left), \stt\ events (upper-right), \stl\ events
(bottom-left) and \nt\ events (bottom-right). The yellow (outer)
histograms show the mass distributions for all combinations and the
blue (inner) histograms show the distributions only for the correct
jet-parton assignments. }
  \end{figure}

Before reconstructing the invariant mass of hadronically decaying $W$
bosons ($\mjj$), we apply the $\eta$-dependent, and absolute
corrections to jet energies.  In addition, corrections specific to
light quark jets from $W$ boson decays in $\ttb$ events are applied.
To reconstruct $\mjj$, one has to know which of the jets in the final
state comes from the $W$ boson decay. This problem is dealt with by
considering all the dijet combinations that can be made using the jets
that are not $b$-tagged. Only the four highest $E_T$ jets are
considered. Consequently, there can be more than one mass per event
that are considered. There are in fact 1, 3, and 6 $\mjj$ per event
for the \dt, \st\ and \nt\ subsamples, respectively.  The distribution
of $\mjj$ for {\tt HERWIG} $\ttb$ events is shown in Fig.
\ref{f_defaultwtemplate} for each event category (with a top quark
mass ($\mtop$) of \gevcc{178}).  The mass resolution improves with the
number of $b$-tagged jets present in the event.

\begin{figure}[htbp]
\center
      \leavevmode
    \resizebox{12.0cm}{10.0cm}{\includegraphics{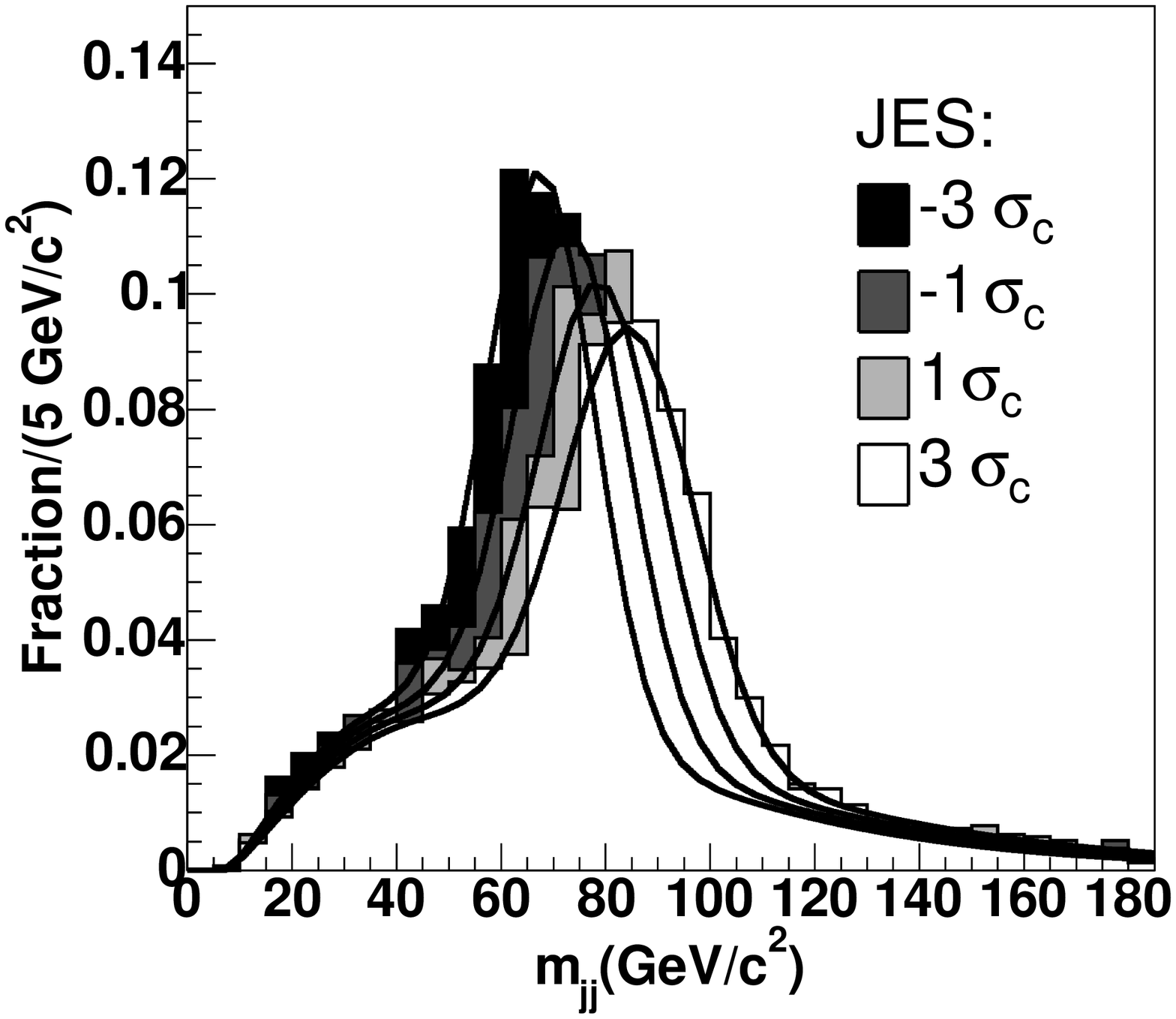}}
\caption{\label{f_templates_prd} Four $\mjj$ signal templates are
shown for jet energy scale (JES) values ranging from -3$\sigma_c$ to
+3$\sigma_c$. Overlaid are the fitted parameterizations at each value
of jet energy scale.}  \end{figure}

Distributions of $\mjj$ are constructed from {\tt HERWIG} $\ttb$ Monte
Carlo with $\mtopdef$ (corresponding to the central value of the
Tevatron Run I average) with jet energy scale values ranging from -3
to +3$\sigma_c$, where $\sigma_c$ is the total jet energy scale
uncertainty defined in Sec. \ref{sec:sumsyst} of this document.
Smooth probability density functions are obtained by fitting the mass
distributions as a function of $\mtop$ and jet energy scale using an
analytical function.  Figure \ref{f_templates_prd} shows the $\mjj$
distribution for various jet energy scale values for the \dt\
subsample with the fitted templates overlaid.  Templates for
background events are obtained from $W$+jets, QCD multijets and
single-top MC events.

The fitted jet energy scale is obtained by comparing the reconstructed
mass distributions obtained in the data with the signal and background
templates using an unbinned likelihood fit. The data used in this
measurement corresponds to the $W\to e\nu_e$ and $W\to \mu\nu_{\mu}$
but additionally requiring at least 4 jets and $\met$ greater than
20~GeV. The systematic uncertainties in this measurement arises from
the MC modeling of signal and background events that we use to create
the templates and thus extracting the jet energy scale.  We consider
uncertainties in the top quark mass ($\pm$ \gevcc{5}), amount of
initial and final state gluon radiation, parton distribution
functions, background mass shape and general MC modeling. The total
systematic uncertainties corresponds to 0.68 $\sigma_c$.

\begin{figure}[t]
\center
      \leavevmode
      \resizebox{14.0cm}{11.0cm}{\includegraphics{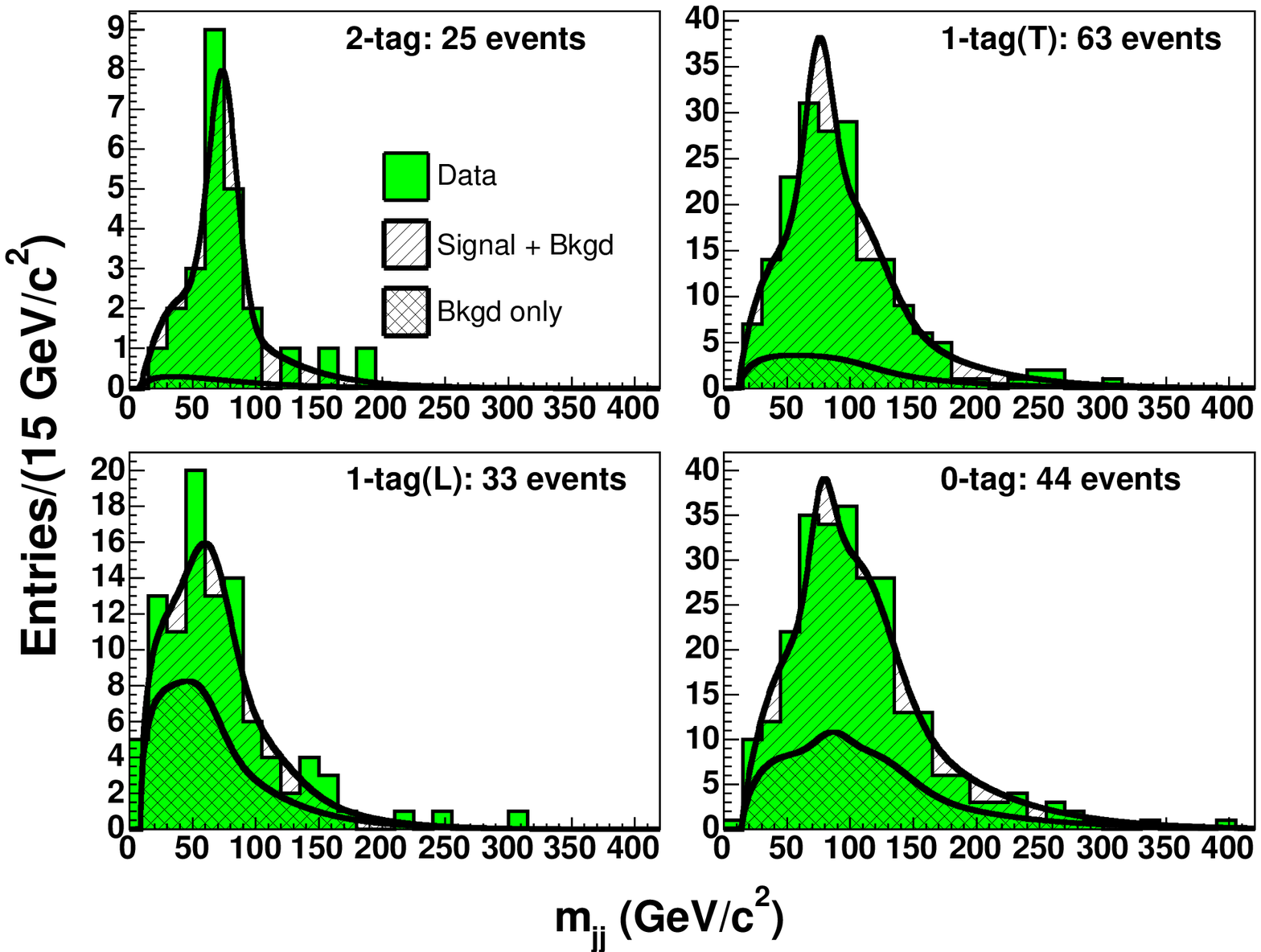}}
\caption{\label{f_data_mjj_xcheck} Data $\mjj$ distributions for the \dt\ (upper-left),
\stt\ (upper-right), \stl\ (lower-left) and \nt\ (lower-right) subsamples. The signal
and background template shapes corresponding to the best fit of the
jet energy scale cross-check are overlaid on the histograms. The value of $\mtop$
has been constrained to \gevcc{178}.}
\end{figure}

This measurement is performed using 318 \pb\ of data that results in a
total of 165 events in the lepton+jets sample. The application of the
likelihood fit to the data yields $-0.76 \pm 1.00\ (stat.)\
\sigma_c$. This means that the data and simulation of $t\bar t$ events is in 
agreement in one $\sigma_c$. By adding the systematic uncertainties,
this result changes to $-0.76 \pm 1.27\ \sigma_c$.  The $\mjj$
distributions reconstructed in the data are shown in
Fig.~\ref{f_data_mjj_xcheck}. The shape of the signal and background
MC templates corresponding to the best fit are overlaid on top of the
histograms. We conclude that the average jet energy scale as
determined by $\wjj$ decays is in good agreement with the nominal jet
energy scale of the CDF MC simulation.

The calibration of the jet energy scale with $\wjj$ decays has been
used to measure precisely the top quark mass in Run II
\cite{mtop_prd_prl}. We note that jet energy scale uncertainties
obtained from this technique are mostly statistical and will improve
as more data is accumulated.

\subsection{Test of the Uncertainties}
\label{sec:systest}

\begin{figure}[htbp]
  \begin{center}
  \includegraphics[width=\linewidth,clip=]{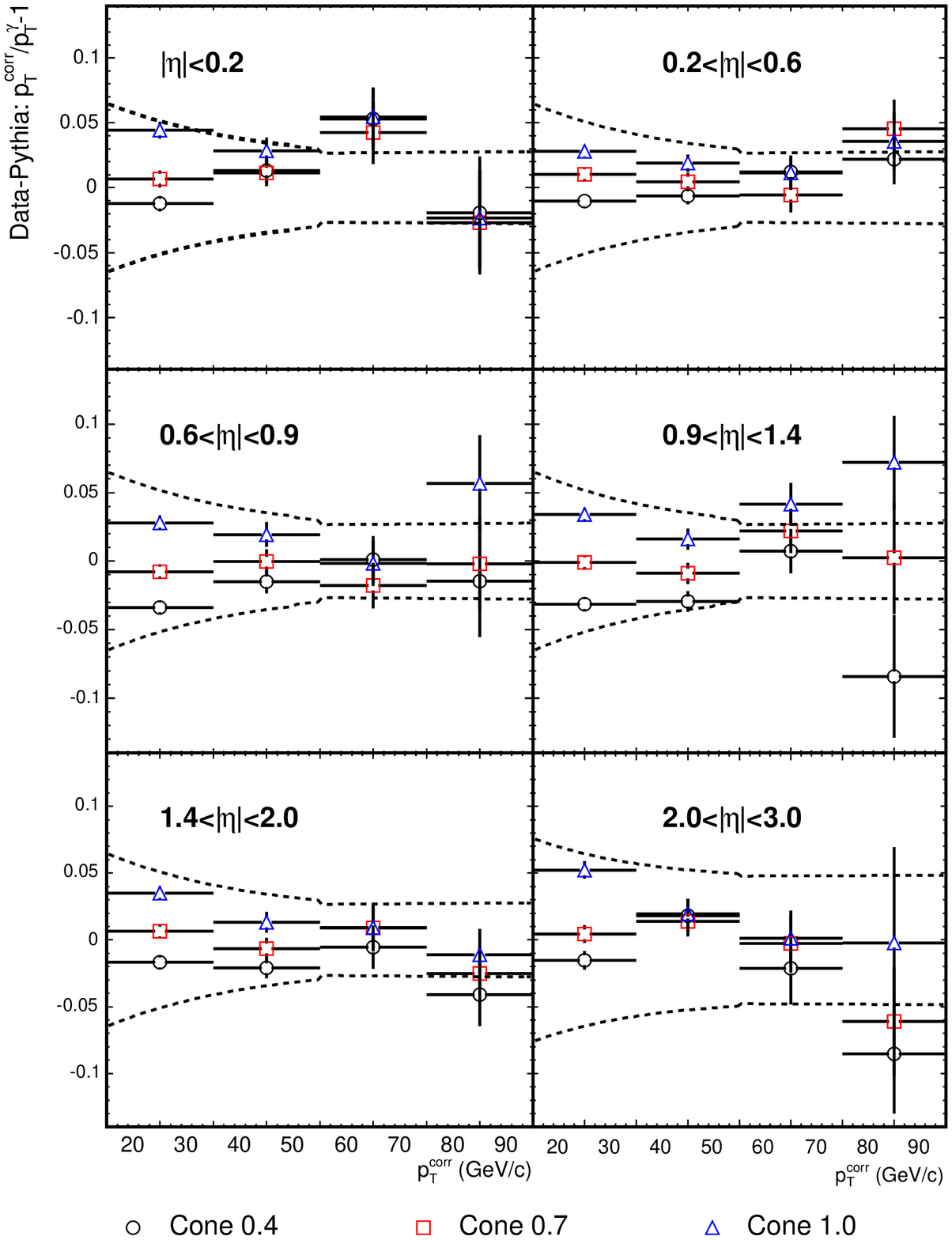}
  \caption{\sl Difference of the $\gamma$-jet balance between data and
  {\tt PYTHIA} as a function of $p_T^\gamma$ in six regions of
  $\eta_{jet}$. All three cone sizes are shown: $0.4$ (blue squares),
  $0.7$ (red open circles) and $1.0$ (black triangles).  The curves
  indicate the total systematic uncertainty in each $\eta$ region.
\label{fig:gamjet_pt_py}}
  \end{center}
\end{figure}

\begin{figure}[htbp]
  \begin{center}
  \includegraphics[width=\linewidth,clip=]{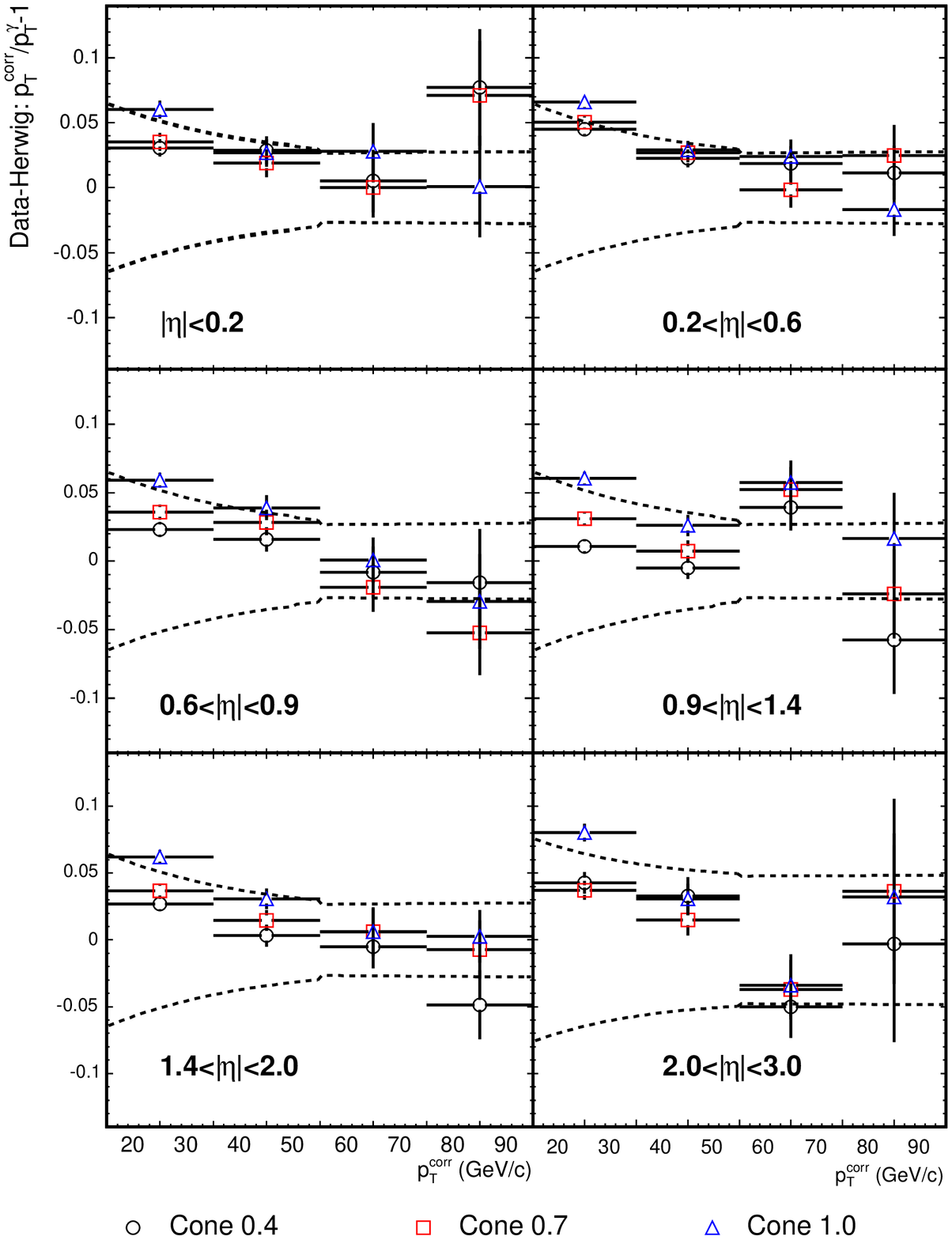}
  \caption{\sl Difference of the $\gamma$-jet balance between data and
  {\tt HERWIG} as a function of $p_T^\gamma$ in six regions of
  $\eta_{jet}$. All three cone sizes are shown: $0.4$ (blue squares),
  $0.7$ (red open circles) and $1.0$ (black triangles).  The curves
  indicate the total systematic uncertainty in each $\eta$ region.
\label{fig:gamjet_pt_hw}}
  \end{center}
\end{figure}

We test whether the agreement of data and MC is within the calculated
uncertainties for all $p_T$ and $\eta$ bins. Figure
\ref{fig:gamjet_pt_py} shows the difference of the mean values of the
$\gamma$-jet balance between data and {\tt PYTHIA} as a function of
$p_T$ and for six regions of pseudo-rapidity. Overlaid is the total
systematic uncertainty on the jet energy scale. It is seen that the
data are modeled well by the simulation at all $\eta$ and $p_{T}$, and
that any differences are covered by the quoted uncertainties. In Fig.
\ref{fig:gamjet_pt_hw}, the same comparison has been made with {\tt HERWIG},
leading to the same conclusion. The $\gamma$-jet balance for data,
{\tt PYTHIA} and {\tt HERWIG} is independent of $p_T^\gamma$ and $\eta_{jet}$
after applying all the corrections.

\subsection{Summary}

We have shown that the corrections and the systematic uncertainties
are valid for several control samples. We have found that the
transverse momentum of the jet, after all corrections, is in balance
with the transverse momentum of the $\gamma$ and the $Z$ boson in both
measured and simulated $\gamma$+jets and $Z$+jets samples,
respectively. We also determined the average jet energy scale using
$W\to jj$ decays in $t\bar t$ events and found good agreement between
data and simulation.

We have also investigated differences between the {\tt PYTHIA} and
{\tt HERWIG} MC generators. With the dijet-balancing technique we
observe rather large differences in the plug calorimeter region which
are also seen for particle jets. For the dijet process, the data do
not support the behavior of {\tt HERWIG}. This problem is only
observed in the dijet process, e.g. not in the $\gamma$-jet process.

\clearpage

\section{Summary of Systematic Uncertainties}
\label{sec:sumsyst}

We have presented the systematic uncertainties associated with the jet
energy response. The systematic uncertainties are largely independent
of the correction applied and mostly arise from the modeling of jets
by the MC simulation and by the knowledge of the response to single
particles.

\begin{figure}[h]
  \begin{center} \includegraphics[width=\linewidth,clip=]{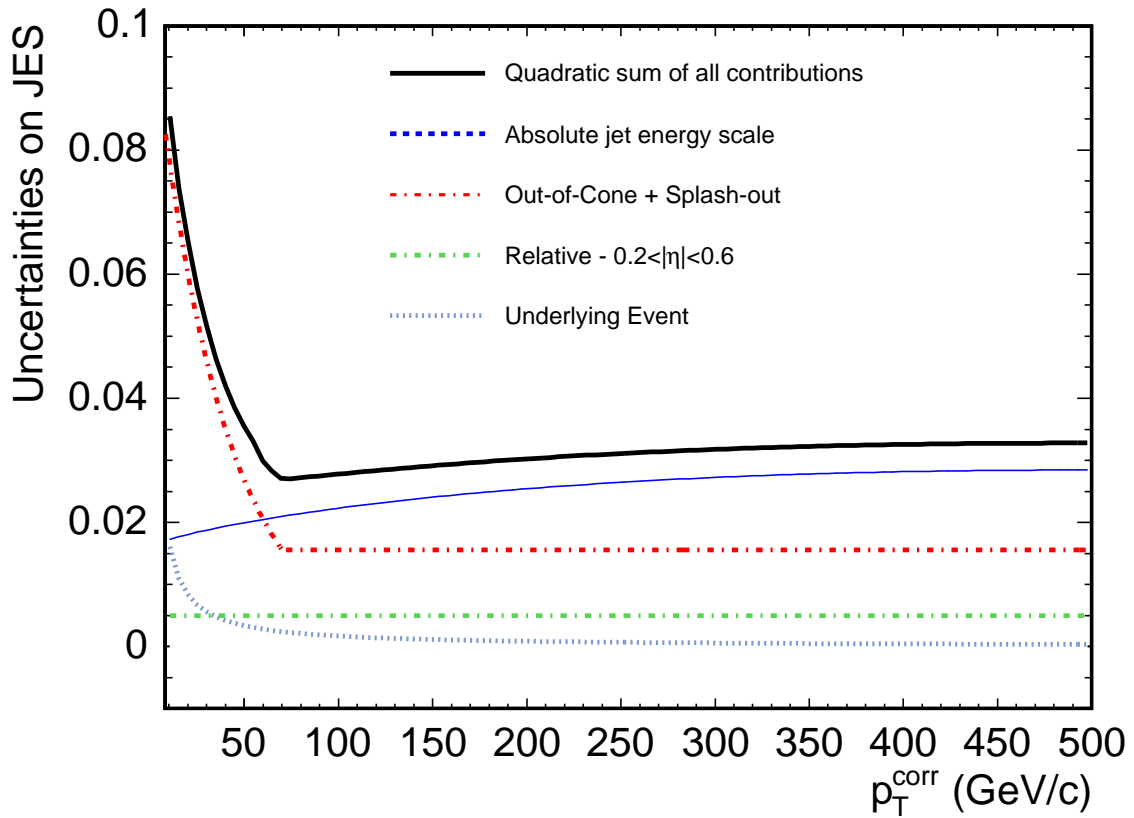}
  \caption{\sl Systematic uncertainties as a function of the corrected
  jet $p_{T}$ in 0.2$<|\eta|<$0.6. \label{tot1}} \end{center}
\end{figure}

Figure \ref{tot1} shows the individual systematic uncertainties as a
function of jet $p_T$ in the central region, of the calorimeter,
$0.2<|\eta|<0.6$, of the calorimeter. They are independent and thus
added in quadrature to derive the total uncertainty.

For $p_T>60$ GeV/$c$ the largest contribution arises from the absolute
jet energy scale which is limited by the uncertainty of the
calorimeter response to charged hadrons. A further reduction of the
systematic uncertainties can be achieved by improving the tuning of
the simulation, and by including {\it in situ} single track data which
recently became available, replacing test beam data used so far in the
momentum region 7-20 GeV/$c$ and probably beyond.

At low $p_T$ the largest uncertainty arises from the out-of-cone
energy which can be improved by further studying differences between
the data and the predictions of {\tt PYTHIA} and {\tt HERWIG}, and by
optimizing the fragmentation and underlying event model of both
generators.

Additional uncertainties arise from the fragmentation models, the
stability of the calorimeter calibration and the underlying event
modeling.

\clearpage

\section{Conclusions}\label{sec:Conclusions}

We have determined a set of corrections to estimate the parton energy
from the jet energy measured in the Collider Detector at Fermilab.
The calibration is based on data taken between 2001 and 2004 at the
Tevatron $p\bar{p}$ collider, corresponding to an integrated
luminosity of about $350$ pb$^{-1}$, and on test beam data.

These corrections involve several steps and for each step a systematic
uncertainty is determined. Both the central and forward components of
the calorimeter are calibrated using test beam and {\it in situ}
data. The response of jets in the forward calorimeter is calibrated
with respect to that of the central calorimeter. The shower simulation
is in particular tuned in detail to the data in the central rapidity
region. Using several MC generators it has been shown to provide a
good description of the energy response of various physics
processes. The MC simulation is used to derive a correction for the
calorimeter jet energy response in the central region. Further
corrections are made for multiple $p\bar p$ interactions, the
underlying event and the fractional energy of the parton that is not
contained within the jet cone. Finally, we have verified that the
corrected jet energy is a good measure of the initial parton energy
using prompt photon and $Z$ events and have shown that the various MC
generators provide a good description of the data within the quoted
systematic uncertainties.

The total systematic uncertainty on the jet energy scale varies
between 8\% at low jet $p_T$ and 3\% at high jet $p_T$. The dominant
sources of systematic uncertainty are the uncertainty on the test beam
measurements at high energy and the uncertainty in modeling the energy
flow around the jet cone.

\end{document}